\UseRawInputEncoding
\documentclass[10pt]{article}
\usepackage{amsmath,amssymb,cite}
\usepackage{appendix}
\usepackage{feynmp}
\usepackage{float}
\usepackage{inputenc}
\usepackage[vcentermath,enableskew]{youngtab}
\usepackage{tabularx}
\usepackage[english]{babel}
\usepackage{graphicx}
\usepackage{indentfirst}
\usepackage{epsfig}
\usepackage{epstopdf}
\usepackage{hyperref}
\usepackage[section]{placeins}
\usepackage[stable]{footmisc}
\usepackage{slashed}
\usepackage{fancyhdr}
\numberwithin{equation}{section}

\fancypagestyle{plain}{\fancyhead{}
}

\begin{document}

\title{
  On the AdS/CFT correspondence and quantum entanglement 
}

\author{Badis Ydri\\
Department of Physics, Faculty of Sciences, Annaba University,\\
 Annaba, Algeria.
}

\maketitle

\begin{abstract}
String theory provides one of the most deepest insights into quantum gravity. Its single most central and profound result is the gauge/gravity duality, i.e. the emergence of gravity from gauge theory. The two examples of M(atrix)-theory and the AdS/CFT correspondence, together with the fundamental phenomena of quantum entanglement, are of paramount importance to many fundamental problems including the physics of black holes (in particular to the information loss paradox), the emergence of spacetime geometry  and to the problem of the reconciliation of general relativity and quantum mechanics. In this article an account of the AdS/CFT correspondence and the role of quantum entanglement in the emergence of spacetime geometry using strictly the language of quantum field theory is put forward.  
\end{abstract}

\tableofcontents

\section{Introduction}
The goal in this chapter is to provide a pedagogical presentation of the celebrated AdS/CFT correspondence adhering mostly to the language of quantum field theory (QFT). This is certainly possible, and perhaps even natural, if we recall that in this correspondence we are positing that quantum gravity in an anti-de Sitter spacetime $AdS_{d+1}$ is nothing else but a conformal field theory ($CFT_d$) at the boundary of AdS spacetime. Some of the reviews of the AdS/CFT correspondence which emphasize the QFT aspects and language include  Kaplan \cite{kaplan}, Zaffaroni \cite{Zaffaroni:2000vh} and Ramallo \cite{Ramallo:2013bua}. 

This chapter contains therefore a thorough introductions to conformal symmetries, anti-de Sitter spacetimes, conformal field theories and the AdS/CFT correspondence. The primary goal however in this chapter is the holographic entanglement entropy. In other words, how spacetime geometry as encoded in Einstein's equations in the bulk of AdS spacetime can emerge from the quantum entanglement entropy of the CFT living on the boundary of AdS. 

A sample of the original literature for the holographic entanglement entropy is \cite{Lashkari:2013koa,Faulkner:2013ica,VanRaamsdonk:2016exw,Casini:2011kv}. However, a very good concise and pedagogical review of the formalism relating spacetime geometry to quantum entanglement due to Van Raamsdonk and collaborators is found in \cite{VanRaamsdonk:2016exw} and \cite{Jaksland:2017nqx}.
\section{Conformal symmetry}

\subsection{The conformal groups $SO(p+1,q+1)$} 

We assume a spacetime with signature  $(-1,...,+1,...)$ with $q$ minus signs and $p$ plus signs where $q+p=d$ is the dimenion of spacetime.

We have then $q=0$ and $p=d$ for Euclidean, and $q=1$ and $p=d-1$ for Lorentzian.

We start with the group of diffeomorphisms, i.e. the group of general coordinate transformations. The conformal group is a subgroup of the diffeomorphism group which preserves the conformal flatness of the metric. The corresponding transformations preserve the angles but change the lengths.

A manifold is called conformally flat if the metric takes the following form
\begin{eqnarray}
ds^2=\exp(\omega(x))dx^{\mu}dx_{\mu}.
\end{eqnarray} 
Conformal transformations are given by
\begin{eqnarray}
x^{\mu}\longrightarrow x^{'\mu}(x)~,~dx^{'\mu}dx^{'}_{\mu}=\Omega^{2}(x)dx_{\mu}dx^{\mu}.\label{conf}
\end{eqnarray} 
Thus if the manifold is conformally flat, it will remain so under conformal transformations. These conformal transformations are generalization of the scale transformation 
\begin{eqnarray}
x^{\mu}\longrightarrow x^{'\mu}=\alpha x^{\mu}~,~dx^{'\mu}dx^{'}_{\mu}=\alpha^{2}dx_{\mu}dx^{\mu}.
\end{eqnarray} 
We consider infinitesimal conformal transformations $x^{'}_{\mu}=x_{\mu}+v_{\mu}$, $\Omega=1+\omega/2$. We get immediately from (\ref{conf}) the conditions
 
\begin{eqnarray}
\omega=\frac{2}{d}\partial_{\mu}v^{\mu}.
\end{eqnarray} 
 \begin{eqnarray}
\partial_{\nu}v^{\mu}+\partial^{\mu}v_{\nu}-\frac{2}{d}\partial_{\rho}v^{\rho}\eta^{\mu}_{\nu}=0.
\end{eqnarray} 
In $d=2$ this equation admits an infinite number of solutions and thus the conformal group in two dimensions is infinite dimensional. 

In $d\neq 2$ there is a finite number of solutions given precisely by 
\begin{eqnarray}
v_{\mu}=\delta x^{\mu}=a^{\mu}+\omega^{\mu}~_{\nu}x^{\nu}+\lambda x^{\mu}+b^{\mu} x^2-2x^{\mu} bx.
\end{eqnarray} 
The conformal group contains therefore:
\begin{enumerate}
\item Lorentz transformations $\Lambda\in SO(p,q)$, i.e. rotations and boosts, with parameters $\omega^{\mu}~_{\nu}$. The finite transformations are $x\longrightarrow x^{\prime}=\Lambda x$. There are $d(d-1)/2$ generators denoted by $M_{\mu\nu}$. They satisfy
\begin{eqnarray}
[M_{\mu\nu},M_{\rho\sigma}]=i\big(\eta_{\nu\rho}M_{\mu\sigma}+\eta_{\mu\sigma}M_{\nu\rho}-\eta_{\mu\rho}M_{\nu\sigma}-\eta_{\nu\sigma}M_{\mu\rho}\big).
\end{eqnarray} 
\item Translations with parameters $a^{\mu}$. There are $d$ generators denoted by $P^{\mu}$.
\item The scale (dilatation) transformation $x^{\mu}\longrightarrow \lambda x^{\mu}$ where $\lambda$ is a constant. The generator of dilatation is denoted by $D$. A field theory which is invariant under scale transformations will also, under mild conditions, be invariant under all conformal transformations.
\item The special conformal transformations with parameters $b^{\mu}$ given by 
 \begin{eqnarray}
\delta x^{\mu}=b^{\mu} x^2-2x^{\mu} bx.
\end{eqnarray} 
A special conformal transformation is obtained from the composition of an inversion, a translation by a vector $b^{\mu}$ and another inversion. The inversion is an element in the conformal group which is not connected to the identity given by
  \begin{eqnarray}
x^{\mu}\longrightarrow \frac{x^{\mu}}{x^2}.
\end{eqnarray} 
There are $d$ generators of special conformal transformations denoted by $K^{\mu}$. The finite form of special conformal transformations is 
 \begin{eqnarray}
x^{\mu}\longrightarrow x^{\prime \mu}=\frac{x^{\mu}+b^{\mu}x^2}{1+2bx+b^2x^2}\Rightarrow x\longrightarrow x^{\prime 2}=\frac{x^2}{1+2bx+b^2x^2}.
\end{eqnarray} 
\end{enumerate}

Altogether we have then $(d+1)(d+2)/2$ conformal transformations. This is exactly the number of generators of the rotation group $SO(d+2)$. However, the conformal group must be a non-compact group. It is therefore given by $SO(p+1,q+1)$. In Lorentzian signature the conformal group is $SO(d,2)$ whereas in Euclidean signature the conformal group is $SO(d+1,1)$.  For $d>2$ the conformal group is actually given by $O(p+1,q+1)$ and consists of two disconnected components since the inversion element is not infinitesimally generated.

\subsection{Differential representation of the conformal algebra}
The conformal group in Lorentzian signature $\eta=(-1,+1,...,+1)$ is $O(d,2)$ with generators $P_{\mu}$, $K_{\mu}$, $M_{\mu\nu}$ and $D$ which satisfy the algebra:

\begin{enumerate} 
\item $M_{\mu\nu}$ generate the algebra of the Lorentz group $SO(d-1,1)$, viz
\begin{eqnarray}
[M_{\mu\nu},M_{\rho\sigma}]=i\bigg(\eta_{\mu\sigma}M_{\rho\nu}-\eta_{\mu\rho}M_{\sigma\nu}+\eta_{\sigma\nu}M_{\mu\rho}-\eta_{\rho\nu}M_{\mu\sigma}\bigg).
\end{eqnarray}
\item $D$ is a scalar under the Lorentz group, viz
\begin{eqnarray}
[M_{\mu\nu},D]=0.
\end{eqnarray}
\item $P_{\mu}$, $K_{\mu}$ are vectors under the Lorentz group, viz
\begin{eqnarray}
[M_{\mu\nu},P_{\rho}]=i(\eta_{\mu\rho}P_{\nu}-\eta_{\nu\rho}P_{\mu})~,~[M_{\mu\nu},K_{\rho}]=i(\eta_{\mu\rho}K_{\nu}-\eta_{\nu\rho}K_{\mu}).
\end{eqnarray}
\item $D$ is the Hamiltonian and $P_{\mu}$, $K_{\mu}$ are the raising and lowering operators since
\begin{eqnarray}
[D,P_{\mu}]=P_{\mu}~,~[D,K_{\mu}]=-K_{\mu}.
\end{eqnarray}
\item $P_{\mu}$, $K_{\mu}$ close on a dilatation and a Lorentz transformation, viz
\begin{eqnarray}
[P_{\mu},K_{\nu}]=2(\eta_{\mu\nu}D+iM_{\mu\nu}).
\end{eqnarray}
\end{enumerate}
We define the new generators ${\cal M}_{AB}$ by the relations
\begin{eqnarray}
{\cal M}_{AB}=M_{\mu\nu}~,~A=\mu=0,...,d-1~,~B=\nu=0,...,d-1.
\end{eqnarray}
\begin{eqnarray}
{\cal M}_{\mu d}=-{\cal M}_{d\mu}=i\frac{K_{\mu}-P_{\mu}}{2}~,~\mu=0,...,d-1.
\end{eqnarray}
\begin{eqnarray}
{\cal M}_{\mu d+1}=-{\cal M}_{d+1\mu}=-i\frac{K_{\mu}+P_{\mu}}{2}~,~\mu=0,...,d-1.
\end{eqnarray}
\begin{eqnarray}
{\cal M}_{dd+1}=-{\cal M}_{d+1d}=iD.
\end{eqnarray}
The algebra becomes 
\begin{eqnarray}
[{\cal M}_{AB},{\cal M}_{CD}]=i\bigg(\eta_{AD}{\cal M}_{CB}-\eta_{AC}{\cal M}_{DB}+\eta_{DB}{\cal M}_{AC}-\eta_{CB}{\cal M}_{AD}\bigg).
\end{eqnarray}
The metric $\eta_{AB}$ is a flat $(d+2)-$dimensional with signature $-1+1+1...-1$.  By going from Lorentzian to the Euclidean spacetime the conformal group $SO(d,2)$ becomes the group $SO(d+1,1)$ while the Lorentz group $SO(d-1,1)$ becomes $SO(d)$ and the metric $\eta_{AB}$ becomes of signature $-1+1+1...+1$. The Poincare and dilatation generators form together a subgroup of the conformal group.

The mass operator $P_{\mu}P^{\mu}$ is a Casimir of the Poincare group but it is not a Casimir of the conformal group. States in a conformal field theory are therefore not classified by their mass, as in the Poincare group, but they are instead classified by their scaling dimension, i.e. by the eigenvalue of the dilatation operator 
\begin{eqnarray}
D|\Delta\rangle=i\Delta|\Delta\rangle.
\end{eqnarray}
The representation of the dilatation operator (which is a hermitian operator) on classical fields is not unitary and hence the factor $i$. Indeed, dilatations are not bounded transformations and as a consequence a finite dimensional representation of a non-compact Lie algebra must be necessarily non-unitary. This happens also for example with boosts in the Lorentz group \cite{Zaffaroni:2000vh,Qualls:2015qjb,Ferrara:1998pr}.

The scaling dimension $\Delta$ of a field $\Phi$ is defined by the action of dilatations on the field given by the equation 
\begin{eqnarray}
\Phi(\lambda x)=\lambda^{-\Delta}\Phi(x).\label{scaling0}
\end{eqnarray}
If we take for example a free scalar field theory in $d$ dimensions given by the action 
\begin{eqnarray}
S&=&\int d^dx \partial_{\mu}\Phi(x)\partial^{\mu}\Phi(x)\nonumber\\
&=&\lambda^{-d} \lambda^{2\Delta}\int  d^d(\lambda x)\partial_{\mu}\Phi(\lambda x)\partial^{\mu}\Phi(\lambda x)\nonumber\\
&=&\lambda^{-d+2\Delta+2}\int  d^d x^{\prime}\partial_{\mu}^{\prime}\Phi(x^{\prime})\partial^{\mu\prime}\Phi(x^{\prime}).
\end{eqnarray}
Hence this invariant under dilatation if and only if 
\begin{eqnarray}
-d+2\Delta+2=0.
\end{eqnarray}
The differential representation of the dilatation operator on scalar fields with scaling dimension $\Delta$ is given by 
\begin{eqnarray}
D\Phi=-i(x^{\mu}\partial_{\mu}+\Delta)\Phi.\label{scalingdiff}
\end{eqnarray}
This is the analogue of the differential representation of the momentum and Lorentz generators given by 
\begin{eqnarray}
P_{\mu}\Phi=-i\partial_{\mu}\Phi.\label{Pact}
\end{eqnarray}
\begin{eqnarray}
M_{\mu\nu}\Phi=i(x_{\mu}\partial_{\nu}-x_{\nu}\partial_{\mu})\Phi+S_{\mu\nu}\Phi.\label{Mact}
\end{eqnarray}
The special conformal generator is represented differentially by 
\begin{eqnarray}
K_{\mu}\Phi=(-2i\Delta x_{\mu}-x^{\nu}S_{\mu\nu}-2ix_{\mu}x^{\nu}\partial_{\nu}+ix^2\partial_{\mu})\Phi.\label{Kact}
\end{eqnarray}
By using (\ref{scalingdiff}) we can rewrite the quantum analogue of the scaling transformation $\Phi(x)\longrightarrow \lambda^{\Delta}\phi(\lambda x)$ as

\begin{eqnarray}
[D,\Phi]=-i(x^{\mu}\partial_{\mu}+\Delta)\Phi.
\end{eqnarray}
Under a coordinate transformation  $x\longrightarrow x^{\prime}$ the metric tensor transforms as $g_{\alpha\beta}(x)\longrightarrow g_{\alpha\beta}^{\prime}(x^{\prime})=({\partial x^{\mu}}/{\partial x^{\prime \alpha}})({\partial x^{\nu}}/{\partial x^{\prime \beta}})g_{\mu\nu}(x)$. The conformal group is the subgroup which satisfies $g_{\alpha\beta}(x)\longrightarrow g_{\alpha\beta}^{\prime}(x^{\prime})=\Omega^2(x) g_{\mu\nu}(x)$. From the invariance of the volume element we can see that the Jacobian of the conformal transformation $x\longrightarrow x^{\prime}$, $dx^{'\mu}dx^{'}_{\mu}=\Omega^{2}(x)dx_{\mu}dx^{\mu}$ is given by 
\begin{eqnarray}
|\frac{\partial x^{\prime}}{\partial x}|=\frac{1}{\sqrt{{\rm det}g_{\mu\nu}^{\prime}}}=\Omega^{-d}.
\end{eqnarray}  
The action of the generators of the conformal group on the scalar field $\Phi$ with scaling dimension $\Delta$ given by the transformations  (\ref{scalingdiff}), (\ref{Pact}), (\ref{Mact}), (\ref{Kact}) translates into the transformation law
\begin{eqnarray}
\Phi(x)\longrightarrow \Phi^{\prime}(x^{\prime})=|\frac{\partial x^{\prime}}{\partial x}|^{-\Delta/d}\Phi(x).
\end{eqnarray} 
The scalar field $\Phi$ is called quasi-primary operator. The covariance of the theory on conformal transformations is then given by the behavior 
 of the correlation functions under conformal transformations given  by 
 \begin{eqnarray}
\langle \Phi_1(x_1)...\Phi_N(x_N)\rangle\longrightarrow \langle \Phi^{\prime}(x^{\prime}_1)...\Phi^{\prime}(x^{\prime}_N)\rangle=|\frac{\partial x^{\prime}}{\partial x}|^{-\Delta_1/d}_{x=x_1}...|\frac{\partial x^{\prime}}{\partial x}|^{-\Delta_N/d}_{x=x_N}\langle \Phi_1(x_1)...\Phi_N(x_N)\rangle.
\end{eqnarray}
For a scalar field we have obviously $\Phi^{\prime}=\Phi$.

\subsection{Constraints of conformal symmetry}
Conformal invariance fully constrains the two- and three-point functions of the conformal field theory. For the two-point function of two quasi-primary operators $\phi_1$ and $\phi_2$ we have
\begin{eqnarray}
\langle \Phi_1(x_1)\Phi_2(x_2)\rangle\longrightarrow \langle \Phi^{}(x^{\prime}_1)\Phi^{}(x^{\prime}_2)\rangle=|\frac{\partial x^{\prime}}{\partial x}|^{-\Delta_1/d}_{x=x_1}|\frac{\partial x^{\prime}}{\partial x}|^{-\Delta_2/d}_{x=x_2}\langle \Phi_1(x_1)\Phi_2(x_2)\rangle.
\end{eqnarray}
In other words, we have
\begin{eqnarray}
\langle \Phi_1(x_1)\Phi_2(x_2)\rangle=
|\frac{\partial x^{\prime}}{\partial x}|^{\Delta_1/d}_{x=x_1}|\frac{\partial x^{\prime}}{\partial x}|^{\Delta_2/d}_{x=x_2}\langle \Phi^{}(x^{\prime}_1)\Phi^{}(x^{\prime}_2)\rangle.
\end{eqnarray}
Invariance under translations and rotations (for which the Jacobian is equal $1$) yields a dependence only on the combination $r_{12}=|x_1-x_2|$ (the difference is due to translations and the modulus is due to rotations). Invariance under the scaling transformations $x\longrightarrow x^{\prime}=\lambda x$ leads to 
\begin{eqnarray}
\langle \Phi_1(x_1)\Phi_2(x_2)\rangle=
\lambda^{\Delta_1+\Delta_2}\langle \Phi^{}(x^{\prime}_1)\Phi^{}(x^{\prime}_2)\rangle.
\end{eqnarray}  
This gives immediately the behavior 
\begin{eqnarray}
\langle \Phi_1(x_1)\Phi_2(x_2)\rangle=\frac{C_{12}}{r_{12}^{\Delta_1+\Delta_2}}.
\end{eqnarray} 
Under special conformal transformation we have 
\begin{eqnarray}
x^{\mu}\longrightarrow x^{\prime \mu}=\frac{x^{\mu}+b^{\mu}x^2}{1+2bx+b^2x^2}\Rightarrow 
 r_{12}^{\prime 2}=\frac{r_{12}^2}{(1+2bx_1+b^2x^2_1)(1+2bx_2+b^2x^2_2)}.
\end{eqnarray} 
\begin{eqnarray}
|\frac{\partial x^{\prime}}{\partial x}|=\frac{1}{(1+2bx+b^2x^2)^d}.
\end{eqnarray} 
We can then easily verify that we must have $\Delta_1=\Delta_2=\Delta$. Hence the two-point function of a conformal field theory must be constrained such that
\begin{eqnarray}
\langle \Phi_1(x_1)\Phi_2(x_2)\rangle=\frac{C_{12}}{r_{12}^{2\Delta}}.
\end{eqnarray} 
However, if $\Delta_1\neq \Delta_2$ then we must have $\langle \Phi_1(x_1)\Phi_2(x_2)\rangle=0$.

Similarly, the three-point function of a conformal field theory must be constrained by the invariance under translations, rotations, scalings and special conformal transformations such that
\begin{eqnarray}
\langle \Phi_1(x_1)\Phi_2(x_2)\Phi_3(x_3)\rangle=\frac{C_{123}}{r_{12}^{\Delta_1+\Delta_2-\Delta_3}r_{13}^{\Delta_1+\Delta_3-\Delta_2}r_{23}^{\Delta_2+\Delta_3-\Delta_1}}.
\end{eqnarray} 
\subsection{Conformal  algebra in two dimensions}
We consider a free boson $\Phi$ in two dimensions with the Euclidean action (with $z=\sigma^1+i\sigma^2$)
\begin{eqnarray}
S=\frac{1}{4\pi}\int d^2\sigma \partial_{\mu}\Phi\partial^{\mu}\Phi=\frac{1}{2\pi}\int dzd\bar{z}\partial\Phi\bar{\partial}\Phi.
\end{eqnarray}
The symmetries of this action are given by the conformal mappings 
\begin{eqnarray}
z\longrightarrow f(z)~,~\bar{z}\longrightarrow \bar{z}=\bar{f}(\bar{z}).
\end{eqnarray}
These are angle-preserving transformations when $f$ and its inverse are both holomorphic, i.e. $f$ is biholomorphic. For example, $z\longrightarrow z+a$ is a translation, $z\longrightarrow \zeta z$ where $|\zeta|=1$ is a rotation, and $z\longrightarrow \zeta z$ where $\zeta$ is real not equal to $1$ is a scale transformation called also dilatation.

We work with the complex coordinates
\begin{eqnarray}
z=\sigma^1+i\sigma^2\longrightarrow z=e^{2(\sigma^2-i\sigma^1)}~,~\bar{z}=\sigma^1-i\sigma^2\longrightarrow \bar{z}=e^{2(\sigma^2+i\sigma^1)}.
\end{eqnarray}
The factor of $2$ in the exponent is included for consistency with the periodicity of the closed string given by $\sigma^1\longrightarrow \sigma^1+\pi$. The worldsheet for a closed string is a cylinder which is topologically an ${\bf R}^2$. It can also be regarded as a Riemann surface, i.e. as a deformation of the complex plane. The Euclidean time $\sigma^2$ on the worldsheet corresponds to the radial distance $r=\exp(2\sigma^2)$ on the complex plane, with the infinite past $\sigma^2=-\infty$ at $r=0$, and the infinite future $\sigma^2=+\infty$ is a circle at $r=\infty$. 

The generators of conformal mappings are given by the infinitesimal transformations 
\begin{eqnarray}
z\longrightarrow z^{'}=z- \epsilon_n z^{n+1}~,~\bar{z}\longrightarrow \bar{z}^{'}=\bar{z}-\bar{\epsilon}_n \bar{z}^{n+1}~,~n\in Z.
\end{eqnarray}
The generators are immediately given by
\begin{eqnarray}
l_n=- z^{n+1}\partial ~,~\bar{l}_{n}=-\bar{z}^{n+1}\bar{\partial}~,~n\in Z.
\end{eqnarray}
The generators with $n<-1$ are defined on the punctured complex plane whereas the generators with  $n>1$ are defined on the complex plane with the point at infinity removed. The generators $l_{-1}$, $l_0$, $l_1$ are defined on the whole Riemann sphere, i.e. the complex plane+the point at infinity. They satisfy the classical Virasoro algebra 
 \begin{eqnarray}
[l_m,l_n]=(m-n)l_{m+n}~,~[\bar{l}_m,\bar{l}_n]=(m-n)\bar{l}_{m+n}~,~[l_m,\bar{l}_n]=0.
\end{eqnarray}
It is also easily seen that the Virasoro algebra is the same as the algebra of infinitesimal diffeomorphisms of the circle ${\bf S}^1$. 

The group $SO(3,1)$ is called the restricted conformal group. The full conformal group in two dimensions is infinite dimensional. The finite dimensional subgroup $SO(3,1)$ is generated by $l_{0}$, $l_{\pm 1}$, $\bar{l}_0$, $\bar{l}_{\pm 1}$. These are given explicitly by
\begin{eqnarray}
&&l_{-1}~:~z\longrightarrow z-\epsilon\nonumber\\
&&l_0~:~z\longrightarrow z-\epsilon z\nonumber\\
&&l_1~:~z\longrightarrow z-\epsilon z^2.\label{lkp}
\end{eqnarray}
And similarly for $\bar{l}_{-1}$, $\bar{l}_0$, $\bar{l}_1$. We have the following interpretation:
\begin{eqnarray}
&&l_{-1}~,~\bar{l}_{-1}~~{\rm translations}\nonumber\\
&&l_0-\bar{l}_0~~{\rm rotations}\nonumber\\
&&l_0+\bar{l}_0~~{\rm scalings}\nonumber\\
&&l_{1}~,~\bar{l}_{1}~~{\rm special~conformal~transformations}.
\end{eqnarray}
The finite or global form of these transformations are:
\begin{eqnarray}
&&l_{-1}~:~z\longrightarrow z+\alpha\nonumber\\
&&l_0~:~z\longrightarrow \lambda z\nonumber\\
&&l_{1}~:~z\longrightarrow \frac{z}{1-\beta z}.
\end{eqnarray}
By combining these transformations we obtain 
\begin{eqnarray}
z\longrightarrow \frac{az+b}{cz+d}~,~ad-bc=1.\label{sl2}
\end{eqnarray}
For infinitesimal $z$ we obtain
\begin{eqnarray}
z\longrightarrow \frac{b}{d}+\frac{1}{d^2}z-\frac{c}{d^3}z^2
\end{eqnarray}
Only three parameters are independent as it should be. We obtain a linear combination of the transformations (\ref{lkp}).

The group given by the relation (\ref{sl2}) is $SO(3,1)=SL(2,C)/{\bf Z}_2$. The division by ${\bf Z}_2$ is to take into account the property that the above transformations remain unchanged if $a,b,c,d\longrightarrow -a,-b,-c,-d$. The Lorentzian analogue is the group $SO(2,2)=SL(2,R)\times SL(2,R)$ where one factor of $SL(2,R)$ stands for left-movers and the other factor stands for right-movers.

\section{The AdS spacetime}
For an extensive discussion of anti-de Sitter spacetimes and their uses see \cite{Gibbons:2011sg,ing}.
\subsection{Maximally symmetric spaces}
The maximally symmetric manifolds of dimension $d$ are those spaces with the maximum number $d(d+1)/2$ of Killing vector fields generating isometries, i.e. symmetries, consisting of $d$ translations and $d(d-1)/2$ rotations/boosts. A space is a maximally symmetric manifold if and only if its Riemann curvature tensor is given by 
\begin{eqnarray}
R_{\mu\nu\alpha\beta}=\frac{R}{d(d-1)}(g_{\mu\alpha}g_{\nu\beta}-g_{\mu\beta}g_{\nu\alpha}),
\end{eqnarray}
where $g$ is of course the metric tensor. This means that in a maximally symmetric space the curvature tensor looks the same everywhere and in every direction and thus it is fully specified locally by the Ricci scalar curvature $R$. Hence, there are only three possible maximally symmetric spaces locally specified by the sign of the Ricci scalar $R$ (since the scale of $R$ specifies only the size of the space) which are given by:
\begin{enumerate}
\item {\bf Positive Curvature:} The spheres ${\bf S}^d$ in Euclidean signature and de Sitter spacetimes $dS_d$ in Lorentzian signature. The de Sitter spacetime plays a crucial role in cosmology in the early universe during inflation (when the cosmological constant was very large and the expansion was exponential) as well as in the final state of the universe which is seen to be dominated by a very small cosmological constant and an accelerated expansion.
\item {\bf Zero Curvature:} The Euclidean spaces ${\bf R}^d$ in Euclidean signature and Minkowski spacetimes ${\bf M}^d$ in Lorentzian signature.
\item {\bf Negative Curvature:} The hyperboloids ${\bf H}^d$ in Euclidean signature and anti-de Sitter spacetimes $AdS_d$ in Lorentzian signature. The anti-de Sitter spacetime is crucial for the holographic principle and the AdS/CFT correspondence. This is primary point of interest to us here in this chapter.
\end{enumerate}
For more detail see chapters $3$ and $8$ in \cite{carroll}.
\subsection{Global and poincare coordinates}
\subsubsection{Global coordinates}
We are interested mostly in $AdS_5$. Thus, we start from a six-dimensional flat spacetime ${\bf R}^{2,4}$ with metric 
\begin{eqnarray}
ds^2_5=-dX_0^2-dX_5^2+dX_1^2+dX_2^2+dX_3^2+dX_4^2.
\end{eqnarray}
The isometry group of this metric is obviously $SO(2,4)$ which is precisely the conformal group in $4$ dimensions. For $AdS_{d+1}$ we need to start from ${\bf R}^{2,d}$ with isometry group $SO(2,d)$. We Lobachevski-like embed in this Minkowski spacetime  ${\bf R}^{2,4}$ the following hyperboloid  
\begin{eqnarray}
-X_0^2-X_5^2+X_1^2+X_2^2+X_3^2+X_4^2=-L^2.
\end{eqnarray}
This hyperboloid is obviously five-dimensional. Now we can induce global coordinates $(\tau,\rho,\hat{x}_i)$ on this hyperboloid where $\hat{x}_i$ define an ${\bf S}^3$, i.e. $\sum_{i=1}^4\hat{x}_i^2=1$, by the relations
\begin{eqnarray}
&&X_0=L\cos\tau\cosh \rho\nonumber\\
&&X_5=L\sin\tau\cosh\rho\nonumber\\
&&X_i=L\sinh\rho\hat{x}_i.
\end{eqnarray}
The metric becomes (with $d\Omega_3$ is the solid angle on ${\bf S}^3$)
\begin{eqnarray}
ds^2_5=L^2(-\cosh^2\rho d\tau^2+d\rho^2+\sinh^2\rho d\Omega_3).
\end{eqnarray}
The coordinate $\rho$ plays the role of a radius in $AdS_5$ since $\rho\in [0,+\infty[$ whereas the coordinate $\tau$ is timelike with $\tau\in [0,2\pi]$. These coordinates cover the whole of $AdS_5$ which is the reason why they are called global. On the other hand, $\tau$ is periodic which signals the existence of closed timelike curves. However, this property is not intrinsic to the $AdS_5$ space but it is an artifact of this system of coordinates. Instead of the above space we will then take its universal cover space, in which we allow $\tau$ to run over the unrestricted range $]-\infty,+\infty[$, as the definition of anti-de Sitter spacetime $AdS_5$. The isometry group becomes a cover of  $SO(2,4)$.

The anti-de Sitter spacetime, as opposed to de Sitter spacetime, is not a globally hyperbolic spacetime. This means that $AdS$ does not admit a well-defined time evolution starting from suitable initial data on a spacelike (Cauchy) hypersurface. Indeed, by specifying the initial data on a spacelike hypersurface in $AdS$ (together with the knowledge of the equations of motion) is not sufficient to determine the future evolution uniquely and deterministically. This is due to the existence of a boundary at timelike infinity (see below) and thus information can flow in from infinity. See \cite{Ishibashi:2004wx} for a systematic discussion of this issue.

The other difference between $dS$ and $AdS$ is with regard to topology. $AdS_5$ spacetime is topologically equivalent to ${\bf R}^4\times{\bf S}^1$ whereas $dS_5$ spacetime is topologically equivalent to ${\bf R}\times {\bf S}^4$. This can be fleshed out using Penrose diagrams. See \cite{carroll,ing} for the explicit discussion.

\subsubsection{Poincare coordinates}
Next set of coordinates is much more important for quantum field theory. We introduce the following so-called Poincare coordinates given by a Minkowski $4-$vector $x_{\mu}$ and a radial coordinate $u$ defined by the relations 
\begin{eqnarray}
&&X_0=\frac{1}{2u}(1+u^2 (L^2+\vec{x}^2-t^2))\nonumber\\
&&X_i=L ux_i\nonumber\\
&&X_4=\frac{1}{2u}(1-u^2 (L^2-\vec{x}^2+t^2))\nonumber\\
&&X_5=Lu t.
\end{eqnarray}
We get immediately the metric 
\begin{eqnarray}
ds_5^2=L^2\bigg(\frac{du^2}{u^2}+u^2(dx_i^2-dt^2)\bigg)=L^2\bigg(\frac{du^2}{u^2}+u^2(dx_{\mu}dx^{\mu})\bigg).
\end{eqnarray}
Thus, for each fixed value of the radial coordinate $u$ we have a $4-$dimensional ordinary Minkowski spacetime, i.e. the ordinary spacetime is foliated over the radial coordinate $u$ which takes the value from $0$ to infinity.  However, because of the overall conformal factor $R^2u^2$ multiplying the Minkowski metric all distances in the $4-$dimensional theory on the Minkowski slices or branes are rescaled by a factor of $Ru$.

The point $u=\infty$ is a conformal boundary since it is the conformally equivalent metric $ds_5^2/u^2$ which is seen to have a boundary ${\bf M}^4$ at $u=\infty$. However, the point $u=0$ is a horizon since the $00$ component of the metric vanishes there and hence the Killing vector $\partial/\partial t$ becomes of zero norm at this point. Furthermore, the metric can be extended beyond the horizon $u=0$ which is only a coordinate singularity and the Poincare coordinates cover only half of the hyperboloid.

Another form of the metric can be obtained by the substitution $u=1/z$ and thus the conformal boundary becomes located at $z=0$ whereas the horizon becomes located at $z=\infty$. The metric takes the form
\begin{eqnarray}
ds_5^2=\frac{L^2}{z^2}\bigg(dz^2+dx_{\mu}dx^{\mu}\bigg).\label{me3}
\end{eqnarray}
Another system of coordinates can be obtained by the substitution $\sinh \rho=2r/(1-r^2)$ in the system of coordinates, i.e. $r$ runs from $0$ to $1$. We obtain the metric 
\begin{eqnarray}
ds^2_5=\frac{L^2}{(1-r^2)^2}\bigg(-(1+r^2)^2d\tau^2+4dr^2+4r^2 d\Omega_3\bigg).\label{me4}
\end{eqnarray}
The metric (\ref{me3}) should be thought of as an approximation of the metric (\ref{me4}) in the vicinity of a point on the boundary located now at $r=1$. In this approximation the $3-$sphere is replaced with the flat coordinates $x_i$, $t$ is replaced with $\tau$, and the radial coordinate $z$ is replaced with $z=1-r$.  In this system of coordinates $r=0$ is the center of anti-de Sitter and there is no horizon since the metric (\ref{me4}) is geodesically complete. However, the metric (\ref{me3}) is  geodesically incomplete because we can reach $z=\infty$ along a timelike geodesic in a finite proper time, i.e. $z=\infty$ is a horizon. 

Furthermore, we can check that we can travel from the center $r=0$ of anti-de Sitter to the boundary $r=1$ (which is an infinite proper distance) and back along a null curve satisfying $(1+r^2)d\tau=2dr$ in a finite proper time given by $\tau=\pi$. This means that anti-de Sitter space is causally finite and it behaves as a finite box of size $L$. See \cite{Susskind:2005js} for more discussion on the relation between the metrics  (\ref{me3}) and  (\ref{me4}). 
\subsubsection{Generalization}
We generalize to anti-de Sitter in $d+1$ dimensions, i.e. $AdS_{d+1}$, with isometry group $SO(2,d)$ which is precisely the conformal group in $d$ dimensions.   $AdS_{d+1}$ spacetime is topologically equivalent to ${\bf R}^d\times{\bf S}^1$ (to be contrasted with the topology of $dS_{d+1}$ spacetime given by ${\bf R}\times {\bf S}^d$). We embed  $AdS_{d+1}$ in the Minkowski spacetime  ${\bf R}^{2,d}$ by   
\begin{eqnarray}
-X_0^2-X_{d+1}^2+X_1^2+X_2^2+X_3^2+...+X_d^2=-L^2.
\end{eqnarray}
Now we can induce global coordinates $(\tau,\rho,\hat{x}_i)$, or the slightly different ones  $(\tau,r,\hat{x}_i)$, on this hyperboloid by the relations
\begin{eqnarray}
&&X_0=L\cos\tau\cosh \rho=L\frac{\cos \tau}{\cos r}\nonumber\\
&&X_{d+1}=L\sin\tau\cosh\rho=L\frac{\sin \tau}{\cos r}\nonumber\\
&&X_i=L\sinh\rho\hat{x}_i=L\tan r\hat{x}_i.
\end{eqnarray}
The range is $\rho\in[0,+\infty[$ or correspondingly $r\in[0,\pi/2[$, and $\tau\in [0,2\pi]\longrightarrow \tau \in[-\infty,\infty]$ as we unwrap to the universal cover, and $\hat{x}_i$ defines a $(d-1)-$dimensional sphere ${\bf S}^{d-1}$, i.e. $\sum_{i=1}^4\hat{x}_i^2=1$. The metric reads 
\begin{eqnarray}
ds^2_{d+1}=L^2(-\cosh^2\rho d\tau^2+d\rho^2+\sinh^2\rho d\Omega_{d-1})=\frac{L^2}{\cos^2 r}(-d\tau^2+dr^2+\sin^2 rd\Omega_{d-1}).
\end{eqnarray}
Thus, $AdS_{d+1}$ can be viewed as a cylinder with bases at $\tau=-\infty$ and $\tau=+\infty$, a center at $\rho=r=0$ whereas the spatial infinity $\rho=\infty$ is at $r=\pi/2$, while going around the cylinder is given by the angular variables $\Omega_{d-1}$. The anti-de Sitter space is therefore not compact both in time and space yet it behaves as a box as we will discuss.

The symmetry group of $AdS_{d+1}$ is given by the conformal group $SO(2,d)$. We have $SO(d)$ rotations among the  $X_i$, $SO(2)$ rotation in the plane $X_0X_{d+1}$, $d$ boosts in the planes $X_0X_i$ and $d$ boosts in the planes $X_{d+1}X_i$. In total we have $d(d-1)/2+1+2d=(d+1)(d+2)/2$ generators. These generators can be represented in terms of the coordinates $X_A$ as
\begin{eqnarray}
L^A_{B}=X^A\frac{\partial}{\partial X^B}-X_B\frac{\partial}{\partial X_A}.
\end{eqnarray}
We will also need later 
\begin{eqnarray}
L^{AB}=X^A\frac{\partial}{\partial X_B}-X^B\frac{\partial}{\partial X_A}~,~L_{AB}=X_A\frac{\partial}{\partial X^B}-X_B\frac{\partial}{\partial X^A}.
\end{eqnarray}
For example, the rotation in the timelike plane  $X_0X_{d+1}$ is given by (recall the metric signature $-++..-$)
\begin{eqnarray}
L_{d+1}^0=X_0\frac{\partial}{\partial X_{d+1}}-X_{d+1}\frac{\partial}{\partial X_0}.
\end{eqnarray}
On the other hand, we have
\begin{eqnarray}
\frac{\partial}{\partial t}=\frac{\partial X_0}{\partial t}\frac{\partial}{\partial X_{0}}+\frac{\partial X_{d+1}}{\partial t}\frac{\partial}{\partial X_{d+1}}=-X_{d+1}\frac{\partial}{\partial X_0}+X_0\frac{\partial}{\partial X_{d+1}}.
\end{eqnarray}
Hence the Hamiltonian in anti-de Sitter spacetime is given by 
\begin{eqnarray}
{\rm Hamiltonian}\equiv -i \frac{\partial}{\partial t}=-iL_{d+1}^0.
\end{eqnarray}
\subsection{Euclidean Poincare patch and RG equation}
\subsubsection{Poincare coordinates revisited}
Let us start with the most general metric in $(d+1)-$dimension which enjoys Poincare invariance in $d$ dimensions given by 
\begin{eqnarray}
ds^2_{d+1}=\Omega^2(z)\bigg(dz^2+dx_{\mu}dx^{\mu}\bigg)=\Omega^2(z)\bigg(dz^2+d\vec{x}^2-dt^{2}\bigg).
\end{eqnarray}
As we will see the extra coordinate $z$ corresponds to an energy scale of a conformally invariant theory. Hence the above metric must be invariant under $z\longrightarrow z$, $\vec{x}\longrightarrow \lambda\vec{x}$ and $t\longrightarrow \lambda t$. This leads immediately to the requirement 
\begin{eqnarray}
ds^2_{d+1}=\Omega^2(z)\bigg(dz^2+d\vec{x}^2-dt^{2}\bigg)\longrightarrow \lambda^2\Omega^2(\lambda z)\bigg(dz^2+d\vec{x}^2-dt^{2}\bigg)=\Omega^2(z)\bigg(dz^2+d\vec{x}^2-dt^{2}\bigg).\nonumber\\
\end{eqnarray}
The function $\Omega$ must then satisfy 
\begin{eqnarray}
\Omega(z)=\frac{L}{z}.
\end{eqnarray}
We get then 
\begin{eqnarray}
ds^2_{d+1}=\frac{L^2}{z^2}\bigg(dz^2+dx_{\mu}dx^{\mu}\bigg)=\frac{L^2}{z^2}\bigg(dz^2+d\vec{x}^2-dt^{2}\bigg).\label{ads}
\end{eqnarray}
This is precisely anti-de Sitter spacetime $AdS_{d+1}$ in Poincare coordinates with boundary at $z=0$ and horizon at $z=\infty$. The constant $L$ is the radius of anti-de Sitter. This metric solves Einstein equations with cosmological constant $\Lambda$, viz
\begin{eqnarray}
R_{\mu\nu}-\frac{1}{2}g_{\mu\nu}R=-\Lambda g_{\mu\nu}.\label{EE}
\end{eqnarray}
The Ricci tensor and Ricci scalar of the metric (\ref{ads}) are given by 
\begin{eqnarray}
R_{\mu\nu}=-\frac{d}{L^2}g_{\mu\nu}\Rightarrow R=-\frac{d(d+1)}{L^2}.
\end{eqnarray}
Thus, the metric (\ref{ads}) defines an Einstein space. However, by contracting both sides of the Einstein equations (\ref{EE}) we obtain the Ricci scalar 
\begin{eqnarray}
R=2\frac{d+1}{d-1}\Lambda.
\end{eqnarray}
By comparing the above two last equations we can determine the cosmological constant in terms of the radius of anti-de Sitter by the relation 
\begin{eqnarray}
\Lambda=-\frac{d(d-1)}{2L^2}.
\end{eqnarray}
\subsubsection{Euclidean Poincare coordinates}
The relationship between the global coordinates $\tau$, $\rho$ and $\hat{x}_i$ (with a manifest $SO(2)\times SO(d)$ symmetry subgroup) and the Poincare coordinates $t$, $z$, $\vec{x}$ (with the $d-$dimensional Poincare symmetry subgroup manifest) is given by 
\begin{eqnarray}
&&X_0=L\cos\tau\cosh \rho=\frac{z}{2}(1+\frac{L^2+\vec{x}^2-t^2}{z^2})\nonumber\\
&&X_i=L\sinh\rho\hat{x}_i=L\frac{x_i}{z}~,~i<d\nonumber\\
&&X_{d}=L\sinh\rho\hat{x}_d=\frac{z}{2}(1-\frac{L^2-\vec{x}^2+t^2}{z^2})\nonumber\\
&&X_{d+1}=L\sin\tau\cosh\rho=L\frac{t}{z}.
\end{eqnarray}
Although these Poincare coordinates cover only half of the $AdS_{d+1}$ spacetime (with conformal group $SO(2,d)$) their Euclidean analogues cover all of the Euclidean $AdS_{d+1}$ space (with conformal group $SO(1,d+1)$). The Euclidean  $AdS_{d+1}$ space is obtained by the Wick rotation $X_{d+1}\longrightarrow -iX_{d+1}$, i.e. it is given by the embedding 
\begin{eqnarray}
-X_0^2+X_{d+1}^2+\sum_{i=1}^dX_i^2=L^2.
\end{eqnarray}
This corresponds to the Wick rotations $t\longrightarrow -it$ and $\tau\longrightarrow -i\tau$. The metric becomes 
\begin{eqnarray}
ds^2_{d+1}=R^2(\cosh^2\rho d\tau^2+d\rho^2+\sinh^2\rho d\Omega_{d-1})=\Omega^2(z)\bigg(dz^2+d\vec{x}^2+dt^{2}\bigg).
\end{eqnarray}
The map between these systems becomes then given by
\begin{eqnarray}
&&X_0=L\cosh\tau\cosh \rho=\frac{z}{2}(1+\frac{L^2+\vec{x}^2+t^2}{z^2})\nonumber\\
&&X_i=L\sinh\rho\hat{x}_i=L\frac{x_i}{z}~,~i<d+1\nonumber\\
&&X_{d+1}=L\sinh\tau\cosh\rho=\frac{z}{2}(1-\frac{L^2-\vec{x}^2-t^2}{z^2}).
\end{eqnarray}  
The boundary $z=0$ in the Minkowski metric becomes ${\bf R}^d$ in the Euclidean metric whereas the horizon $z=\infty$ in the Minkowski metric shrinks to a point in the Euclidean metric. By adding the point $z=\infty$ to the boundary ${\bf R}^d$ we obtain a sphere ${\bf S}^d$. This compactified Euclidean $AdS_{d+1}$ is thus the solid $(d+1)-$dimensional ball.
\subsubsection{Renormalization group equation}
As we said the extra coordinate $z$ corresponds to an energy scale of a conformal field theory, i.e. it defines a lattice spacing $a$. The $d-$dimensional slices or branes, defined by the points $x=(x_{0},x_1,...,x_d)$ in the higher dimensional $AdS_{d+1}$ spacetime, should then be regarded as lattices of increasing size in a Kadanoff-Wilson renormalization group approach \cite{Kadanoff:1966wm,Wilson:1973jj}. 

\begin{figure}[htbp]
\begin{center}
  \includegraphics[width=10.0cm,angle=0]{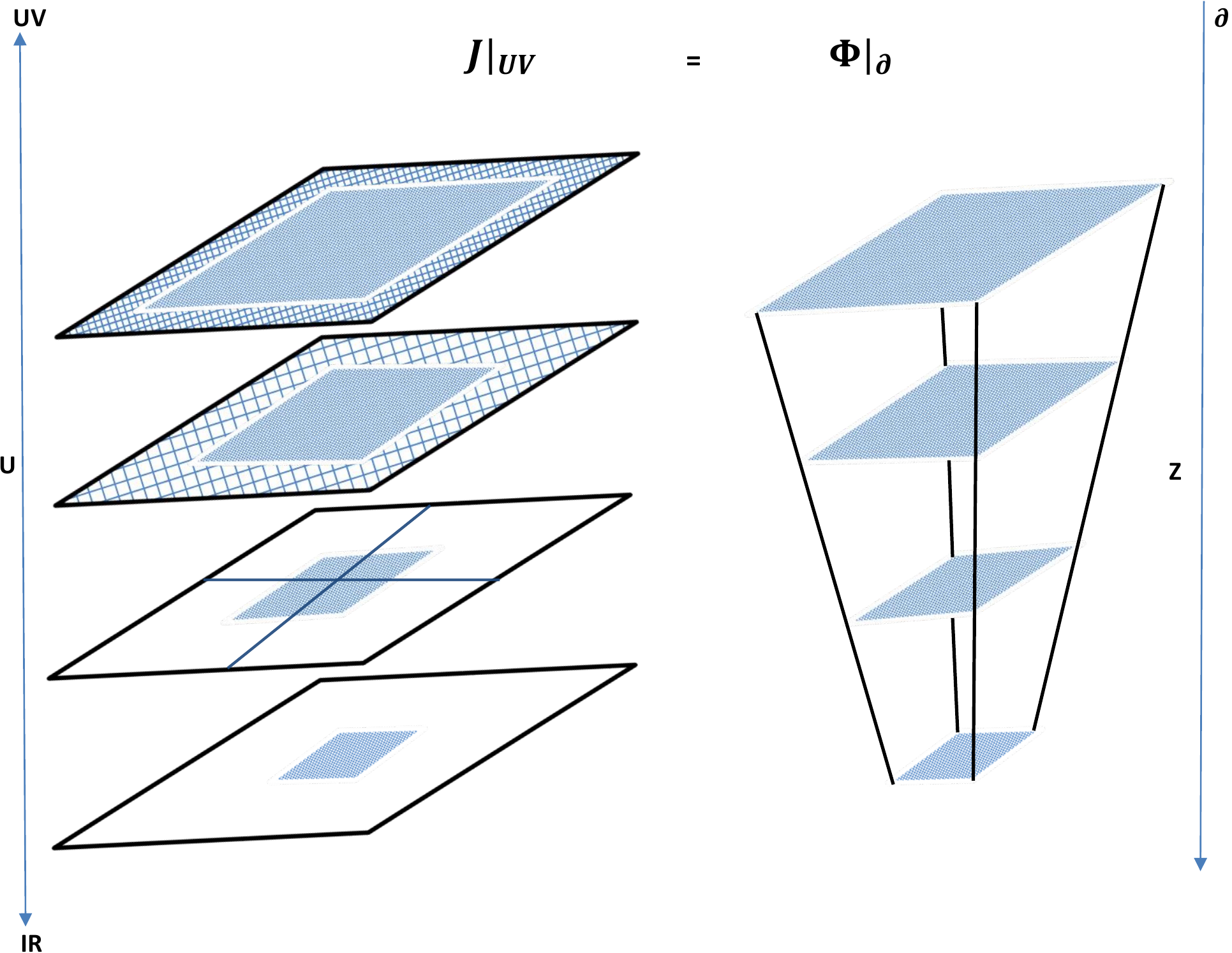}
\end{center}
\caption{$AdS_{d+1}$ as a foliation of $M_d$ over the energy scale.}\label{adsrg}
\end{figure}

To exhibit this crucial point in some detail we start with a field theory Hamiltonian on a lattice $a$, with coupling constants or sources $J_i(x,a)$ and field operators ${\cal O}_i$, given by \cite{Ramallo:2013bua}
\begin{eqnarray}
H=\sum_{i,x}J_i(x,a){\cal O}_i(x).
\end{eqnarray}
Then under the Kadanoff-Wilson renormalization group approach the lattice is coarse grained, i.e. we increase the lattice spacing successively as $\mu=a\longrightarrow 2a\longrightarrow 4a...$ and replace at each step the spins (fields) by block spins (averages of fields) \cite{Kadanoff:1966wm,Wilson:1973jj}. As a result the Hamiltonian will change in such a way that its form remains invariant but only the coupling constants $J_i(x,\mu)$ change or flow, i.e. the weightings of the field operators  ${\cal O}_i$ flow, according to the renormalization group equation 
\begin{eqnarray}
\mu\frac{\partial}{\partial \mu}J_i(x,\mu)=\beta_i(J_j(x,\mu),\mu).
\end{eqnarray}
In the AdS/CFT proposal we view the lattice scale $\mu$ as an extra dimension and as a consequence the collection of the lattices $\mu=a,2a,4a,...$ should be viewed as slices of a higher dimensional space.  The coupling constants $J_i(x,\mu)$ should then be reinterpreted as fields $\Phi_i(x,\mu)$ in this new higher dimensional space with the equation of motion in the extra dimension given by the above renormalization group equation.  The dynamics of these bulk fields $\Phi_i(x,\mu)$ is determined by a gravitational action with their boundary values equated with the microscopic/continuum values of the coupling constants $J_i(x,\mu)$ of the field theory (corresponding to the operators  ${\cal O}_i$) in the UV \cite{Ramallo:2013bua}.  See figure (\ref{adsrg}).

\section{Scalar field in $AdS_{d+1}$}

\subsection{The  Klein-Gordon equation}
Recall that $AdS_{d+1}$ in this system can be viewed as a cylinder with bases at $\tau=-\infty$ and $\tau=+\infty$, a center at $\rho=r=0$ whereas the spatial infinity is at $r=\pi/2$, while going around the cylinder is given by the angular variables $\Omega_{d-1}$. The embedding is given explicitly by (the radius is denoted here by $R$)
\begin{eqnarray}
&&X_0=R\frac{\cos \tau}{\cos r}\nonumber\\
&&X_{d+1}=R\frac{\sin \tau}{\cos r}\nonumber\\
&&X_{\mu}=R\tan r\hat{x}_{\mu}.
\end{eqnarray}
Recall also the ranges $r\in[0,\pi/2[$, and $\tau \in[-\infty,\infty]$ as we unwrap to the universal cover and $\hat{x}_{\mu}$ defines a $(d-1)-$dimensional sphere ${\bf S}^{d-1}$. 
We will consider global coordinates on $AdS_{d+1}$ given by 
\begin{eqnarray}
ds^2_{d+1}=\frac{R^2}{\cos^2 r}(-d\tau^2+dr^2+\sin^2 rd\Omega_{d-1}).
\end{eqnarray}

We start by writing the action of a scalar field $\phi$ in $AdS_{d+1}$ with $SO(2,d)$ invariance given by 
\begin{eqnarray}
S=\int d^{d+1}x\sqrt{g}\bigg[-\frac{1}{2}g^{MN}\partial_M\phi\partial_N\phi-\frac{1}{2}m^2\phi^2\bigg].
\end{eqnarray}
The Klein-Gordon equation is the Euler-Lagrange equation derived from this action given by 
\begin{eqnarray}
\frac{1}{\sqrt{g}}\partial_M\big(\sqrt{g}g^{MN}\partial_N\phi\big)-m^2\phi=0.
\end{eqnarray}

\subsection{Example of $AdS_2$} For simplicity we consider $AdS_{2}$ with isometry given by $SO(2,1)$ and metric $ds_2^2=R^2(-d\tau^2+dr^2)/\cos^2 r$. We obtain the equation 
 \begin{eqnarray}
-\partial_{\tau}^2\phi=\bigg(-\partial_r^2+\frac{R^2m^2}{\cos^2 r}\bigg)\phi.\label{KG}
\end{eqnarray}
We pull the time dependence as $\phi=\exp(i\Delta \tau)\chi(r)$ where $E$ is the energy. We get immediately 
 \begin{eqnarray}
\Delta^2\chi=\bigg(-\partial_r^2+\frac{R^2m^2}{\cos^2 r}\bigg)\chi.
\end{eqnarray}
A solution is given by the ansatz 
 \begin{eqnarray}
\chi(r)=\cos^{\alpha}r.
\end{eqnarray}
The equation of motion becomes algebraic given by
\begin{eqnarray}
(\Delta^2-\alpha^2)\chi=\bigg(-\alpha(\alpha-1)+R^2m^2\bigg)\chi\Rightarrow \Delta^2-\alpha^2=-\alpha(\alpha-1)+R^2m^2=0.
\end{eqnarray}
In other words, $\chi(r)=\cos^{\alpha}r$ is a solution iff the mass $m$ of the scalar field is related to its energy $\Delta$ (in the ground state) by the relation 
\begin{eqnarray}
R^2m^2=\Delta(\Delta-1).
\end{eqnarray}
This crucial result can also be found using group theoretic method as follows.

Recall the signature $-++...-$ and that the number of generators of $SO(2,d)$ is $(d+1)(d+2)/2$. Thus, the number of generators of  $SO(2,1)$ is $3$ and they are given by $L_2^0$, $L_1^0$ and $L_2^1$. Explicitly we have 
 \begin{eqnarray}
L_2^0=X^0\frac{\partial}{\partial X^2}-X_2\frac{\partial}{\partial X_0}=\partial_{\tau}.
\end{eqnarray}
 \begin{eqnarray}
L_1^0=X^0\frac{\partial}{\partial X^1}-X_1\frac{\partial}{\partial X_0}=\sin\tau\sin r \partial_{\tau}-\cos\tau \cos r\partial_r.
\end{eqnarray}
\begin{eqnarray}
L_2^1=X^1\frac{\partial}{\partial X^2}-X_2\frac{\partial}{\partial X_1}=-\cos\tau\sin r \partial_{\tau}-\sin\tau \cos r\partial_r.
\end{eqnarray}
They satisfy the algebra 
\begin{eqnarray}
[L_2^0,L_1^0]=-L_2^1~,~[L_2^0,L_2^1]=L_1^0~,~[L_1^0,L_2^1]=L_2^0.
\end{eqnarray}
Or equivalently 
\begin{eqnarray}
[D,P]=P~,~[D,K]=-K~,~[K,P]=-2D.
\end{eqnarray}
The operators $D$, $P$ and $K$ are given in terms of the operators $L_B^A$ by the relations:
\begin{itemize}
\item The dilatation generator or Hamiltonian operator:
\begin{eqnarray}
D=-iL_2^0.
\end{eqnarray}
The representations of the $SO(2,1)$ algebra are then labeled by the eigenvalues $\Delta$ of $D$, viz
\begin{eqnarray}
D|\psi\rangle=\Delta |\psi\rangle.
\end{eqnarray} 
\item The special conformal generator or lowering operator
\begin{eqnarray}
K=L_1^0-iL_2^1.
\end{eqnarray}
Thus, by acting on the ground state $|\psi_0\rangle$ with the lowering operator $K$, we must have
\begin{eqnarray}
K|\psi_0\rangle=0.
\end{eqnarray} 
\item The momentum generator or raising operator
\begin{eqnarray}
P=L_1^0+iL_2^1.
\end{eqnarray}
All other states above the ground state $|\psi_0\rangle$ are obtained by acting successively with the raising operator $P$ on  $|\psi_0\rangle$. 
\end{itemize}
The conditions $D|\psi_0\rangle=\Delta |\psi_0\rangle$ and  $K|\psi_0\rangle=0$ read in the global coordinates $\tau$, $r$ as follows

\begin{eqnarray}
-i\partial_{\tau}\psi_0(t,r)=\Delta \psi_0(t,r)\Rightarrow \psi_0(t,r)=\exp(i\Delta t)\chi(r).
\end{eqnarray}
\begin{eqnarray}
\big(i \sin r \partial_{\tau}-\cos r\partial_r\big)\psi_0(t,r)=0\Rightarrow \chi(r)=\cos^{\Delta}r.
\end{eqnarray}
All other states $|\psi_n\rangle$ can now be obtained by the action of the raising operator $P$ on this ground state $|\psi_0\rangle$, i.e. $|\psi_n\rangle=P^n|\psi_0\rangle$. The energy is found to be quantized as 
\begin{eqnarray}
E_n=\Delta+n.
\end{eqnarray}
In other words, the energy levels are integer spaced analogously to the harmonic oscillator motion and hence all $AdS_2$ orbits have the same period with respect to the $AdS_2$ time $\tau$. 
\subsection{Generalization}

The generators of the conformal algebra $SO(d,2)$ are: 
\begin{itemize}
\item The dilatation generator $D=-iL^0_{d+1}$ which plays the role of the Hamiltonian.
\item The momentum generators  $P_{\mu}=L_{\mu}^0+iL_{\mu}^{d+1}$ which play the role of raising operators. Note that $L_{\mu}^{d+1}=L_{d+1}^{\mu}$.
\item The special conformal generators  $K_{\mu}=L_{\mu}^0-iL_{\mu}^{d+1}$ which play the role of lowering operators.
\item The rotation generators $M_{\mu\nu}=iL_{\mu}^{\nu}$ which generate the Lie algebra $SO(d)$.
\end{itemize}
The generators $L_A^B$ satisfy the algebra
\begin{eqnarray}
[L^B_A,L^D_C]=\eta^D_AL^B_C-\eta_{AC}L^{BD}-\eta^{DB}L_{AC}-\eta^B_CL^D_A.
\end{eqnarray}
The rotation generators were absent in $SO(1,2)$.
Also, we have $d$ momentum generators and $d$ special conformal generators in the case of $SO(d,2)$, i.e. $P_{\mu}$ and $K_{\mu}$ transform as vectors under $SO(d)$, viz
\begin{eqnarray}
[M_{\mu\nu},P_{\rho}]=i(\eta_{\mu\rho}P_{\nu}-\eta_{\nu\rho}P_{\mu})~,~[M_{\mu\nu},K_{\rho}]=i(\eta_{\mu\rho}K_{\nu}-\eta_{\nu\rho}K_{\mu}).
\end{eqnarray}
The generators $P_{\mu}$ and $K_{\mu}$ are indeed the raising and lowering operators respectively, viz
\begin{eqnarray}
[D,P_{\mu}]=P_{\mu}~,~[D,K_{\mu}]=-K_{\mu}.
\end{eqnarray}
Also we have 
\begin{eqnarray}
[P_{\mu},K_{\nu}]=2(\eta_{\mu\nu}D+iM_{\mu\nu}).
\end{eqnarray}
The $SO(d)$ rotation generators satisfy 
\begin{eqnarray}
[M_{\mu\nu},M_{\rho\sigma}]=i\bigg(\eta_{\mu\sigma}M_{\rho\nu}-\eta_{\mu\rho}M_{\sigma\nu}+\eta_{\sigma\nu}M_{\mu\rho}-\eta_{\rho\nu}M_{\mu\sigma}\bigg).
\end{eqnarray}
The dilatation generator $D$ is a scalar under $SO(d)$ rotations, viz
\begin{eqnarray}
[M_{\mu\nu},D]=0.
\end{eqnarray}
This means that the Hamiltonian and the angular momentum operators can be diagonalized simultaneously.  

The ground state $|\psi_0\rangle$ will correspond to the smallest possible eigenvalue $\Delta$ of the Hamiltonian $D$ and it must be annihilated by all the lowering operators $K_{\mu}$, viz
\begin{eqnarray}
D|\psi_0\rangle=\Delta |\psi_0\rangle~,~K_{\mu}|\psi_0\rangle=0.
\end{eqnarray}
The ground state or highest weight state $|\psi_0\rangle$ is called a primary state. All other states (descendant states) can be obtained by acting successively with the raising operators $P_{\mu}$ (which commute among themselves), viz
\begin{eqnarray}
|\psi_{nlm}\rangle\sim (P_{\mu}^2)^n P_{\mu_1}...P_{\mu_l}|\psi_0\rangle.\label{nlm}
\end{eqnarray}
The energy of this state is clearly (since each $P_{\mu}$ raises the energy by a single unit) is
\begin{eqnarray}
E_{nl}=\Delta+2n+l.
\end{eqnarray}
These states carry $SO(d)$ indices and hence they are characterized by an angular momentum quantum number equal exactly to the integer $l$, i.e. to the number of uncontracted vector indices. If $l=0$ then the state does not carry a free $SO(d)$ index, i.e. it is a scalar, and as consequence its spin is given directly by $l=0$. Whereas if $l=1$ the state carries a single free vector index, i.e. it transforms as a vector under $SO(d)$, and as a consequence its spin must be $l=1$. The quantum number $l$ is therefore the angular momentum or spin quantum number. 

The quantum numbers $m$\footnote{It should not be confused with the mass $m$. But it should also be clear from the context which one is which.} appearing in (\ref{nlm}) relate to the spherical harmonics $Y_{lm}(\Omega)$ on $AdS_{d+1}$ which, by rotational symmetry, carry the angular dependence of the wave functions $\psi_{nlm}(\tau,r,\Omega)=\langle\tau,r,\Omega|\psi_{nlm}\rangle\ $. For example for $d=3$ we have a single (integer) number $m$ which is the usual magnetic quantum number and $Y_{lm}(\Omega)=Y_{lm}(\theta,\phi)$ are the usual spherical harmonics.  Thus, the wave functions $\psi_{nlm}(\tau,r,\Omega)$ are of the general form \cite{kaplan}
\begin{eqnarray}
\psi_{nlm}(\tau,r,\Omega)=\exp(iE_{nl}\tau)Y_{lm}(\Omega)\psi_{nl}(r).
\end{eqnarray}
The radial part $\psi_{nl}(r)$ is proportional to a hypergeometric function which for the ground state reduces to 
\begin{eqnarray}
\psi_{00}(r)\sim \cos^{\Delta} r.
\end{eqnarray} 
This can be shown as follows. We start from the identity 
\begin{eqnarray}
K_{\mu}&=&-(X_0-iX_{d+1})\frac{\partial}{\partial X_{\mu}}+X_{\mu}(\frac{\partial}{\partial X_0}-i\frac{\partial}{\partial X_{d+1}})\nonumber\\
&=&-\frac{\exp(-i\tau)}{r}\frac{\cos^2r\hat{x}_{\mu}}{\cos r}\frac{\partial}{\partial \hat{x}_{\mu}}+X_{\mu}(\frac{\partial}{\partial X_0}-i\frac{\partial}{\partial X_{d+1}}).
\end{eqnarray} 
The ground state $|\psi_0\rangle$ is characterized by zero angular momentum, i.e. $l=0$, and hence it does not depend on the angles $\hat{x}_{\mu}$ or equivalently $\Omega$. The condition $K_{\mu}|\psi_0\rangle=0$ reduces then to 
 \begin{eqnarray}
(\frac{\partial}{\partial X_0}-i\frac{\partial}{\partial X_{d+1}})|\psi_0\rangle=0.
\end{eqnarray} 
We compute (by dropping the  derivative with respect to $\hat{x}_{\mu}$ in $\partial/\partial r$)
\begin{eqnarray}
\frac{\partial}{\partial X_0}=\frac{\cos r}{R}\big(-\sin\tau\frac{\partial}{\partial \tau}+\tan^{-1} r\cos\tau\frac{\partial}{\partial r}\big).
\end{eqnarray} 
\begin{eqnarray}
\frac{\partial}{\partial X_{d+1}}=\frac{\cos r}{R}\big(\cos\tau\frac{\partial}{\partial \tau}+\tan^{-1} r\sin\tau\frac{\partial}{\partial r}\big).
\end{eqnarray} 
The above condition becomes then 
 \begin{eqnarray}
(-i\sin r\frac{\partial}{\partial \tau}+\cos r\frac{\partial}{\partial r})|\psi_0\rangle=0.
\end{eqnarray} 
This is immediately solved by 
\begin{eqnarray}
\psi_0\sim \exp(i\Delta t)\cos^{\Delta}r.
\end{eqnarray} 
This fundamental result can be found from another route. The generalization of the Klein-Gordon equation (\ref{KG}) to $AdS_{d+1}$ is given by
\begin{eqnarray}
-\partial_{\tau}^2\phi=\bigg(-\partial_r^2+\frac{1-d}{\cos r\sin r}\partial_r+\frac{l(l+d-2)}{\sin^2\rho}+\frac{R^2m^2}{\cos^2 r}\bigg)\phi.
\end{eqnarray}
In this equation $l(l+d-2)$ are the eigenvalues of the Laplacian on the sphere ${\bf S}^{d-1}$ with corresponding eigenfunctions given by the $d-$dimensional spherical harmonics $Y_{lm}$ where $m$ is the corresponding set of magnetic quantum numbers on $AdS_{d+1}$. In other words, the complete scalar field on $AdS_{d+1}$ is actually $Y_{lm}(\Omega)\phi(\tau,r)$. We separate the remaining variables as $\phi(\tau,r)=\exp(iE\tau)\chi(\rho)$. For the ground state we must have $E=\Delta$ and $l=0$ and the Klein-Gordon equation reduces to
 \begin{eqnarray}
-\Delta^2\chi=\bigg(\partial_r^2+\frac{d-1}{\cos r\sin r}\partial_r-\frac{R^2m^2}{\cos^2 r}\bigg)\chi.
\end{eqnarray}
We make the change of variables 
 \begin{eqnarray}
\chi=\sin^{\alpha}r\cos^{\beta}r\hat{\chi}.
\end{eqnarray}
The exponents $\alpha$ and $\beta$ are determined from the requirement that the linear derivative vanishes. We have
\begin{eqnarray}
\sin^{\alpha-1}r\cos^{\beta-1}r\partial_r\hat{\chi}\bigg(-2\beta+d-1+(2\beta+2\alpha)\cos^2r\bigg)=0\Rightarrow \beta=-\alpha=\frac{d-1}{2}.
\end{eqnarray}
Hence 
\begin{eqnarray}
\partial_r^2+\frac{d-1}{\cos r\sin r}\partial_r=\sin^{\alpha}r\cos^{\beta}r\partial_r^2\hat{\chi}+\sin^{\alpha}r\cos^{\beta}r\hat{\chi}\bigg[\frac{\beta(\beta -d)}{\cos^2r}+\frac{\beta(\beta -d+2)}{\sin^2r}\bigg].
\end{eqnarray}
The equation of motion becomes 
\begin{eqnarray}
-\Delta^2\hat{\chi}=\partial_r^2\hat{\chi}-\frac{1}{4}\bigg[\frac{d^2-1+4R^2m^2}{\cos^2 r}+\frac{(d-1)(d-3)}{\sin^2r}\bigg]\hat{\chi}.
\end{eqnarray}
We propose the solution 
\begin{eqnarray}
\hat{\chi}=\sin^{\alpha_1}r\cos^{\beta_1}r.
\end{eqnarray}
We compute immediately the second derivative 
\begin{eqnarray}
-(\alpha_1+\beta_1)^2\hat{\chi}=\partial_r^2\hat{\chi}-\bigg[\frac{\beta_1(\beta_1-1)}{\cos^2 r}+\frac{\alpha_1(\alpha_1-1)}{\sin^2r}\bigg]\hat{\chi}.
\end{eqnarray}
By comparing the above two final equations we obtain
\begin{eqnarray}
\alpha_1(\alpha_1-1)=\frac{1}{4}(d-1)(d-3)\Rightarrow \alpha_1=\frac{d-1}{2}.
\end{eqnarray}
\begin{eqnarray}
\beta_1(\beta_1-1)=\frac{1}{4}(d^2-1+4R^2m^2)\Rightarrow \beta_1=\frac{1}{2}+\frac{1}{2}\sqrt{d^2+4R^2m^2}.
\end{eqnarray}
\begin{eqnarray}
\alpha_1+\beta_1=\Delta\Rightarrow \frac{d-1}{2}+\frac{1}{2}+\frac{1}{2}\sqrt{d^2+4R^2m^2}=\Delta\Rightarrow R^2m^2=\Delta(\Delta-d).
\end{eqnarray}

\section{Representation theory of the conformal group}

\subsection{More on dilatation operator and primary/descendant operators}
Hamiltonian quantization of quantum field theory involves foliation of the $d-$dimensional spacetime by equal time $(d-1)-$dimensional surfaces characterized by the same Hilbert space. The unitary evolution operator $U=\exp(iH(t_2-t_1)$ allows us to advance from the  surface $t=t_1$ to the surface $t=t_2$. The theory in this case is covariant under the Poincare group and the states of the Hilbert space are specified by two quantum numbers: mass $m$ and spin $s$.
  
In conformal field theory the dilatation operator is what plays the role of the Hamiltonian and the scaling dimension $\Delta$ is what plays the role of the momentum. In this case Euclidean spacetime is foliated using spheres ${\bf S}^{d-1}$ characterized by the same Hilbert space. This is called radial quantization. By the action of the dilatation operator we move from one sphere to another. States of the Hilbert space are specified now by the scaling dimension $\Delta$ and the spin $s$, viz
\begin{eqnarray}
D|\Delta\rangle=i\Delta|\Delta\rangle.
\end{eqnarray}
\begin{eqnarray}
M_{\mu\nu}|\Delta,s\rangle=\Sigma_{\mu\nu}|\Delta,s\rangle.
\end{eqnarray}
The metric reads (with $\Omega$ the solid angle on ${\bf S}^{d-1}$ and $\tau=\log r$)
\begin{eqnarray}
ds^2&=&dr^2+r^2d\Omega^2\nonumber\\
&=&e^{2\tau}(d\tau^2+d\Omega^2).
\end{eqnarray}
This metric is conformally equivalent to the metric on the cylinder. In other words, the transformation $r\longrightarrow \tau=\log r$ maps ${\bf R}\times {\bf S}^{d-1}$ to ${\bf R}^{d}$. The parameter $\tau$ plays then the role of the time parameter. The lower base of the cylinder is at the infinite past $\tau\longrightarrow -\infty$ ($r=0$) whereas the upper base of the cylinder is at the infinite future $\tau\longrightarrow +\infty$ ($r=+\infty$). Going around the cylinder is given by the solid angle    $\Omega$. The evolution operator is given by 
\begin{eqnarray}
U|\Delta\rangle=\exp(i\tau D)|\Delta\rangle=r^{-\Delta}|\Delta\rangle.
\end{eqnarray}
There is a unique vacuum state $|0\rangle$ which is invariant under the global conformal group. This corresponds to no operator insertion in the cylinder which would create a state at a given time $\tau$ (corresponding to a given radius $r$).

Let ${\cal O}_{\Delta}(x)$ be some operator with scaling dimension $\Delta$. The insertion of this operator at the origin $r=0$ (or infinite past $\tau=-\infty$) creates the state $|\Delta\rangle={\cal O}_{\Delta}(0)|0\rangle$ with scaling dimension $\Delta$. By inserting the operator ${\cal O}_{\Delta}(x)$ at an arbitrary point $x$ will create the state 
 \begin{eqnarray}
|\chi\rangle&=&{\cal O}_{\Delta}(x)|0\rangle\nonumber\\
&=&\exp(iPx) {\cal O}_{\Delta}(0)\exp(-iPx)|0\rangle\nonumber\\
&=&\exp(iPx) |\Delta\rangle.
\end{eqnarray}
The momentum operator $P_{\mu}$ is a raising operator with respect to the eigenvalues of the dilatation operator, i.e. it raises the scaling dimension $\Delta$ by unity. Thus, by expanding the exponential $\exp(iPx)$ and acting with the momentum operator we obtain a linear superposition of states with different eigenvalues $\Delta$. 

Similarly, the special conformal generator  $K_{\mu}$ is a lowering operator with respect to the eigenvalues of the dilatation operator, i.e. it lowers the scaling dimension $\Delta$ by unity. An operator annihilated by  $K_{\mu}$ is called a primary operator. By acting on this primary operator with  $P_{\mu}$ we obtain the so-called descendant operators. The primary operator and its descendant operators form a conformal family.

Each state then corresponds to an operator and vice versa. This state-operator correspondence is one-to-one. For example, by inserting a primary operator with scaling dimension $\Delta$ at the origin we obtain a state with scaling dimension $\Delta$ annihilated by  $K_{\mu}$. Conversely, given a state with a scaling dimension $\Delta$  annihilated by  $K_{\mu}$ we can construct a local primary operator at the origin by constructing its correlators with other operators, viz
 \begin{eqnarray}
\langle\phi(x_1)\phi(x_2)...{\cal O}_{\Delta}(0)\rangle=\langle 0|\phi(x_1)\phi(x_2)...|\Delta\rangle
\end{eqnarray}

\subsection{Representation theory of $O(4,2)$}
A concise description of this topic can be found in \cite{Ferrara:1998pr,Zaffaroni:2000vh} and references therein. See also \cite{Mack:1969rr,Gunaydin:1998sw,Dirac:1936fq,Ferrara:1973yt,Ferrara:1998jm}.

The conformal group in $4$ dimensions in Lorentzian signature is given by $SO(4,2)$ or more precisely $O(4,2)$. There are $15$ generators ${\cal M}_{AB}=-{\cal M}_{BA}$. The irreducible representations of the conformal group are infinite dimensional. They are characterized by the eigenvalues of the three Casimir (quadratic, cubic and quartic) operators
  \begin{eqnarray}
C_I={\cal M}_{AB}{\cal M}^{AB}.
\end{eqnarray}
 \begin{eqnarray}
C_{II}=\epsilon_{ABCDEF}{\cal M}^{AB}{\cal M}^{CD}{\cal M}^{EF}.
\end{eqnarray}
\begin{eqnarray}
C_{III}={\cal M}_A^{B}{\cal M}_B^{C}{\cal M}_C^{D}{\cal M}_D^{A}.
\end{eqnarray}
An infinite dimensional irreducible representation of the conformal group is determined by an irreducible representation of the Lorentz group with definite conformal dimension and annihilated by the special conformal operators $K_{\mu}$. The stability algebra at the origin consists of the generators $D$, $K_{\mu}$ and $M_{\mu\nu}$. A primary conformal operator $O$ (the lowest weight state) in a given representation of the Lorentz group is defined by 
\begin{eqnarray}
[D,O(0)]=i\Delta O(0).
\end{eqnarray}
\begin{eqnarray}
[K,O(0)]=0.
\end{eqnarray}
The descendants $\partial...\partial O(0)$ are obtained by the repeated action of the momentum operators $P_{\mu}$. The eigenvalues of the Lorentz operators $M_{\mu\nu}$ on the primary operator  $O$ are spin quantum numbers denoted for example by $j_L$ and $j_R$. This defines an irreducible representation of the conformal group characterized by $\Delta$, $j_L$ and $j_R$.

The three Casimirs of the stability algebra are 
\begin{eqnarray}
{\cal D}=\Delta.
\end{eqnarray}
\begin{eqnarray}
\frac{1}{2}M_{\mu\nu}M^{\mu\nu}=j_L(j_L+1)+j_R(j_R+1).
\end{eqnarray}
\begin{eqnarray}
\frac{1}{2}\epsilon_{\mu\nu\alpha\beta}M_{\mu\nu}M^{\alpha\beta}=j_L(j_L+1)-j_R(j_R+1).
\end{eqnarray}
The eigenvalues of the conformal group $O(4,2)$ are then given by 
 \begin{eqnarray}
C_I=\Delta(\Delta-4)+2j_L(j_L+1)+2j_R(j_R+1).
\end{eqnarray}
 \begin{eqnarray}
C_{II}=(\Delta-2)\bigg(j_L(j_L+1)-j_R(j_R+1)\bigg).
\end{eqnarray}
\begin{eqnarray}
C_{III}=(\Delta -2)^4-4(\Delta-2)^2\bigg(j_L(j_L+1)+j_R(j_R+1)+1\bigg)+16j_Lj_R(j_L+1)(j_R+1).\nonumber\\
\end{eqnarray}
In particular for tensor representations of spin $s$ associated with the quantum numbers $(\Delta,j_L=s/2,j_R=s/2)$ we get 
\begin{eqnarray}
C_I=\Delta(\Delta-4)+s(s+2).\label{ci}
\end{eqnarray}
 \begin{eqnarray}
C_{II}=0.
\end{eqnarray}
\begin{eqnarray}
C_{III}=\bigg(\Delta(\Delta -2)-s(s+2)\bigg)\bigg((\Delta-2)(\Delta-4)-s(s+2)\bigg).
\end{eqnarray}
The requirement of unitarity imposes the following constraints on the possible values of $\Delta$, $j_L$ and $j_R$. We have 
\begin{eqnarray}
\Delta\geq j+1~,~j_Lj_R=0.\label{fc1}
\end{eqnarray}
\begin{eqnarray}
\Delta\geq j_L+j_R+2~,~j_Lj_R\neq0.\label{fc2}
\end{eqnarray}
The first constraint (\ref{fc1}) is saturated by massless fields satisfying $\partial^2\Phi_{(0,j)}=0$. Indeed, this wave equation is conformally covariant only if $\Delta=j+1$. Similarly, the second constraint (\ref{fc2}) is saturated by conserved tensor fields satisfying $\partial^{\alpha_1\dot{\alpha}_1}O_{\alpha_1...\alpha_{2j_L},\dot{\alpha}_1...\dot{\alpha}_{2j_R}}=0$ which is a conformally covariant equation only if $\Delta=2+j_L+j_R$. 
\subsection{$O(4,2)$ and isometries of $AdS_5$}
The conformal group $O(d,2)$ is the isometry group of $AdS_{d+1}$. By the AdS/CFT correspondence there is a $d-$dimensional conformal field theory (${\rm CFT}_d$) living on the boundary of $AdS_{d+1}$ where gauge invariant composite operators in the ${\rm CFT}_d$ are associated with fields in $AdS_{d+1}$.   

This means in particular that the scaling dimension $\Delta$ can be re-interpreted as the energy of a particle moving in anti-de Sitter space. Indeed, particles in $AdS_5$ are characterized by the quantum numbers $(E,j_L,j_R)$ where the energy $E$ is identified with the scaling dimension of the corresponding conformal primary operator living on the boundary. 

Indeed, the covariant wave equation of a particle in $AdS_5$ can be re-expressed in terms of the Casimir of the conformal group and hence the mass of the particle can be determined in terms of the quantum numbers 
$(E,j_L,j_R)$. For example, for a scalar field with $j_L=j_R=0$ the Laplacian operator in $AdS_5$ is precisely the Casimir operator $C_I$ and hence from (\ref{ci}) we obtain the mass squared $m^2=\Delta(\Delta-4)$ which will be obtained more directly in due course. By using also equation (\ref{ci}) we obtain for fermions of spin $1/2$ with $j_L=1/2$, $j_R=0$ or $j_L=0$, $j_R=1/2$  the mass $m=\Delta-2$, and for vector fields of spin $1$ with $j_L=j_R=1/2$ we get the mass squared $m^2=\Delta(\Delta-4)+3=(\Delta-1)(\Delta-3)$, whereas for symmetric tensor fields of spin $2$ with $j_L=j_R=1$ we obtain the mass squared $m^2=\Delta(\Delta-4)$. See \cite{Ferrara:1998pr} for more detail.

The generators $D$ and $M_{\mu\nu}$, giving the quantum numbers $(\Delta,j_L,j_R)$,  correspond to the non-compact subgroup $O(3,1)$ of the conformal group $O(4,2)$. However, we observe that the operators in the representations $(\Delta,j_L,j_R)$ yields, when applied to the vacuum,  non-normalizable states. This is because  these states can not furnish a finite dimensional unitary representation of a non-compact group.

Another set of good quantum numbers corresponds to the maximal compact subgroup $O(2)\times O(4)$ of the conformal group $O(4,2)$. This compact group $O(2)\times O(4)$ allows us to obtain finite dimensional unitary representations of the conformal group using states with finite norm. Indeed, the quantum numbers $(E,j_L,j_R)$ can now be viewed as the eigenvalues of the Cartan generators of $O(2)\times O(4)$. In particular, the quantum number $E$ is associated with the $O(2)$ generator $H=(K_0+P_0)/2$ which is called the conformal energy. This can also be seen by going to the Euclidean spacetime ${\bf R}^4$ which can be mapped via a conformal transformation to ${\bf R}\times{\bf S}^3$ (radial quantization). In this case the group $O(2)$ is seen acting on ${\bf R}$ and hence $H$ is the Hamiltonian corresponding to translations in this direction whereas the factor $O(4)$ acts on ${\bf S}^3$. 

Furthermore, we remark that since $P_0$ is a raising operator and $K_0$ is a lowering operator the eigenvalue $E$ seems to be integer-valued. But in the quantum theory it is the covering space of the conformal group that is being realized and is obtained by unwinding the factor $O(2)$ giving rise to a continuous spectrum of $E$. Hence the identification of $E$ with $\Delta$.

\subsection{The fields and operators in ${\rm CFT}_d$}
Although the scaling dimension $\Delta$ in the ${\rm CFT}_4$ can be identified with the conformal energy $E$ in the $AdS_5$ we strictly speaking do not have particle states in a conformal field theory. The first obvious reason is that the mass operator $P_{\mu}P^{\mu}$ is not a Casimir of the conformal group, i.e. it does not commute with the dilatation operator. Hence if a state in a given representation of the conformal group has an energy $E_0$, then by the action of the dilatation operator we can obtain states in this representation with any other value of the energy between $0$ and $\infty$. 

This can be understood more precisely by means of the Kallen-Lehmann  spectral representation of the two-point function of a general interacting ${\rm QFT}_d$ given in terms of the free propagator $\Delta(p,\mu^2)$ and the spectral density $\rho(\mu^2)$ by the relation \cite{Kallen:1952zz,Lehmann:1954}

\begin{eqnarray}
\Delta(p)=\int_0^{\infty}\rho(\mu^2)\Delta_0(p,\mu^2) d\mu^2~,~\Delta(p,\mu^2)=\frac{1}{p^2-\mu^2+i\epsilon}.
\end{eqnarray}
The spectral density $\rho(\mu^2)$ encodes the contributions of the states $|s\rangle$ with momenta $p_s$ to the two-point function and it is given explicitly by
\begin{eqnarray}
\rho(p^2)=\sum_s\delta^d(p-p_s)|\langle s|\phi\rangle|^2.
\end{eqnarray}
For a free scalar field $\phi$ we have 
\begin{eqnarray}
\rho(p^2)=\delta^d(p^2-m^2).
\end{eqnarray}
This corresponds to a single massive excitation. If $\phi$ overlaps with heavier states then there will be other terms with $p^2> m^2$.

On the other hand, for a conformal field theory in $4-$dimension we know that the two-point function should behave as 
\begin{eqnarray}
\Delta(x)\sim\frac{1}{x^2}\Rightarrow \rho(p^2)=\delta^4(p^2).
\end{eqnarray}
This also corresponds to a single massless excitation. But for a generic scalar operator in  ${\rm CFT}_4$ characterized by a non-trivial scaling dimension (anomalous dimension) the behavior of the two-point function is altered as 
\begin{eqnarray}
\Delta(x)\sim\frac{1}{x^{2+2\delta}}\Rightarrow \rho(p^2)=(p^2)^{\delta-1}~,~\delta >0.
\end{eqnarray}
This is a continuous power-law spectrum characterized by $\delta$. In other words, there is no mass scale nor a discrete set of particles but the operator just creates a scale-invariant continuous set of states.

Hence, fields in an ordinary ${\rm QFT}_d$ are local operators which furnish a representation of the Lorentz group, generate the Hilbert space, and create particle states. But in a ${\rm CFT}_d$ the basic objects are operators which are not necessarily  fields since they do not create particle-like excitations. Thus, the formalism of scattering theory and the $S-$matrix does not apply for conformal field theory.

\subsection{Unitary bounds revisited}
The states which saturate the unitarity bound (\ref{fc1}) are called singleton and they are topological configurations living at the boundary of $AdS_5$ associated with fundamental fields (and not gauge invariant operators)  of the ${\rm CFT}_4$.

On the other hand, the constraint equation (\ref{fc2}) enjoys a profound physical meaning. Recall that this inequality is saturated by conserved tensor fields satisfying $\partial^{\alpha_1\dot{\alpha}_1}O_{\alpha_1...\alpha_{2j_L},\dot{\alpha}_1...\dot{\alpha}_{2j_R}}=0$ which is a conformally covariant equation only if $\Delta=2+j_L+j_R$. These conserved tensor fields are precisely the conserved currents in the ${\rm CFT}_4$ which are associated with massless fields in $AdS_5$ with local gauge invariance. In other words, global symmetries in ${\rm CFT}_4$ corresponds to local symmetries in ${\rm AdS}_5$. For example, the energy-momentum tensor $T_{\mu\nu}$ is associated with the graviton field $g_{\mu\nu}$ and the global current $J_{\mu}$ is associated with the gauge field $A_{\mu}$, etc. The equation $\partial^{\alpha_1\dot{\alpha}_1}O_{\alpha_1...\alpha_{2j_L},\dot{\alpha}_1...\dot{\alpha}_{2j_R}}=0$ satisfied by these conserved currents means that the number of degrees of freedom contained in the conserved tensor in ${\rm CFT}_4$ is precisely the number of degrees of freedom contained in the massless field in $AdS_5$. Obviously, the tensor contains $(2j_L+1)(2j_R+1)-(2j_L)(2j_R)=2(j_L+j_R)+1$ degrees of freedom, i.e. a massless particle of spin $j_L+j_R$. For $\Delta>2+j_L+j_R$ the fields in $AdS_5$ are massive and the corresponding tensor fields are not conserved. See \cite{Ferrara:1998pr} for more detail.

\section{Holography}
The number of degrees of freedom, or equivalently the amount of information, contained in a quantum system is measured as we know by thermodynamic entropy. In quantum mechanics the entropy is an extensive quantity and thus the entropy of a $d-$dimensional spatial region ${\bf R}^d$ is proportional to its volume  $V_d$. 

However, in quantum gravity the entropy is sub-extensive, i.e. the entropy of a $d-$dimensional spatial region ${\bf R}^d$ is actually proportional to the surface area $S_{d-1}=\partial V_d$ which bounds its volume $V_d$ and not proportional to the volume $V_d$ itself. In other words, the entropy of a $d-$dimensional spatial region, in a gravitational theory, is bounded by the entropy of the black hole which fits inside that spatial region. This is essentially what is called the holographic principle introduced first by 't Hooft \cite{tHooft:1993dmi} and then extended to string theory by Susskind \cite{Susskind:1994vu} (see also \cite{Bousso:2002ju}). As we can see this principle is largely inspired by the Bekenstein-Hawking formula which  states that the entropy of a black hole ${\cal S}_{BH}$ is proportional to the surface area $A_H$ of the black hole horizon with the constant of proportionality equal $1/4G_N$ where $G_N$ is Newton's constant, viz 
\begin{eqnarray}
{\cal S}_{BH}=\frac{A_H}{4G_N}.
\end{eqnarray} 
The holographic principle provides therefore a partial answer to the question of how could a higher dimensional gravity theory (${\rm AdS}_5$) contain the same number of degrees of freedom, the same amount information, and have the same entropy as a lower dimensional quantum field theory (${\rm CFT}_4$), i.e. it lies at the heart of the celebrated AdS/CFT correspondence \cite{Maldacena:1997re}.

This can be seen more explicitly as follows. By the AdS/CFT correspondence, the AdS space $AdS_{d+1}$ is the gravity dual of a $d-$dimensional conformal field theory ${\rm CFT}_d$ living on the boundary $z=\epsilon$ of AdS space. The  radial coordinate $z$ should be thought of as a lattice spacing, i.e. as a UV cutoff. Thus,  the boundary theory is a quantum field theory on a $d-$dimensional lattice with lattice spacing $\epsilon$. At any given instant of time, the boundary theory is also regulated by placing it in a spatial box of size $R$ (IR cutoff). Hence, the number of cells in the box is given by $(R/L)^{d-1}$.

The central charge $c_{\rm QFT}$ of the CFT is by definition equal to the number of degrees freedom per lattice site. Thus the total number of degrees of freedom contained in the box is given by
\begin{eqnarray}
N_{\rm QFT}=(\frac{R}{\epsilon})^{d-1}c_{\rm QFT}.\label{qft}
\end{eqnarray}
From the AdS space side the estimation of the degrees of freedom can be carried out as follows. The metric at $z=\epsilon$ is
\begin{eqnarray}
  ds^2_{d}=\frac{L^2}{\epsilon^2}dx_{\mu}dx^{\mu}.
\end{eqnarray}
By using the holographic principle, i.e. the Bekenstein-Hawking formula, the number of degrees of freedom at a given instant of time contained in the spatial volume of AdS space is given by the maximum entropy given by
\begin{eqnarray}
  N_{\rm AdS}=\frac{A_H}{4G_N}.
\end{eqnarray}
Here $A_H$ is the area of the spatial boundary of $AdS_{d+1}$ delimiting the spatial volume of AdS space. We use the metric to compute this area as follows
\begin{eqnarray}
A_H=\int_{z=\epsilon}d^{d-1}x\sqrt{g}=(\frac{L}{\epsilon})^{d-1}\int d^{d-1}x=(\frac{L}{\epsilon})^{d-1}R^{d-1}.
\end{eqnarray}
By using also the fact that for gravity in $d+1$ dimensions the Newton constant is given by $G_N=l_P^{d-1}=1/M_P^{d-1}$ we arrive at the result
\begin{eqnarray}
  N_{\rm AdS}=\frac{1}{4}(\frac{R}{\epsilon})^{d-1}(\frac{L}{l_P})^{d-1}.\label{ads}
\end{eqnarray}
By comparing (\ref{qft}) and (\ref{ads}) we obtain the central charge
\begin{eqnarray}
  c_{\rm QFT}=\frac{1}{4}(\frac{L}{l_P})^{d-1}.\label{rm}
\end{eqnarray}
Hence semi-classical gravity corresponding to $L>>l_P$ is dual to a CFT with a large central charge. If the conformal field theory is an $SU(N)$ gauge theory then the central charge is proportional to $N^2$ and as a consequence semi-classical gravity is dual in this case to a large $N$ gauge theory.

\section{The AdS/CFT correspondence}
In this section we follow the presentation of \cite{Ramallo:2013bua}.
\subsection{Approaching the AdS boundary}
We go back to Euclidean $AdS_{d+1}$ in the Poincare patch:
\begin{eqnarray}
ds^2_{d+1}=\frac{L^2}{z^2}\bigg(dz^2+d\vec{x}^2+dt^{2}\bigg).
\end{eqnarray}
The action of a scalar field in $AdS_{d+1}$ is given by 
\begin{eqnarray}
S=\int d^{d+1}x\sqrt{g}\bigg[-\frac{1}{2}g^{MN}\partial_M\phi\partial_N\phi-\frac{1}{2}m^2\phi^2\bigg].
\end{eqnarray}
The Klein-Gordon equation in the $AdS_{d+1}$ background reads explicitly 
\begin{eqnarray}
z^{d+1}\partial_z(z^{1-d}\partial_z\phi)+z^2\partial^2\phi-m^2L^2\phi=0.
\end{eqnarray}
We perform Fourier transform in the $x-$space, viz
\begin{eqnarray}
\phi(z,x)=\int \frac{d^dk}{(2\pi)^d}\exp(ikx)f_k(z).
\end{eqnarray}
The Klein-Gordon equation reduces to
\begin{eqnarray}
z^{d+1}\partial_z(z^{1-d}\partial_zf_k)-k^2z^2f_k-m^2L^2f_k=0.\label{KGF}
\end{eqnarray}
Near the conformal boundary $z=0$ we may expect $f_k\sim z^{\beta}$. This gives immediately 
\begin{eqnarray}
\beta(\beta-d)-m^2L^2=0\Rightarrow \beta=\frac{d}{2}\pm\sqrt{\frac{d^2}{4}+m^2L^2}.
\end{eqnarray}
The solution near the boundary is therefore of the general form
\begin{eqnarray}
f_k(z)\longrightarrow A(k)z^{d-\Delta}+B(k)z^{\Delta}~,~z\longrightarrow 0.\label{KGF1}
\end{eqnarray}
The exponent $\Delta$ is the so-called scaling dimension of the field and it is given by
\begin{eqnarray}
\Delta=\frac{d}{2}+\sqrt{\frac{d^2}{4}+m^2L^2}.
\end{eqnarray}
The scaling dimension $\Delta$ is real iff the mass $m$ satisfies the Breitenlohner-Freedman (BF) bound
\begin{eqnarray}
m^2>-\frac{d^2}{4L^2}.
\end{eqnarray}
In this case $d-\Delta \leq \Delta$ and hence $z^{d-\Delta}$ is the dominant term as $z\longrightarrow 0$. We place the boundary at $z=\epsilon\longrightarrow 0$. Then the behavior of the scalar field on the boundary is given by
\begin{eqnarray}
\phi(z=\epsilon,x)= A(x)\epsilon^{d-\Delta}.
\end{eqnarray}
For $m^2>0$ the exponent $d-\Delta$ is negative and hence this field is divergent. The quantum field theory source $\varphi(x)$, i.e. the scalar field living on the boundary, is then identified with $A(x)$, viz
\begin{eqnarray}
  \varphi(x)={\rm lim}_{\epsilon\longrightarrow 0}\epsilon^{\Delta-d}\phi(\epsilon,x).
\end{eqnarray}
This is the scalar field representing (or dual to) the anti-de Sitter scalar field $\phi(z,x)$ at the boundary $z=0$. The field $\varphi$ is the holographic dual of the AdS field $\phi$. The scaling dimension of the source $\varphi$ is given by $d-\Delta$.  

Let ${\cal O}(z,x)$ and ${\cal O}(x)$ be the dual operators to the scalar fields $\phi(z,x)$ and $\varphi(x)$ respectively. Their coupling is a boundary term of the form
\begin{eqnarray}
  S_{\rm bound}&=&\int d^dx\sqrt{\gamma}\phi(\epsilon,x){\cal O}(\epsilon,x)\nonumber\\
  &=&\int d^dx\frac{L^d}{\epsilon^d} \epsilon^{d-\Delta}\varphi(x){\cal O}(\epsilon,x)\nonumber\\
  &=&L^d \int d^dx\varphi(x){\cal O}(x),
\end{eqnarray}
where
\begin{eqnarray}
   {\cal O}(\epsilon,x)=\epsilon^{\Delta}{\cal O}(x).
\end{eqnarray}
This is the wave function renormalization of the operator ${\cal O}$ as we move into the bulk. This also shows explicitly that $\Delta$ is the scaling dimension of the dual operator ${\cal O}$ since going from $z=0$ to $z=\epsilon$ is a dilatation operation in the quantum field theory.

In summary, the scaling dimensions of the source $\varphi$ and its dual operator ${\cal O}$ are given by $d-\Delta$ and $\Delta$ respectively. Any deformation of the conformal field theory with the operator ${\cal O}$ takes then the form
\begin{eqnarray}
  \Delta S=\int d^dx M^{d-\Delta}{\cal O}.
\end{eqnarray}
We distinguish the usual three cases:
\begin{itemize}
\item For $m^2>0$ we have $\Delta>d$ and thus corrections to various amplitudes are of the form $(E/M)^{\alpha}$ for some positive $\alpha$. In other words, these corrections are negligible for low energies and the operator ${\cal O}$ does not change the IR behavior of the theory. It is an irrelevant operator.
\item For $m^2=0$ we have $\Delta=d$ and the operator ${\cal O}$ is marginal.
  \item For $m^2<0$ we have $\Delta<d$ and  the operator ${\cal O}$ is relevant since it changes the IR behavior of the theory.
\end{itemize}
\subsection{Correlation functions}
We have now an operator ${\cal O}(x)$ living on the AdS boundary. We are interested in computing  the correlation functions
 \begin{eqnarray}
\langle{\cal O}(x_1)...{\cal O}(x_n)\rangle.
 \end{eqnarray}
These will determine the CFT living on the boundary completely.  On the conformal field theory side with lagrangian ${\cal L}$ the calculation of these correlation functions  proceeds as usual by introducing the generating functional
 \begin{eqnarray}
Z_{\rm CFT}[J]=\int ~\exp({\cal L}+\int J(x){\cal O}(x))=\langle\exp(\int J(x){\cal O}(x))\rangle .
 \end{eqnarray}
 Then we have immediately 
 \begin{eqnarray}
\langle{\cal O}(x_1)...{\cal O}(x_n)\rangle=\frac{\delta^n\log Z_{CFT}[J]}{\delta J(x_1)...\delta J(x_n)}|_{J=0}.
 \end{eqnarray}
 The operator ${\cal O}(x)$ is sourced by a scalar field $\varphi (x)$ living on the boundary which is related to the boundary value of the bulk scalar field $\phi(z,x)$ by the relation
 \begin{eqnarray}
  \varphi(x)={\rm lim}_{z\longrightarrow 0}z^{\Delta-d}\phi(z,x).
 \end{eqnarray}
The boundary scalar field $\phi_0(x)$ is actually divergent and it is simply defined by 
  \begin{eqnarray}
  \phi_0(x)=\phi(0,x).
  \end{eqnarray}
  The AdS/CFT correspondence states then that the CFT generating functional with source $J=\phi_0$ is equal to the path integral on the gravity side evaluated over a bulk field which has the value $\phi_0$ at the boundary of AdS  \cite{Gubser:1998bc,Witten:1998qj}. We write
  \begin{eqnarray}
Z_{\rm CFT}[\phi_0]=\langle\exp(\int \phi_0(x){\cal O}(x))\rangle =\int_{\phi\longrightarrow \phi_0} {\cal D}\phi\exp(S_{\rm grav})=Z_{\rm grav}[\phi\longrightarrow\phi_0].
 \end{eqnarray}
  In the limit in which classical gravity is a good approximation the gravity path integral can be replaced by the classical amplitude given by the classical on-shell gravity action, i.e.

  \begin{eqnarray}
    Z_{\rm CFT}[\phi_0]=\exp(S_{\rm grav}^{\rm on-shell}[\phi\longrightarrow\phi_0]). 
 \end{eqnarray}
  Typically the on-shell gravity action is divergent requiring thus regularization and renormalization using the so-called holographic renormalization  \cite{Henningson:1998gx,deHaro:2000vlm,Skenderis:2002wp}. The on-shell action gets renormalized and the above prescription becomes
  \begin{eqnarray}
    Z_{\rm CFT}[\phi_0]=\exp(S_{\rm grav}^{\rm renor}[\phi\longrightarrow\phi_0]). 
  \end{eqnarray}
  The correlation functions are then renormalized as
  \begin{eqnarray}
\langle{\cal O}(x_1)...{\cal O}(x_n)\rangle=\frac{\delta^n S_{\rm grav}^{\rm renor}[\phi\longrightarrow\phi_0]}{\delta\varphi(x_1)...\delta\varphi(x_n)}|_{\varphi=0}.
  \end{eqnarray}
  Let us compute the one-point correlation function, viz
   \begin{eqnarray}
\langle{\cal O}(x)\rangle_{\varphi}=\frac{\delta S_{\rm grav}^{\rm renor}[\phi\longrightarrow\phi_0]}{\delta\varphi(x)}={\rm lim}_{z\longrightarrow 0}z^{d-\Delta}\frac{\delta S_{\rm grav}^{\rm renor}[\phi\longrightarrow\phi_0]}{\delta \phi(z,x)}.\label{rt2}
   \end{eqnarray}
   We can rewrite this in terms of an action of the form
   \begin{eqnarray}
S_{\rm grav}=\int_{\cal M}dz d^dx{\cal L}(\phi,\partial\phi).
   \end{eqnarray}
   Under $\phi\longrightarrow\phi+\delta\phi$, and assuming the Euler-Lagrange equations of motion, and also by assuming that the variation $\delta\phi$ vanishes whenever  $x\longrightarrow\pm \infty$ or $z\longrightarrow \infty$, we obtain the on-shell variation
   \begin{eqnarray}
\delta S_{\rm grav}^{\rm on-shell}=\int_{\partial \cal M}d^dx\Pi.\delta\phi|_{z=\epsilon}.
   \end{eqnarray}
   The canonical momentum with respect to $z$ is given by
    \begin{eqnarray}
      \Pi=\frac{\delta S_{\rm grav}^{\rm on-shell}}{\delta\phi(z,x)}=-\frac{\delta {\cal L}}{\delta\partial_z \phi(z,x)}.
   \end{eqnarray}
   The renormalized action is obtained by adding counter terms, viz
    \begin{eqnarray}
S_{\rm grav}^{\rm renor}=  S_{\rm grav}^{\rm on-shell}+S_{\rm ct}.                          
    \end{eqnarray}
    We define in this case the renormalized canonical momentum as
     \begin{eqnarray}
       \Pi^{\rm renor}=\frac{\delta S_{\rm grav}^{\rm renor}}{\delta\phi(z,x)}.
   \end{eqnarray}
     Hence
      \begin{eqnarray}
        \langle{\cal O}(x)\rangle_{\varphi}
        ={\rm lim}_{z\longrightarrow 0}z^{d-\Delta} \Pi^{\rm renor}(z,x).
      \end{eqnarray}
      From the one-point function we can obtain the two-point function as follows. In the quantum field theory with fields $\psi$ the one-point function is given by the path integral
      \begin{eqnarray}
 \langle{\cal O}(x)\rangle_{\varphi}=\int [d\psi]{\cal O}(x)\exp(S_E[\psi]+\int d^dy\varphi(y){\cal O}(y)).
      \end{eqnarray}
      By expanding in power series of the source $\varphi$ we obtain
      \begin{eqnarray}
        \langle{\cal O}(x)\rangle_{\varphi}=  \langle{\cal O}(x)\rangle_{\varphi=0}+\int d^dyG_E(x-y)\varphi(y)+...
      \end{eqnarray}
      $G_E(x-y)$ is obviously the two-point function defined by
       \begin{eqnarray}
G_E(x-y)=         \langle{\cal O}(x){\cal O}(y)\rangle_{\varphi=0}.
       \end{eqnarray}
       By assuming that the observable ${\cal O}(x)$ has been normal-ordered we have $\langle{\cal O}(x)\rangle_{\varphi=0}=0$. In other words, $ \langle{\cal O}(x)\rangle_{\varphi}$ measures really fluctuations of the observable away from its expectation value without source. Hence
       \begin{eqnarray}
        \langle{\cal O}(x)\rangle_{\varphi}=  \int d^dyG_E(x-y)\varphi(y)+...
       \end{eqnarray}
       In momentum space this reads
        \begin{eqnarray}
        \langle{\cal O}(k)\rangle_{\varphi}=G_E(k)\varphi(k).  
        \end{eqnarray}
        Therefore the two-point function in momentum space is given immediately by
         \begin{eqnarray}
           G_E(k)=\frac{\langle{\cal O}(k)\rangle_{\varphi}}{\varphi(k)}.\label{rt3}
         \end{eqnarray}
         \subsection{The two-point function}
We return to the action of a scalar field in Euclidean $AdS_{d+1}$  given by (including an overall constant $\eta$) 
\begin{eqnarray}
S_{\rm grav}=\eta \int d^{d+1}x\sqrt{g}\bigg[-\frac{1}{2}g^{MN}\partial_M\phi\partial_N\phi-\frac{1}{2}m^2\phi^2\bigg].
\end{eqnarray}
By using the Euler-Lagrange equations of motion we obtain the on-shell action and variation as
\begin{eqnarray}
S^{\rm on-shell}_{\rm grav}=\frac{\eta}{2}\int d^{d}x\big(\sqrt{g}\phi g^{zz}\partial_z\phi\big)_{z=\epsilon}.
\end{eqnarray}
\begin{eqnarray}
\delta S^{\rm on-shell}_{\rm grav}=\eta\int d^{d}x\big(\sqrt{g}\phi g^{zz}\partial_z\phi\big)_{z=\epsilon}.
\end{eqnarray}
The canonical momentum with respect to $z$ is then given by
\begin{eqnarray}
      \Pi=\frac{\delta S_{\rm grav}^{\rm on-shell}}{\delta\phi(z,x)}=-\frac{\delta {\cal L}}{\delta\partial_z \phi(z,x)}=\eta \sqrt{g} g^{zz}\partial_z\phi.
   \end{eqnarray}
Thus the on-shell action is of the form
\begin{eqnarray}
  S^{\rm on-shell}_{\rm grav}&=&\frac{1}{2}\int_{z=\epsilon} d^{d}x\Pi(z,x)\phi(z,x)\nonumber\\
  &=&\frac{1}{2}\int_{z=\epsilon} \frac{d^{d}k}{(2\pi)^d}\pi_{-k}(z) f_k(z).
\end{eqnarray}
The field $f_k(z)$ satisfies the equation of motion (\ref{KGF}). We introduce another function $g_k(z)$ by $f_k(z)=z^{d/2}g_k(z)$ and substitute in (\ref{KGF}) to obtain the differential equation  
\begin{eqnarray}
z^2\partial_z^2g_k+z\partial_zg_k-(\nu^2+k^2z^2)g_k=0~,~\nu^2=\frac{d^2}{4}+m^2L^2=(\Delta-\frac{d}{2})^2.
\end{eqnarray}
This is the modified Bessel equation. The solutions are therefore the modified Bessel functions $g_k(z)=I_{\pm\nu}(kz)$, viz $f_k(z)=z^{d/2}I_{\pm\nu}(kz)$. Recall the small and large $z$ limits of the modified Bessel functions
\begin{eqnarray}
I_{\pm\nu}(kz)\sim \frac{1}{\Gamma(1\pm\nu)}(\frac{kz}{2})^{\pm\nu}~,~z\longrightarrow 0.
\end{eqnarray}
\begin{eqnarray}
I_{\pm\nu}(kz)\sim \frac{e^{kz}}{\sqrt{2\pi kz}}~,~z\longrightarrow \infty.
\end{eqnarray}
The two independent solutions of (\ref{KGF}) are taken to be given by
\begin{eqnarray}
\phi_1(z,k)=\Gamma(1-\nu)(\frac{k}{2})^{\nu}z^{d/2}I_{-\nu}(kz)\longrightarrow z^{d-\Delta}~,~z\longrightarrow 0
\end{eqnarray}
\begin{eqnarray}
\phi_2(z,k)=\Gamma(1+\nu)(\frac{k}{2})^{-\nu}z^{d/2}I_{\nu}(kz)\longrightarrow z^{\Delta}~,~z\longrightarrow 0
\end{eqnarray}
Remark that the small $z$ behavior agrees with the previously found one in (\ref{KGF1}). Equation (\ref{KGF1}) should then be generalized as
\begin{eqnarray}
f_k(z)&=&A(k)\phi_1(z,k)+B(k)\phi_2(z,k)\nonumber\\
&=&z^{d/2}\bigg[\Gamma(1-\nu)(\frac{k}{2})^{\nu}A(k)I_{-\nu}(kz)+\Gamma(1+\nu)(\frac{k}{2})^{-\nu}B(k)I_{\nu}(kz)\bigg].
\end{eqnarray}
The large $z$ limit of this solution is given by
\begin{eqnarray}
f_k(z)&=&z^{d/2} \frac{e^{kz}}{\sqrt{2\pi kz}}\bigg[\Gamma(1-\nu)(\frac{k}{2})^{\nu}A(k)+\Gamma(1+\nu)(\frac{k}{2})^{-\nu}B(k)\bigg]~,~z\longrightarrow\infty.
\end{eqnarray}
This diverges in the limit $z\longrightarrow\infty$ unless the term between brackets vanishes. Thus we obtain a regular field in the IR limit $z\longrightarrow\infty$ iff
\begin{eqnarray}
\frac{B(k)}{A(k)}=-\frac{\Gamma(1-\nu)}{\Gamma(1+\nu)}(\frac{k}{2})^{2\nu}.\label{rt}
\end{eqnarray}
Now in the small $z$ limit the field behaves as
\begin{eqnarray}
  f_k(z)&=&A(k) z^{d-\Delta}+B(k) z^{\Delta}~,~z\longrightarrow 0.
\end{eqnarray}
We compute immediately the small $z$ limit of the conjugate field as
\begin{eqnarray}
  \pi_{-k}(z)&=&\eta L^{d-1}\bigg[(d-\Delta)A(-k) z^{-\Delta}+\Delta B(-k) z^{\Delta-d}\bigg]~,~z\longrightarrow 0.
\end{eqnarray}
The on-shell action becomes

\begin{eqnarray}
  S^{\rm on-shell}_{\rm grav}
  &=&\frac{\eta}{2}L^{d-1}\int_{} \frac{d^{d}k}{(2\pi)^d}\bigg((d-\Delta)A(k)A(-k)\epsilon^{-2\nu}+dA(k)B(-k)\bigg).\label{rt1})
\end{eqnarray}
The first term is divergent and thus this action requires a renormalization. The required counter term is a quadratic local term living on the  boundary of AdS space given by \cite{Ramallo:2013bua}

\begin{eqnarray}
  S^{}_{\rm ct}
  &=&\frac{\eta}{2}\eta_1 \int_{} \sqrt{\gamma}d^dx \phi^2.
\end{eqnarray}
The metric $\gamma$ is the induced metric on the boundary, viz $ds^2=L^2 dx_{\mu}dx_{\mu}/\epsilon^2$. Indeed, we compute
\begin{eqnarray}
  S^{}_{\rm ct}
  &=&\frac{\eta}{2}\eta_1 L^{d}\int_{} \frac{d^{d}k}{(2\pi)^d}\bigg(A(k)A(-k)\epsilon^{-2\nu}+2 A(k)B(-k)\bigg).
\end{eqnarray}
The divergence is canceled iff $\eta_1=-(d-\Delta)/L$. The renormalized action is therefore
\begin{eqnarray}
  S^{\rm renor}_{\rm grav}&=& S^{\rm on-shell}_{\rm grav}+S^{}_{\rm ct}\nonumber\\
  &=&\frac{\eta}{2}L^{d-1}(2\Delta -d)\int_{} \frac{d^{d}k}{(2\pi)^d} A(k)B(-k)\nonumber\\
  &=&-\frac{\eta}{2}L^{d-1}(2\Delta -d)\frac{\Gamma(1-\nu)}{\Gamma(1+\nu)}\int_{} \frac{d^{d}k}{(2\pi)^d} A(k)(\frac{k}{2})^{2\nu}A(-k)\nonumber\\
   &=&-\frac{\eta}{2}L^{d-1}(2\Delta -d)\frac{\Gamma(1-\nu)}{\Gamma(1+\nu)}\int_{} \frac{d^{d}k}{(2\pi)^d} \varphi(k)(\frac{k}{2})^{2\nu}\varphi(-k),
\end{eqnarray}
where we have used (\ref{rt}) and
\begin{eqnarray}
\varphi(x)={\rm lim}_{z\longrightarrow 0}z^{\Delta-d}\phi(z,x)=A(x).
\end{eqnarray}
The one-point correlator of the dual operator on the boundary is given by (\ref{rt2}) or equivalently 
 \begin{eqnarray}
   \langle{\cal O}(k)\rangle_{\varphi}=(2\pi)^d\frac{\delta S_{\rm grav}^{\rm renor}[\phi\longrightarrow\phi_0]}{\delta\varphi(-k)}.
   \end{eqnarray}
   We get 
 \begin{eqnarray}
   \langle{\cal O}(k)\rangle_{\varphi}=-\eta L^{d-1}(2\Delta -d)\frac{\Gamma(1-\nu)}{\Gamma(1+\nu)}(\frac{k}{2})^{2\nu}\varphi(k).
 \end{eqnarray}
 The two-point function is then given in terms of the one-point function by the formula (\ref{rt3}), viz
\begin{eqnarray}
G_E(k)=\frac{\langle{\cal O}(k)\rangle_{\varphi}}{\varphi(k)}&=&-\eta L^{d-1}(2\Delta -d)\frac{\Gamma(1-\nu)}{\Gamma(1+\nu)}(\frac{k}{2})^{2\nu}.\label{rt5}
\end{eqnarray}
In position space this two-point function reads
\begin{eqnarray}
\langle{\cal O}(x){\cal O}(0)\rangle=\frac{2\nu L^{d-1}\eta}{\pi^{d/2}}\frac{\Gamma(\frac{d}{2}+\nu)}{\Gamma(-\nu)}\frac{1}{|x|^{2\Delta}}.
\end{eqnarray}
This is the correct behavior of a conformal field of scaling dimension $\Delta$, i.e. the exponent $\Delta$ is indeed the scaling dimension of the boundary operator ${\cal O}(x)$.

\section{Conformal field theory on the torus}
\subsection{Modular invariance}
We will follow \cite{Ginsparg:1988ui,Schellekens}.

We go from the complex plane $z$ to the cylinder $w$ via the conformal exponential mapping, viz $w\longrightarrow z=\exp(w)$ or equivalently $r\exp(i\theta)=\exp(\sigma^2)\exp(-i\sigma^1)$. Thus, the Euclidean time $\sigma^2=-\infty$ on the worldsheet corresponds to $r=0$ on the plane while $\sigma^2=+\infty$ corresponds to $r=+\infty$. The spacelike worldsheet coordinate $\sigma^1$ is identified with the angle $-\theta$ on the plane, i.e. it is periodic.

A primary field $\Phi(z,\bar{z})$ of conformal dimension $(h,\bar{h})$ transforms under  the conformal mapping  $z\longrightarrow w$ covariantly as 
 \begin{eqnarray}
\Phi(z,\bar{z})\longrightarrow (\frac{\partial w}{\partial z})^h(\frac{\partial \bar{w}}{\partial \bar{z}})^{\bar{h}}\Phi(w,\bar{w}).
\end{eqnarray}
Thus, $\Phi(z\bar{z})dz^hd\bar{z}^{\bar{h}}$ is invariant under the conformal mapping. On the cylinder the field is given by 
\begin{eqnarray}
\Phi_{\rm cyl}(w,\bar{w})= (\frac{\partial z}{\partial w})^h(\frac{\partial \bar{z}}{\partial \bar{w}})^{\bar{h}}\Phi(z,\bar{z})=z^h\bar{z}^{\bar{h}}\Phi(z,\bar{z}).
\end{eqnarray}
Laurent expansion on the complex plane becomes Fourier expansion on the cylinder under the conformal mapping. Indeed, we have (for holomorphic fields $\Phi(z)$ and $\Phi_{\rm cyl}(w)$)
 \begin{eqnarray}
\Phi(z)=\sum_n\frac{\phi_n}{z^{n+h}}\Rightarrow \Phi_{\rm cyl}(w)=z^h\Phi(z)=\sum_n\phi_n\exp(-nw).
\end{eqnarray}
It is obvious that under a full rotation in the plane $z\longrightarrow \exp(2\pi i) z$, $\bar{z}\longrightarrow \exp(-2\pi i)\bar{z}$ the field on the cylinder $\Phi_{\rm cyl}$ rotates as $\Phi_{\rm cyl}(w,\bar{w})\longrightarrow \exp(2\pi i(h-\bar{h}))\Phi_{\rm cycl}(w,\bar{w})$. The spin of the field $\Phi$ is precisely $s=h-\bar{h}$ if you recall that $l_0-\bar{l_0}$ generates rotation. Thus a bosonic field $\Phi$ will satisfy the same boundary condition on the plane and on the cylinder whereas for a fermionic field on the plane $\Phi$ with periodic (anti-periodic) boundary condition the field on the cylinder $\Phi_{\rm cyl}$ satisfies anti-periodic (periodic) boundary condition.

Next we construct the torus via discrete identification on the cylinder. Let $H$ and $P$ be the energy and momentum operators which generate translation in the time  ${\rm Re} ~w=\sigma^2$ and space ${\rm Im} ~w=-\sigma^1$  directions respectively. On the plane we know that $l_0+ \bar{l}_0$ and $l_0-\bar{l}_0$ are the generators of dilatations and rotations respectively and thus on the cylinder we must have $H=(l_0)_{\rm cyl}+(\bar{l}_0)_{\rm cyl}$ and $P=(l_0)_{\rm cyl}-(\bar{l}_0)_{\rm cyl}$.    

\begin{figure}[htbp]
\begin{center}
  \includegraphics[width=10.0cm,angle=0]{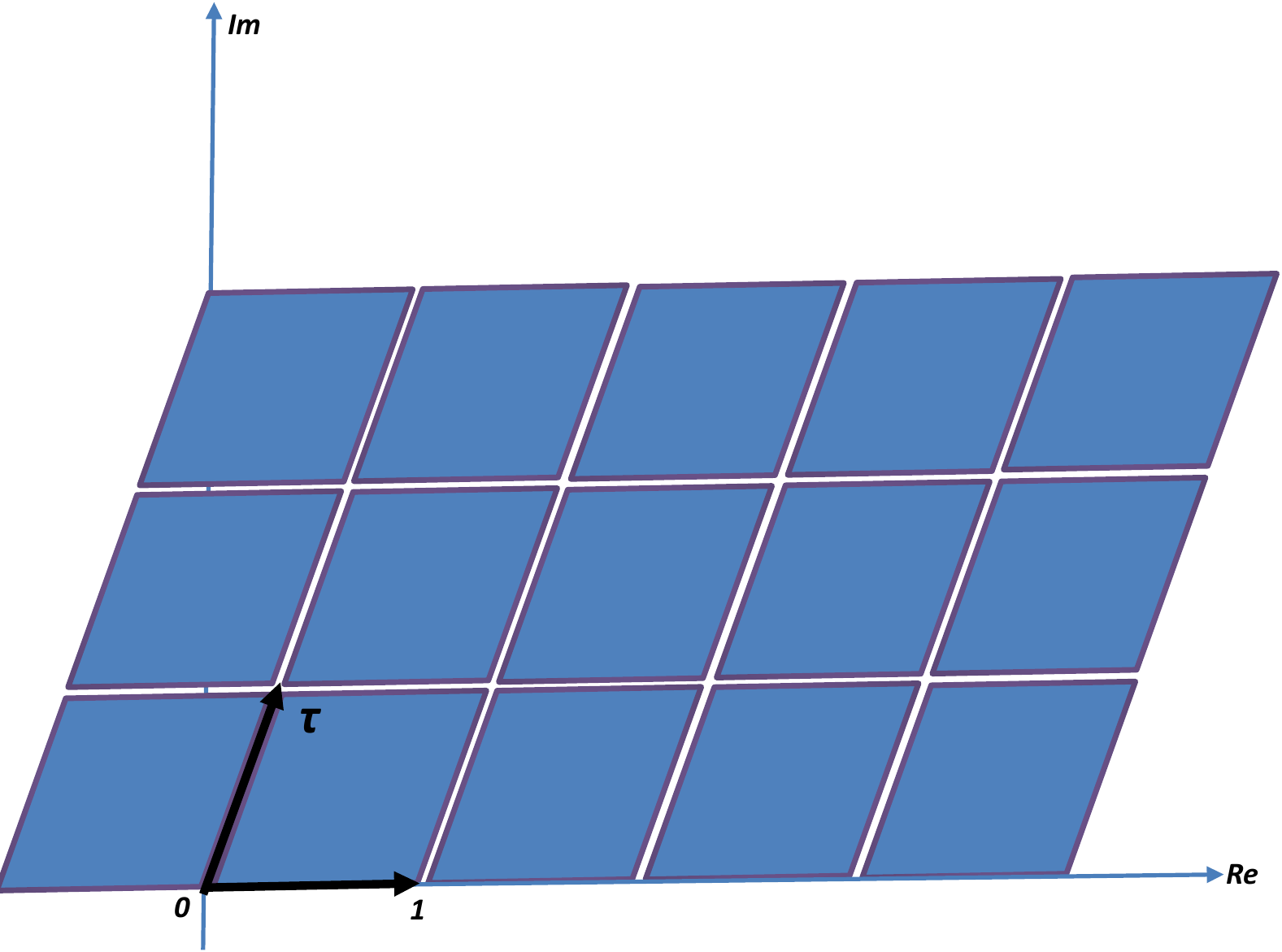}
\end{center}
\caption{The torus.}\label{torus}
\end{figure}

The torus is defined by two periods in $w$. First we redefine $w$ as $w\longrightarrow iw$, i.e. $z=\exp(iw)$ and  ${\rm Re} ~w=\sigma^1$, ${\rm Im} ~w=\sigma^2$. The first period is then $1$ since $w\longrightarrow w+2\pi$ leaves $z$ invariant. This corresponds to an identification along the real axis and rolling up the complex plane to a cylinder. The second period is taken to be equal to $\tau=\tau_1+i\tau_2$ where $\tau$ is called the modular parameter. Thus the identification $w\equiv w+2\pi\tau$ along the vector $\tau$ in the complex plane rolls up the cylinder to a torus. See figure (\ref{torus}). This torus is then effectively a lattice described by the complex number $\tau$. Any field on this torus must then satisfy 
\begin{eqnarray}
\Phi(w+1)=\Phi(w+\tau)=\Phi(w).
\end{eqnarray}
The complex number $w$ can be rewritten conveniently in terms of two real numbers $\tilde{\sigma}^i\in[0,1[$ each of period $1$ as 
\begin{eqnarray}
w=\tilde{\sigma}^1-\tau \tilde{\sigma}^2.  
\end{eqnarray}
The torus is then defined by the more transparent equivalence relation 
\begin{eqnarray}
(\tilde{\sigma}^1, \tilde{\sigma}^2)\sim (\tilde{\sigma}^1+n, \tilde{\sigma}^2+m)~,~n,m\in {\bf Z}.  \label{eqre}
\end{eqnarray}
This torus admits diffeomorphisms which are not connected continuously to the identity given by \cite{Green:1987mn}
\begin{eqnarray}
(\tilde{\sigma}^1, \tilde{\sigma}^2)\longrightarrow (a\tilde{\sigma}^1+b \tilde{\sigma}^2,c\tilde{\sigma}^1+d \tilde{\sigma}^2).\label{sl2Z}
\end{eqnarray}
The points which are equivalent under (\ref{eqre}) are mapped under the transformations (\ref{sl2Z}) to other points which are also equivalent under (\ref{eqre}) if and only of $a$, $b$, $c$ and $d$ are integers. The transformation (\ref{sl2Z}) is invertible and one-to-one if and only if
 \begin{eqnarray}
ad-bc=1.
\end{eqnarray}
This means that the transformation (\ref{sl2Z}) defines an element of $SL(2,Z)$ which is the group of reparametrizations of the torus. This is what is called the modular group of the torus. These transformations can not be reached continuously from the identity by exponentiating infinitesimal reparametrizations. This    $SL(2,Z)$ is the part of the reparametrization invariance that is not taken into account in the Fadeev-Popov procedure \cite{Green:1987mn}. 

Under the modular transformation  (\ref{sl2Z}) we also have, similarly to (\ref{sl2}), the transformation law
\begin{eqnarray}
\tau\longrightarrow \tau^{\prime}&=&\frac{a\tau +b}{c\tau+d}.
\end{eqnarray}
And
\begin{eqnarray}
w\longrightarrow w^{\prime}&=&\tilde{\sigma}^{1\prime}-\tau^{\prime} \tilde{\sigma}^{2\prime}\nonumber\\
&=&(a+c\tau^{\prime})\tilde{\sigma}^1+(b+d\tau^{\prime})\tilde{\sigma}^2\nonumber\\
&=&\frac{\tilde{\sigma}^{1}-\tau^{} \tilde{\sigma}^{2}}{c\tau+d}\nonumber\\
&=&\frac{w}{c\tau+d}.
\end{eqnarray}
The modular group can be generated by two special modular transformations $T$ and $S$. The modular transformation $S$ consists of an inversion in the unit circle $r^{\prime}=1/r$ followed by a reflection with respect to the imaginary axis $\theta^{\prime}=\pi-\theta$. Indeed, the action on the modular parameter $\tau$ is given by
\begin{eqnarray}
S~:~\tau\longrightarrow \tau^{\prime}&=&-\frac{1}{\tau}.
\end{eqnarray}
This corresponds explicitly to the $SL(2,Z)$ transformation  
\begin{eqnarray}
(\tilde{\sigma}^1, \tilde{\sigma}^2)\longrightarrow ( \tilde{\sigma}^2,-\tilde{\sigma}^1).
\end{eqnarray}
This transformation interchanges then the two basis vectors with a change of sign. 

The modular transformation $T$ is a translation given explicitly by 
\begin{eqnarray}
T~:~\tau\longrightarrow \tau^{\prime}&=&\tau+1.
\end{eqnarray}
This corresponds explicitly to the $SL(2,Z)$ transformation  
\begin{eqnarray}
(\tilde{\sigma}^1, \tilde{\sigma}^2)\longrightarrow ( \tilde{\sigma}^1+\tilde{\sigma}^2,\tilde{\sigma}^2).
\end{eqnarray}
The transformations $S$ and $T$ generate the whole modular group. They satisfy $S^2=(ST)^3=1$. The modular group is actually isomorphic to $SL(2,Z)/{\bf Z}_2$ since the two elements ${\bf 1}$ and $-{\bf 1}$ are  indistinguishable.
\subsection{Free bosons on a torus}
Let us then consider the action 
\begin{eqnarray}
S=\frac{1}{2\pi}\int \partial\Phi\bar{\partial}\Phi.
\end{eqnarray}
The measure is normalized such that (with $z=\sigma^1+\tau\sigma^2$ and $\sigma^{1,2}\in[0,1[$)
\begin{eqnarray}
\int 1=\int 2i dz\wedge d\bar{z}=\int 4\tau_2d\sigma^1\wedge d\sigma^2=4\tau_2.
\end{eqnarray}
We consider compactification on a circle of radius $r$, viz
\begin{eqnarray}
\Phi=\Phi+2\pi r.
\end{eqnarray}   
The torus has actually two periods $\tau$ and $1$. Thus the periodic boundary conditions are given explicitly by
\begin{eqnarray}
\Phi_0(z+\tau,\bar{z}+\bar{\tau})=\Phi_0(z,\bar{z})+2\pi r n^{\prime}~,~\Phi_0(z+1,\bar{z}+1)=\Phi_0(z,\bar{z})+2\pi r n.
\end{eqnarray}  
The field space decomposes then into instanton sectors characterized by the pair $(n^{\prime},n)$. The solution of the equation of motion $\partial\bar{\partial}\Phi_0=0$ in the instanton sector  $(n^{\prime},n)$ is given explicitly by 
\begin{eqnarray}
\Phi_0(z,\bar{z})=\frac{2\pi r}{2i\tau_2}\bigg[n^{\prime}(z-\bar{z})+n(\tau\bar{z}-\bar{\tau}z)\bigg].
\end{eqnarray} 
We expand the field as $\Phi=\Phi_0+Y$ where the fluctuation $Y$ is assumed to vanish at infinity. The action becomes 
\begin{eqnarray}
S[\Phi]=S[\Phi_0]+\frac{1}{2\pi}\int Y\Box Y,
\end{eqnarray} 
where $\Box=-\partial\bar{\partial}$. The relevant path integral is 
\begin{eqnarray}
\int {\cal D}\Phi\exp(-S[\Phi])&=&\exp(-S[\Phi_0])\int{\cal D}Y\exp(-\frac{1}{2\pi}\int Y\Box Y).
\end{eqnarray} 
The fluctuation $Y$ is split into a constant part $\tilde{Y}$ and a fluctuation $Y^{\prime}$ containing no zero mode. Thus, we have
\begin{eqnarray}
\int {\cal D}\Phi\exp(-S[\Phi])&=&\exp(-S[\Phi_0])\int d\tilde{Y}{\cal D}Y^{\prime}\exp(-\frac{1}{2\pi}\int Y^{\prime}\Box Y^{\prime})\nonumber\\
&=&2\pi r \exp(-S[\Phi_0])\int {\cal D}Y^{\prime}\exp(-\frac{1}{2\pi}\int Y^{\prime}\Box Y^{\prime}).\label{pto}
\end{eqnarray}  
The path integral is normalized such that 
\begin{eqnarray}
1&=&\int {\cal D}X\exp(-\frac{1}{2\pi}\int X^{2})\nonumber\\
&=&\int d\tilde{X}\exp(-\frac{1}{2\pi}(4\tau_2) \tilde{X}^{2})\int {\cal D}X^{\prime}\exp(-\frac{1}{2\pi}\int X^{\prime 2})\nonumber\\
&=&\frac{\pi}{\sqrt{2\tau_2}}\int {\cal D}X^{\prime}\exp(-\frac{1}{2\pi}\int X^{\prime 2})\nonumber\\
\end{eqnarray}  
Therefore the partition function (\ref{pto}) on the torus becomes (by taking also the sum over all instanton sectors)
\begin{eqnarray}
\int {\cal D}\Phi\exp(-S[\Phi])
&=&2\pi r \frac{\sqrt{2\tau_2}}{\pi}\frac{1}{{\rm det}^{\prime 1/2}\Box}\sum_{n^{\prime},n=-\infty}^{+\infty}\exp(-S[\Phi_0^{(n^{\prime},n)}])\nonumber\\
&=&2\pi r \frac{\sqrt{2\tau_2}}{\pi}\frac{1}{{\rm det}^{\prime 1/2}\Box}\sum_{n^{\prime},n=-\infty}^{+\infty}\exp\bigg(4\tau_2\frac{1}{2\pi}(\frac{2\pi r}{2i\tau_2})^2(n^{\prime}-\bar{\tau}n)(n^{\prime}-{\tau}n)\bigg).\nonumber\\
\end{eqnarray}  
The crucial piece is the determinant. We use the single-valued (under both $z\longrightarrow z+\tau$ and $z\longrightarrow z+1$) eigenfunctions 
\begin{eqnarray}
\psi_{n,m}(z,\bar{z})=\exp\bigg(\frac{2\pi r}{2i\tau_2}\bigg[n(z-\bar{z})+m(\tau\bar{z}-\bar{\tau}z)\bigg]\bigg).
\end{eqnarray}  
Indeed, we have 
\begin{eqnarray}
\Box\psi_{n,m}(z,\bar{z})=\frac{\pi^2}{\tau_2^2}(n-m\bar{\tau})(n-m\tau)\psi_{n,m}(z,\bar{z}).
\end{eqnarray}  
The determinant ${\rm det}^{\prime}\Box$, which does not involve the zero mode, is then given by 
\begin{eqnarray}
{\rm det}^{\prime }\Box&=&\prod_{(m,n)\neq (0,0)}\frac{\pi^2}{\tau_2^2}(n-m\bar{\tau})(n-m\tau)\nonumber\\
&=&\prod_{(m,n)\neq (0,0)}\frac{\pi^2}{\tau_2^2}\prod_{n\neq 0}n^2\prod_{m\neq 0, n\in Z}(n-m\bar{\tau})(n-m\tau).\nonumber\\
\end{eqnarray}  
We use zet-function regularization $\zeta(s)=\sum_{n=1}^{\infty}1/n^s$, $\zeta(-1)=-1/12$, $\zeta(0)=-1/2$, $\zeta^{\prime}(0)=-\ln\sqrt{2\pi}$. Then
\begin{eqnarray}
\prod_{n=1}^{\infty}a=a^{\zeta(0)}=a^{-1/2}.
\end{eqnarray}  
\begin{eqnarray}
\prod_{n=1}^{\infty}n^{\alpha}=e^{\alpha \sum_{n=1}^{\infty}\ln n}=e^{-\alpha\zeta^{\prime}(0)}=(2\pi)^{\alpha/2}.
\end{eqnarray}  
The determinant becomes 
\begin{eqnarray}
{\rm det}^{\prime }\Box
&=&\frac{\tau_2^2}{\pi^2}(2\pi)^2\prod_{m> 0, n\in Z}(n-m\bar{\tau})(n+m\bar{\tau})(n-m\tau)(n+m\tau)\nonumber\\
&=&\frac{\tau_2^2}{\pi^2}(2\pi)^2\prod_{m> 0}\bigg(e^{i\pi m\tau}-e^{-i\pi m\tau}\bigg)^2\bigg(e^{i\pi m\bar{\tau}}-e^{-i\pi m\bar{\tau}}\bigg)^2,
\end{eqnarray}  
where we have used the identity
\begin{eqnarray}
\prod_{n\in Z}(n+a)=a\prod_{n=1}^{\infty}(-n^2)(1-a^2/n^2)=2i\sin\pi a.
\end{eqnarray}  
We get finally
\begin{eqnarray}
{\rm det}^{\prime }\Box
&=&\frac{\tau_2^2}{\pi^2}(2\pi)^2\prod_{m> 0}(q\bar{q})^{-m}(1-q^m)^2(1-\bar{q}^m)^2\nonumber\\
&=&4\tau_2^2(q\bar{q})^{1/12}\prod_{m> 0}(1-q^m)^2(1-\bar{q}^m)^2\nonumber\\
&=&4\tau_2^2\eta^2\bar{\eta}^2.
\end{eqnarray}  
We have defined 
\begin{eqnarray}
\eta=q^{1/24}\prod_{m> 0}(1-q^m)~,~\bar{\eta}=\bar{q}^{1/24}\prod_{m> 0}(1-\bar{q}^m)~,~q=\exp(2i\pi\tau)~,~\bar{q}=\exp(-2i\pi\bar{\tau}).
\end{eqnarray}
The partition function on the torus becomes 
\begin{eqnarray}
\int {\cal D}\Phi\exp(-S[\Phi])
&=&r\sqrt{\frac{2}{\tau_2}} \frac{1}{\eta\bar{\eta}}\sum_{n^{\prime},n=-\infty}^{+\infty}\exp\bigg(-2\pi\bigg[\frac{1}{\tau_2}(n^{\prime}r-\tau_1 nr)^2+r^2n^2\tau_2\bigg]\bigg).\nonumber\\
\end{eqnarray}   
The summation over the winding $n^{\prime}$ in the time direction (corresponding to the period $\tau$) will be converted into a summation over a conjugate momentum by means of  Poisson resummation formula 
\begin{eqnarray}
\sum_{n^{\prime}=-\infty}^{+\infty}f(n^{\prime}r)=\frac{1}{r}\sum_{m=-\infty}^{+\infty}\tilde{f}(\frac{m}{r})~,~\tilde{f}(p)=\int_{-\infty}^{+\infty}dx e^{2\pi i px}f(x).
\end{eqnarray}  
We take the function $f$ and its Fourier transform $\tilde{f}$ to be 
\begin{eqnarray}
f(n^{\prime}r)=\exp\bigg(-2\pi\frac{1}{\tau_2}(n^{\prime}r-\tau_1 nr)^2\bigg)\Rightarrow \tilde{f}(p)=\sqrt{\frac{\tau_2}{2}}\exp\bigg(2\pi i\tau_1 n rp -\frac{1}{2}\pi\tau_2 p^2\bigg).
\end{eqnarray}  
The partition function becomes 
\begin{eqnarray}
\int {\cal D}\Phi\exp(-S[\Phi])
&=&\frac{1}{\eta\bar{\eta}}\sum_{m,n=-\infty}^{+\infty}\exp\bigg(-2\pi r^2n^2\tau_2+2\pi i\tau_1 nm-\frac{1}{2}\pi\tau_2(\frac{m}{r})^2\bigg)\nonumber\\
&=&\frac{1}{\eta\bar{\eta}}\sum_{m,n=-\infty}^{+\infty}q^{\frac{1}{2}(\frac{m}{2r}+nr)^2}\bar{q}^{\frac{1}{2}(\frac{m}{2r}-nr)^2}.
\end{eqnarray}  
As we know compactification on a circle leads to a  momentum and a winding defined by $p=m/r$ and $w=nr$. The zero modes are given by
\begin{eqnarray}
\alpha_0=p_L=\frac{p}{2}+w~,~\bar{\alpha}_0=p_R=\frac{p}{2}-w.
\end{eqnarray} 
Hence we get the partition function 
\begin{eqnarray}
\int {\cal D}\Phi\exp(-S[\Phi])
&=&\frac{1}{\eta\bar{\eta}}\sum_{m,n=-\infty}^{+\infty}q^{\frac{1}{2}p_L^2}\bar{q}^{\frac{1}{2}p_R^2}.\label{pft}
\end{eqnarray}  
We recall that in the quantum theory the generators $L_0$ and $\bar{L}_0$ are given by
\begin{eqnarray}
L_0=\sum_{l=1}\alpha_{-l}\alpha_l+\frac{1}{2}\alpha_0^2~,~\bar{L}_0=\sum_{l=1}\bar{\alpha}_{-l}\bar{\alpha}_l+\frac{1}{2}\bar{\alpha_0}^2.
\end{eqnarray}  
The string ground states $|n,m\rangle$ are characterized by the fact that the oscillators $\alpha_l$ and $\bar{\alpha}_l$ are in their ground state, viz $\alpha_l|n,m\rangle=\bar{\alpha}_l|n,m\rangle=0$, whereas the string center of mass has precisely a non-zero momentum equal $m$ and non-zero winding equal $n$. In other words, we have
\begin{eqnarray}
L_0|n,m\rangle=\frac{1}{2}(\frac{m}{2r}+nr)^2|n,m\rangle~,~\bar{L}_0|n,m\rangle=\frac{1}{2}(\frac{m}{2r}-nr)^2|n,m\rangle.
\end{eqnarray}  
The Hamiltonian and momentum eigenvalues of the  string ground states $|n,m\rangle$ are then given by
 \begin{eqnarray}
H|n,m\rangle=(L_0+\bar{L_0})|n,m\rangle=(\frac{m^2}{4r^2}+n^2r^2)|n,m\rangle~,~P|n,m\rangle=(L_0-\bar{L_0})|n,m\rangle=nm|n,m\rangle.\nonumber\\
\end{eqnarray}  
The partition function (\ref{pft}) can be rewritten in terms of the generators $L_0$ and $\bar{L}_0$ as 
\begin{eqnarray}
\int {\cal D}\Phi\exp(-S[\Phi])&=&(q\bar{q})^{-c/24}tr q^{L_0}\bar{q}^{\bar{L}_0}.\label{form1}
\end{eqnarray}   
This is the most general form of the partition function of the conformal field theory of free bosons on the torus. We can verify modular invariance by studying the effect of the modular transformation $\tau\longrightarrow -1/\tau$ on $\eta$. 
\section{Holographic entanglement entropy}

\subsection{Entanglement entropy}

       In quantum mechanics it is shown by the EPR experiment for example that entanglement is at odd with locality. The action (due to a measurement) seems to propagate with an infinite velocity and although it can not carry any energy we are left in an uncomfortable position. Entanglement as opposed to energy is not conserved and there are degrees of entanglement. Mathematically, entanglement means that the vector state is not separable, i.e. it can not be written as a tensor product.
       
\begin{figure}[H]
\begin{center}
  \includegraphics[angle=-0,scale=0.4]{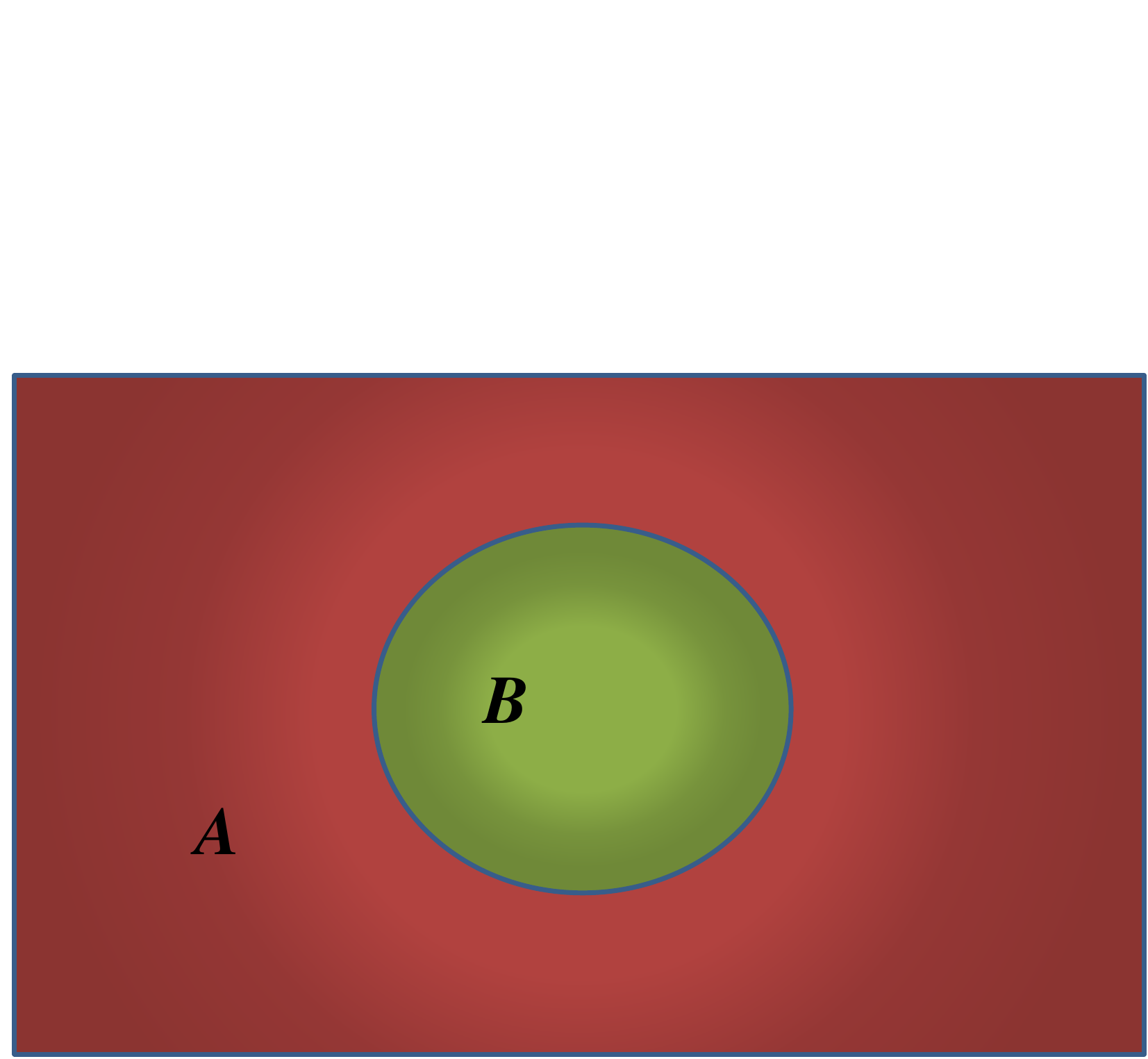}
\end{center}
\caption{The accessible and inaccessible regions.}\label{sketch1}
            \end{figure}

       Quantum entanglement is measured by entropy or more precisely by entanglement entropy. However, entropy has actually two sources: statistical and quantum.
            \begin{enumerate}
\item {\bf Statistical/Thermal Entropy}: The thermal or Boltzmann entropy of a macroscopic state is the logarithm of the number $n$ of microscopic states consistent with this state. Thus this entropy measures the lack of resolution, i.e. the fact that a large number of microscopic configurations correspond to the same macroscopic thermodynamical state. The thermal entropy is defined in terms of the Blotzmann density matrix $\rho_{\rm ther}=\exp(-\beta E)/Z$ by
\begin{eqnarray}
  S_{\rm ther}&=&-tr \rho_{\rm ther}\log\rho_{\rm ther}\nonumber\\
  &=&\log n.
\end{eqnarray}
The second equality holds if the microstates are equally probable.
\item {\bf Entanglement Entropy:}        
\begin{itemize}
          \item {\bf Measurement}: In quantum mechanics, there is another source of entropy associated with the restriction of observers, who are performing the experiments, to finite volume. Indeed, a typical observer performing an experiment on a closed system, which is supposed to be in a pure ground state $|\Psi\rangle$, will only be able to access a particular subsystem, i.e. a partial set of the relevant observables such as those with support in a restricted volume.
            
            We will denote the accessible subsystem by $A$ (where the observers are restricted) and the inaccessible subsystem is $B$. The total system $\Sigma=A\cup B$ is in a pure ground state $|\Psi\rangle$.  See figure (\ref{sketch1}).
            
          \item   {\bf Reduced Density Matrix:}       The state of the system will be given by a mixed density matrix $\rho$ and the entropy will measure the correlation between the inaccessible subsystem $B$ and the accessible part $A$ of the closed system. The total Hilbert space is  ${\cal H}_{\rm Tot}={\cal H}_A\otimes{\cal H}_B$.

            The observer who can not access the subsystem $B$ will describe the total system by the reduced density matrix (obtained by tracing over the inaccessible degrees of freedom) 
\begin{eqnarray}
  \rho_{\rm Red}\equiv \rho_A=tr_B\rho_{\rm Tot}.
\end{eqnarray}
In other words, we trace (integrate) over the inaccessible subsystem $B$, i.e. we take average over the inaccessible degrees of freedom.

  \item {\bf Mixed versus Pure:} The reduced density matrix is an incoherent (mixed) superposition (statistical ensemble, classical probabilities, no interference terms, random relative phases). It is not an idempotent and it satisfies  $Tr\rho^2<1$.

            In contrast, a pure state is a vector in the Hilbert space which is a coherent superposition (interference terms, coherent relative phases) represented by a projector.

            Mixed states are relevant if the exact initial state vector is unknown.

\item {\bf Entanglement Entropy:} The entropy of the subsystem $A$  which measures the correlation between the inaccessible subsystem and the accessible part of the closed system is defined by the von Neumann entropy of this reduced density matrix, viz
\begin{eqnarray}
  S_{\rm Red}\equiv S_A=-tr_A\rho_A\log\rho_A=-\sum_i\rho_i\log \rho_i.
\end{eqnarray}
Thus, entanglement entropy is the logarithm of the number of microscopic states of the inaccessible subsystem $B$ which are consistent with observations restricted to the accessible subsystem $A$, together with the assumption that the total system is in a pure state. It measures the degree of entanglement between $A$ and $B$.  This is different from the thermodynamic Boltzmann entropy.

\item {\bf Examples:}
For a pure (separable) state, i.e. when all eigenvalues with the exception of one vanish, we get
  $S=0$. For mixed states we have $S>0$.

  In the case of a totally incoherent mixed density matrix in which all the eigenvalues are equal to $1/N$ where $N$ is the dimension of the Hilbert space we get the maximum value of the Von Neumann entropy given by 
\begin{eqnarray}
S_{\rm Red}=S_{\rm max}=\log N.
\end{eqnarray}
In the case that $\rho$ is proportional to a projection operator onto a subspace of dimension $n$ we find 
\begin{eqnarray}
S_{\rm Red}=\log n.
\end{eqnarray}
In other words, the Von Neumann entropy measures the number of important states in the statistical ensemble, i.e. those states which have an appreciable probability. This entropy is also a measure of the degree of entanglement between subsystems $A$ and $B$ and hence its other name entanglement entropy.
\item {\bf Information:} The von Neumann entropy $S_V\equiv S_A$ is not additive as opposed to the thermal entropy $S_B$ defined with respect to Boltzmann distribution. We have $S_B\geq S_V$, i.e. the Boltzmann thermal entropy (coarse grained, macroscopic) is always greater or equal to von Neumann entanglement (fine grained,microscopic) entropy.

  The amount of information is the difference:
\begin{eqnarray}
  I=S_B-S_V.
  \end{eqnarray}
            If $\Sigma=A$ then there is no entanglement and the amount of information is maximal, i.e. $S_V=0\Rightarrow I=S_B$. If $A<<B$ then in this case the amount of information is zero, i.e. $S_V=S_B$. Equivalently, if $A<<B$ then the entanglement entropy becomes maximal equal to the thermal entanglement. 

Remark that the von Neumann entropy of the total system is zero, viz $S_{\rm Tot}=tr\rho_{\rm Tot}\log\rho_{\rm Tot}=0$ since there is no inaccessible part here.
        \end{itemize}
  \end{enumerate}

\subsection{Entanglement entropy in quantum mechanics}
For detail of the formalism used here we refer to \cite{Bombelli:1986rw}. We will consider a Hamiltonian of the form
\begin{equation}
  \label{1} H=\frac{1}{2}\sum_{A,B} (\delta_{A,B} \pi^A\pi^B+V_{AB}\varphi^A\varphi^B).
\end{equation}
In this equation $V$ is a real symmetric matrix with positive definite eigenvalues. The normalized ground state of this model is given in the Schrodinger representation by
\begin{equation}
  \langle {\varphi^A}|0\rangle = \big[\det \frac{W}{\pi}\big]^{1/4}\exp \big [-\frac{1}{2}W_{AB}\varphi^A\varphi^B\big].
\end{equation}
$W$ is the square root of the matrix $V$. The corresponding density matrix is
\begin{equation}
  \label{3} \rho(\varphi,\varphi')= \big[\det \frac{W}{\pi}\big]^{1/2}\exp\big [-\frac{1}{2}W_{AB}(\varphi^A\varphi^B+\varphi'^A\varphi'^B)\big]
\end{equation}
If we suppose that the field degrees of freedom $ \varphi^\alpha, \alpha=\overline{1,n}$, are inaccessible then the correct description of the state of the system will be given by the reduced density matrix in which we integrate out these inaccessible degrees of freedom, viz
\begin{equation}
  \label{4} \rho_{\rm red}({\varphi}^{n+1},{\varphi}^{n+2},...,{\varphi}^{'n+1},{\varphi}^{'n+2},...)= \int \prod_{\alpha=1}^{n} d{\varphi}^\alpha \rho(\varphi,{\varphi}')
\end{equation}
The entanglement entropy is the associated Von Newman entropy of $ \rho_{\rm red}$ defined by $ S= -{Tr} \rho_{\rm red}\log\rho_{\rm red}$. The entanglement entropy for any Hamiltonian of the form (\ref{1}) can be shown to be given by \cite{Bombelli:1986rw}
\begin{equation}
  \label{5} S_{\rm ent}= \sum_i\bigg[ \log\big(\frac{1}{2}\sqrt{{\lambda}_i}\big)+ \sqrt{1+{\lambda}_i}\log\bigg(\frac{1}{\sqrt{{\lambda}_i}}+ \sqrt{1+\frac{1}{{\lambda}_i}}\bigg)\bigg].
\end{equation}
The ${\lambda}_i$ are the eigenvalues of the following matrix
\begin{equation}
  \label{6} \Lambda_{i,j}= -\sum_{\alpha=1}^{n}W^{-1}_{i\alpha}W_{\alpha j}
\end{equation}
$W_{\alpha j}$ and $ W^{-1}_{i\alpha}$ are elements of $W$ and $W^{-1}$ respectively with $i,j$ running from $n+1$ to ${\cal N}$ and $\alpha$ from $1$ to $n$, i.e. $\Lambda$ is an $({\cal N}-n)\times({\cal N}-n)$ matrix and $i,j$ run from $n+1$ to $N$.

\subsection{Entanglement entropy in conformal field theory}
In this section we follow mostly \cite{Holzhey:1994we} and \cite{Nishioka:2009un}. See also \cite{Calabrese:2004eu}.

The entropy of a macroscopic state, in statistical mechanics, is defined by the logarithm of the number of microscopic states which are consistent with it. Thus this entropy measures the lack of resolution, i.e. the fact that a large number of microscopic configurations correspond to the same macroscopic thermodynamical state.

However, in quantum mechanics, there is another source of entropy associated with the restriction of observers, who are performing the experiments, to finite volume. Indeed, a typical observer performing an experiment on a closed system, which is supposed to be in a pure ground state $|\Psi\rangle$, will only be able to access a particular subsystem, i.e. a partial set of the relevant observables such as those with support in a restricted volume.

We will denote the accessible subsystem by $A$ and the inaccessible subsystem by $B$ (see (\ref{sketch1})). In this case the state of the system will be given by a mixed density matrix $\rho$ and the entropy will measure the correlation between the inaccessible subsystem and the accessible part of the closed system. The total Hilbert space is clearly given by ${\cal H}_{\rm Tot}={\cal H}_A\otimes{\cal H}_B$. The observer who can not access the subsystem $B$ will describe the total system not by the ground state $|\Psi\rangle$ (or its corresponding density matrix $\rho_{\rm Tot}=|\Psi\rangle\langle\Psi|$) but by the reduced density matrix
\begin{eqnarray}
  \rho_{\rm Red}\equiv \rho_A=tr_B\rho_{\rm Tot}.
\end{eqnarray}
In other words, we trace (integrate) over the inaccessible subsystem $B$, i.e. we take average over the inaccessible degrees of freedom. The entanglement entropy of the subsystem $A$ is defined by the von Neumann entropy of this reduced density matrix, viz
\begin{eqnarray}
  S_{\rm Red}\equiv S_A=-tr_A\rho_A\log\rho_A.
\end{eqnarray}
The entanglement entropy is then essentially the logarithm of the number of microscopic states of the inaccessible part of the system which are consistent with the observations restricted to the accessible subsystem, together with the assumption that the total system is in a pure state. It measures as we said the degree of correlation (entanglement) between the accessible subsystem and the inaccessible part of the total system.

Remark that the von Neumann entropy of the total system is zero, viz $S_{\rm Tot}=tr\rho_{\rm Tot}\log\rho_{\rm Tot}=0$ since there is no inaccessible part here.

We assume now a conformal field theory in two dimensions with complex coordinate $z=\sigma+i\tau$. The spatial dimension $\sigma$ is given by $C=[0,L[$ where $L$ is an infrared cutoff and we will assume periodic boundary condition, i.e. $L\equiv 0$. The subsystem playing the role of the accessible region (where measurements are performed) is $A=[0,\Sigma[$ whereas the unavailable region (to be traced over) is $B=[\Sigma,L[$. The ultraviolet cutoffs are introduced by considering instead the intervals $A=[\epsilon_1,\Sigma-\epsilon_2[$ and $B=[\Sigma+\epsilon_2,L-\epsilon_1[$. We perform the conformal mapping
                   \begin{eqnarray}
                      z\longrightarrow w=-\frac{\sin\frac{\pi}{L}(z-\Sigma)}{\sin\frac{\pi}{L}z}.
                    \end{eqnarray}
                    The regularized intervals $A$ and $B$ become (with the assumption $\Sigma<< L$) the positive half-axis and the negative half-axis respectively, viz
                    \begin{eqnarray}
                      A= ]{\epsilon_2}/{\Sigma},{\Sigma}/{\epsilon_1}]~,~B=]-{\Sigma}/{\epsilon_1},-{\epsilon_2}/{\Sigma}].
                    \end{eqnarray}
                    This is spatial section at $\tau=0$. Extrapolation into the past $\tau\longrightarrow -\infty$ corresponds to extrapolation to the lower half–plane with inner and outer radii $\epsilon_2/\Sigma$ and $\Sigma/\epsilon_1$ respectively.

                    Lastly we perform the conformal mapping
                    \begin{eqnarray}
                      w\longrightarrow y=\frac{1}{\kappa}\ln w.
                    \end{eqnarray}
                    Our points from $\tau=-\infty$ to $\tau=0$ are given by $w=R\exp(i\theta)$ where $R$ ranges from $\epsilon_2/\Sigma$ to $\Sigma/\epsilon_1$ and $\theta$ ranges from $-\pi$ to $0$. Hence $y=\ln R/\kappa+i\theta/\kappa$. We have then
                    \begin{eqnarray}
                      \theta={\rm fixed}~,~\Delta y=\frac{1}{\kappa}\ln \frac{\Sigma}{\epsilon_1}-\frac{1}{\kappa}\ln\frac{\epsilon_2}{\Sigma}=\frac{2}{\kappa}\ln \frac{\Sigma}{\sqrt{\epsilon_1\epsilon_2}}\equiv L.\label{Lkappa}
                    \end{eqnarray}
                    \begin{eqnarray}
                      R={\rm fixed}~,~\Delta y=i\frac{\Delta\theta}{\kappa}=i\frac{\pi}{\kappa}\equiv i h.
                    \end{eqnarray} This is a finite strip of length $L$ and width $h$. The accessible region $A$ corresponds to fixed $\theta$ between $-\pi/2$ and $0$ and thus corresponds to the upper side of the strip whereas the inaccessible region $B$ corresponds to fixed $\theta$ between $-\pi$ and $-\pi/2$ and thus to the lower side of the strip. The "upper" and "lower" are with respect to the width direction $h$.

                    The ground state wave functional $\Psi(\phi_0(x))$ can be defined via a path integral with an appropriate boundary conditions specifying the field on the Cauchy surface $C=A\cup B$. Explicitly we have
                    \begin{eqnarray}
                      \Psi(\phi_0(x))=\int_{\tau=-\infty}^{\tau=0} {\cal D}\phi\exp(-S(\phi))~,~\phi_0(x)=\phi(0,x).
                    \end{eqnarray} The field is also assumed to vanish in the limit $\tau\longrightarrow -\infty$. The complex conjugate wave functional $\bar{\Psi}(\phi_0^{\prime}(x))$ is given similarly by
                    \begin{eqnarray}
                      \bar{\Psi}(\phi_0^{\prime}(x))=\int_{\tau=0}^{\tau=+\infty} {\cal D}\phi\exp(-S(\phi))~,~\phi_0^{\prime}(x)=\phi^{\prime}(0,x).
                    \end{eqnarray}
                    We can write $\phi_0=XY$ where $\phi_0=X$ on $A$ (upper side of the first copy of strip) and $\phi_0=Y$ on $B$ (lower side of this strip) and similarly we can write $\phi_0^{\prime}=X^{\prime}Y^{\prime}$ where $\phi_0^{\prime}=X^{\prime}$ on $A$ (lower side of another copy of the strip) and $\phi_0^{\prime}=Y^{\prime}$ on $B$ (upper side of the second copy of the strip).

                    The total density matrix is then given by
                    \begin{eqnarray}
                      \rho_{\phi_0\phi_0^{\prime}}=\Psi(\phi_0(x))\bar{\Psi}(\phi_0^{\prime}(x)).
                    \end{eqnarray}
                    However, the reduced density matrix $\rho_A$, which describes the observations from the subsystem $A$, is obtained by tracing $\rho$ over the inaccessible subsystem $B$. Thus we need to integrate $\phi_0$ on the region $B$ with the condition $\phi_0(x)=\phi_0^{\prime}(x)$ when $x\in B$. Then the reduced density matrix is given by
                    \begin{eqnarray}
                      (\rho_A)_{XX^{\prime}}=\int {\cal D}Y \Psi(X,Y)\bar{\Psi}(Y,X^{\prime}).
                    \end{eqnarray} This is generally a mixed density matrix as opposed to the total density matrix $\rho$ which is a pure density matrix. This integral involves pasting together two copies of the strip along the inaccessible region $B$. Thus it corresponds to a functional integral over a strip of height $2\pi/\kappa$ with boundary conditions given by $\phi_0=X$ on the upper side of the strip and $\phi_0=X^{\prime}$ on the lower side of the strip given by
                    \begin{eqnarray}
                      (\rho_A)_{XX^{\prime}}=\frac{1}{Z_1}\int {\cal D}\phi \exp(-S(\phi)).
                    \end{eqnarray}
                    The path integral $Z_1$ is determined by the normalization condition $tr \rho_A=1$. This is clearly periodic in the height direction $h$ since the fields are identified by the tracing. If we also impose periodic boundary condition in the length direction $L$ then $Z_1$ is nothing else but the partition function on a torus.

                    The entropy is actually going to be calculated using the so-called replica trick given by the relation
                    \begin{eqnarray}
                      S_A=-tr\rho_A\log\rho_A=-\frac{\partial}{\partial n}tr_A\rho_A^n|_{n=1}.
                    \end{eqnarray} The trace $tr_A\rho_A^n$ is computed by pasting together $n$ copies of the strip along the inaccessible region $B$. We start thus from $n$ copies of the reduced density matrix, viz \begin{eqnarray} (\rho_A)_{X_{1+}X_{1-}}(\rho_A)_{X_{2+}X_{2-}}...(\rho_A)_{X_{n+}X_{n-}}. \end{eqnarray} The pasting or gluing is done by the conditions $X_{i-}(x)=X_{i+1+}(x)$, $\forall i=1,...,n-1$, and then integrating over $X_{i+}$. If we choose $X_{1+}=X$ and $X_{n-}=X^{\prime}$ then we obtain the matrix element \begin{eqnarray} (\rho_A^n)_{XX^{\prime}}=\frac{1}{Z_1^n}\int {\cal D}\phi \exp(-S(\phi)). \end{eqnarray} The functional integral is over a strip of width $2\pi n/\kappa$ with boundary conditions given by $\phi_0=X$ on the upper side of the strip and $\phi_0=X^{\prime}$ on the lower side of the strip. By setting $X=X^{\prime}$ and integrating over $X$ we obtain the desired trace as \begin{eqnarray} tr_A\rho_A^n=\frac{1}{Z_1^n}\int {\cal D}X\int {\cal D}\phi \exp(-S(\phi))=\frac{Z_n}{Z_1^n}. \end{eqnarray} The $Z_n$ is the partition function on the torus of lengths $2\pi n/\kappa$ and $L$ around its two cycles. The entanglement entropy $S_A$ takes finally the form
                    \begin{eqnarray} S_A=\bigg(\bigg[1-n\frac{\partial}{\partial n}\bigg]\ln Z_n\bigg)|_{n=1}. \end{eqnarray}

\subsection{Ryu-Takayanagi formula}
In this section we follow mainly \cite{Nishioka:2009un}.
                    
The Bekenstein-Hawking formula states that the entropy of a black hole ${\cal S}_{BH}$ is proportional to the surface area of the event horizon $A_H$, viz 
\begin{eqnarray}
{\cal S}_{BH}=\frac{A_H}{4G_N}.
\end{eqnarray} 
On the other hand, the entanglement entropy $S_A$  for observers accessible to a subsystem $A$ (outside event horizon) who can not receive any signals from the subsystem $B$ (inside the black hole) is a given by

\begin{eqnarray}
S_A=-tr_A\rho_A\log\rho_A.
\end{eqnarray}
The entanglement entropy satisfies the following properties:
\begin{enumerate}
\item For three subsystems $A$, $B$ and $C$ which do not intersect each other we have the so-called strong subadditivity relations
\begin{eqnarray}
  &&S_{A+B+C}+S_B\leq S_{A+B}+S_{B+C}\nonumber\\
  &&S_{A}+S_C\leq S_{A+B}+S_{B+C}.
\end{eqnarray}
\item By choosing $B$ empty in the above relations we obtain
  \begin{eqnarray}
    &&S_{A+B}\leq S_{A}+S_{B}.
  \end{eqnarray}
  The mutual information is defined by
   \begin{eqnarray}
    && I(A,B)=S_{A}+S_{B}-S_{A+B}\geq 0.
  \end{eqnarray}
  \item If we choose $B$ to be the complement of $A$ then
\begin{eqnarray}
S_A=S_B\Rightarrow S_{A+B}\leq 2S_A.    
\end{eqnarray}
Hence the entanglement entropy is not an extensive quantity.
\end{enumerate}
In a QFT on a $(d+1)-$dimensional manifold ${\bf R}\times {\bf N}$ where $d\geq 2$ and ${\bf N}=A\cup B$ it is found that the entanglement entropy $1)$ depends only on the geometry of $A$  (this is why entanglement entropy is also called geometric entropy), $2)$ is UV divergent and hence the continuum theory should be regularized by a lattice $a$ , and $3)$ it is proportional to the area of the boundary $\partial A$ of $A$ since the entanglement between $A$ and $B$ occurs strongly obviously on the boundary. We have explicitly  \cite{Bombelli:1986rw,Srednicki:1993im}   
\begin{eqnarray}
S_A=\gamma.\frac{{\rm Area}(\partial A)}{a^{d-1}}+{\rm subleading~terms}.
\end{eqnarray} 
This entanglement entropy formula (includes UV divergences, proportional to the number of matter fields) is very similar to the Bekenstein-Hawking formula (does not include UV divergences, is not proportional to the number of matter fields). In fact the quantum corrections to the  Bekenstein-Hawking black hole entropy in the presence of matter fields is given by the entanglement entropy \cite{Susskind:1994sm,Fiola:1994ir,Jacobson:1994iw,Solodukhin:1994st}.

The Ryu-Takayanagi formula is a generalization of the  Bekenstein-Hawking formula, based on the ${\rm AdS}_{d+2}/{\rm CFT}_{d+1}$ correspondence, in which we identify the entanglement entropy in $(d+1)-$dimensional QFT with a geometric quantity in $(d+2)-$dimensional gravity.

We consider the metric in $AdS_{d+2}$ in Poincare patch given by
\begin{eqnarray}
  ds^2=g^{\mu\nu}dx_{\mu}dx_{\nu}=\frac{R^2}{z^2}(-dt^2+\sum_{i=1}^ddx_i^2+dz^2).
  \end{eqnarray}
The dual $(d+1)-$dimensional CFT lives on the boundary located at $z=0$. The radial coordinate $z$ as we have seen is a lattice spacing and the theory on the boundary should be properly understood as the continuum limit (in the sense of RG) of a regularize CFT, i.e. with a cutoff $\Lambda$.  This regularized CFT lives on a surface $z=a$ where $a=1/\Lambda$ and $a\longrightarrow 0$.

\begin{figure}[H]
\begin{center}
  \includegraphics[angle=-0,scale=0.6]{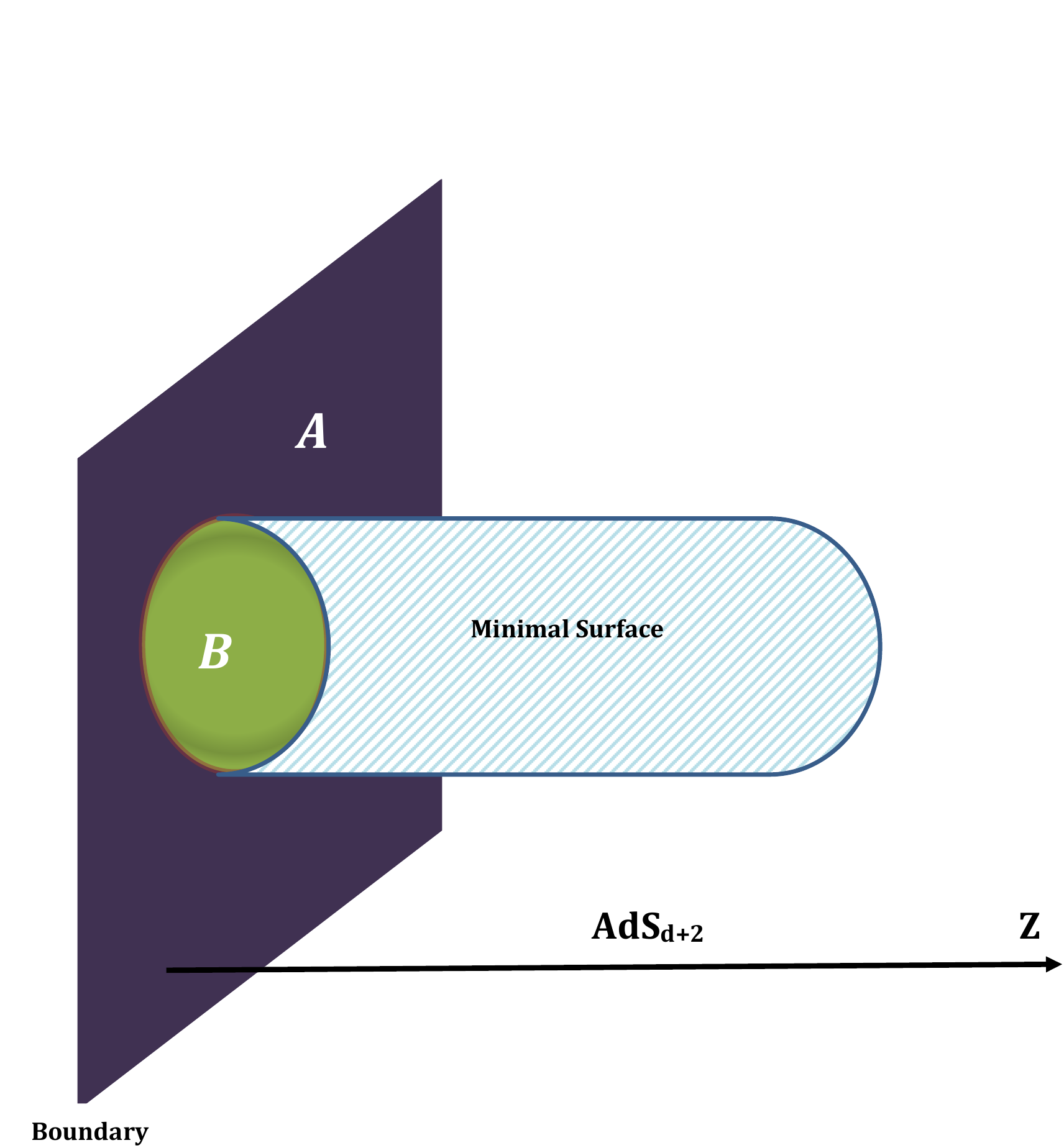}
\end{center}
\caption{The accessible and inaccessible regions in AdS.}\label{sketch2}
            \end{figure}

Our observers live on the boundary $z=a$. The accessible region $A$ and the inaccessible region $B$ are both on this boundary $z=a$. The entanglement entropy in the   ${\rm CFT}_{d+1}$ which lives on this boundary can be compute from the  gravity theory which lives in the bulk ${\rm AdS}_{d+2}$ as follows. We extend  ${\bf N}=A\cup B$ to  the time slice $M$ of the bulk spacetime and we extend $\partial A$ to a $d-$dimensional surface $\Gamma_A\in M$ such that $\partial\Gamma_A=\partial A$ (figure (\ref{sketch2})).  The time slice $M$ is the $(d+1)-$dimensional hyperbolic space ${\bf H}_{d+1}$ which is an Euclidean manifold whereas $\Gamma_A$ is a minimal area surface. The entanglement entropy in the   ${\rm CFT}_{d+1}$ is then given by the formula \cite{Ryu:2006bv,Ryu:2006ef}

\begin{eqnarray}
S_A=\frac{{\rm Area}(\Gamma_A)}{4G_N^{(d+2)}}.
\end{eqnarray}

As an example we consider the case of $AdS_3$. The dual field theory is a $2-$dimensional conformal field theory with central charge $c$. The corresponding partition function is given by the formula (\ref{form1}). This is the partition function of the conformal field theory of free bosons on the torus with periods $1$ (real axis) and $\tau$ (imaginary axis). We have then
\begin{eqnarray}
Z(\tau,\bar{\tau})&=&(q\bar{q})^{-c/24}tr q^{L_0}\bar{q}^{\bar{L}_0}.
\end{eqnarray}
The variables $q$ and $\bar{q}$ are given in terms of the modular parameter $\tau$ by the relations
\begin{eqnarray}
q=\exp(2i\pi\tau)~,~\bar{q}=\exp(-2i\pi\bar{\tau}).
\end{eqnarray}
On the other hand, the entanglement entropy is given in terms of the partition function $Z_n$ on the torus of lengths $2\pi n/\kappa$ and $L$ around its two cycles by the relation
\begin{eqnarray}
  S_A=\bigg(\bigg[1-n\frac{\partial}{\partial n}\bigg]\ln Z_n\bigg)|_{n=1}.
\end{eqnarray}
We can take the modular parameter $\tau$ to be $\tau =(i 2\pi n/\kappa)/L=2i\pi n/\kappa L$. But the partition function is invariant under the modular transformation $\tau\longrightarrow -1/\tau$ and hence we can take $\tau$ to be $\tau=i\kappa L/2\pi n$. In other words, $q=\bar{q}=\exp(-\kappa L/n)$. The entanglement entropy becomes
\begin{eqnarray}
  S_A&=&\bigg(1+\ln q\frac{\partial}{\partial \ln q}\bigg)\ln Z(\tau,\bar{\tau})\nonumber\\
  &=&\bigg(1+\ln q\frac{\partial}{\partial \ln q}\bigg)\bigg(-\frac{c}{12}\ln q+\ln tr q^{L_0+\bar{L}_0}\bigg)\nonumber\\
  &=&-\frac{c}{6}\ln q+\bigg(1+\ln q\frac{\partial}{\partial \ln q}\bigg)\ln tr q^{L_0+\bar{L}_0}\nonumber\\
  &=&-\frac{c}{6}(-\kappa L)\nonumber\\
  &=&\frac{c}{3}\ln \frac{\Sigma}{\epsilon}.
\end{eqnarray}
In going from the third to the fourth lines we have assumed that $\ln tr q^{L_0+\bar{L}_0}$ is exponentially suppressed  whereas in the last line we have used the result (\ref{Lkappa}).

This result can be re-derived from the gravity side as follows.

The metric in $AdS_3$ can be given by (Poincare coordinates)
\begin{eqnarray}
  ds^2=g^{\mu\nu}dx_{\mu}dx_{\nu}=\frac{R^2}{z^2}(-dt^2+dx^2+dz^2).
\end{eqnarray}
The boundary lies at $z=0$. We will regularize by taking the restriction $z\geq a$. On the boundary $z=a$ we are interested in the line segment $x\in [-l/2,l/2]$. This segment is extended in the bulk (the plane $xz$) to a line of minimal length, i.e. a spacelike geodesic, which can be found as follows. The length is written as (fixed time)
\begin{eqnarray}
  L=\int \frac{R}{z}\sqrt{\dot{z}^2+\dot{x}^2}ds.
\end{eqnarray}
Then we write Euler-Lagrange equations for $x$ and $z$ as (with $c$ a constant)
\begin{eqnarray}
&&x\longrightarrow \frac{\dot{x}}{z}=c \sqrt{\dot{z}^2+\dot{x}^2}\nonumber\\
  &&z\longrightarrow \frac{\dot{x}^2}{z^2}=-\frac{\sqrt{\dot{z}^2+\dot{x}^2}}{z} \frac{d}{ds}\bigg(\frac{\dot{z}}{\sqrt{\dot{z}^2+\dot{x}^2}}\bigg).
\end{eqnarray}
The solution is given by the circle
\begin{eqnarray}
  z=A\sin s~,~x=A\cos s.
\end{eqnarray}
By imposing the boundary conditions that the circle start at $(a,+l/2)$ and terminates at $(a,-l/2)$ we obtain
\begin{eqnarray}
  z=\frac{l}{2}\sin s~,~x=\frac{l}{2}\cos s~,~\epsilon \leq s\leq \pi-\epsilon~,~\epsilon=\frac{2a}{l}\longrightarrow 0.
\end{eqnarray}
We compute now the actual length of the minimal line in the bulk as
\begin{eqnarray}
  L=\int_{\epsilon}^{\pi-\epsilon} \frac{R}{\sin s}ds=2R\ln \frac{l}{a}.
  \end{eqnarray}
The central charge $c$ of ${\rm CFT}_2$ is related to the radius $R$ of ${ AdS}_3$ by the relation (\ref{rm}). The proportionality factor is given precisely by \cite{Brown:1986nw}
\begin{eqnarray}
  c=\frac{3R}{2G_N^{(3)}}.
  \end{eqnarray}
The entanglement entropy becomes
\begin{eqnarray}
S_A=\frac{L}{4G_N^{(3)}}= \frac{c}{3}\ln \frac{l}{a}.
\end{eqnarray}

\section{Einstein's gravity from quantum entanglement}

A sample of the original literature for this section is \cite{Lashkari:2013koa,Faulkner:2013ica,VanRaamsdonk:2016exw,Casini:2011kv}. However, a very good concise and pedagogical review of the formalism relating spacetime geometry to quantum entanglement due to Van Raamsdonk and collaborators is found in \cite{Jaksland:2017nqx}.

\subsection{The CFT/black hole correspondence}
The starting point is the statement that Einstein's theory of general relativity on anti-de Sitter spacetime $AdS_{d+1}$ is dual to a conformal field theory ${\rm CFT}_d$ on its boundary $M^d$. Let $|\psi(0)\rangle$ be the vacuum state of the ${\rm CFT}_d$. This state is dual to the Poincare patch of the pure AdS spacetime given by the metric
\begin{eqnarray}
  ds^2= \frac{L^2}{z^2}(dz^2+dx_{\mu}dx^{\mu}).
\end{eqnarray}
Let $|\psi(\zeta)\rangle$ be a one-parameter family of ${\rm CFT}_d$ excited states which are dual to the perturbed metrics 
\begin{eqnarray}
  ds^2= \frac{L^2}{z^2}(dz^2+\Gamma_{\mu\nu}(z,x)dx^{\mu}dx^{\nu}).
\end{eqnarray}
This corresponds to a spacetime $M_{\zeta}$ with boundary at $z\longrightarrow 0$ denoted by $\partial M_{\zeta}$ where the state  $|\psi(\zeta)\rangle$ is living.Thus, $\Gamma_{\mu\nu}\longrightarrow \eta_{\mu\nu}$ and $M_{\zeta}\longrightarrow AdS_{d+1}$ when $\zeta\longrightarrow 0$. For small $z$ (near the boundary) the metric behaves as
\begin{eqnarray}
\Gamma_{\mu\nu}(z,x)=\eta_{\mu\nu}+z^d\bar{h}_{\mu\nu}(z,x)\Rightarrow  ds^2= \frac{L^2}{z^2}(dz^2+dx_{\mu}dx^{\mu}+z^d\bar{h}_{\mu\nu}(z,x)dx^{\mu}dx^{\nu}).
\end{eqnarray}
This is called the Fefferman-Graham coordinates.

However, this metric can also be understood as corresponding to a spacetime  $M_{\zeta}$, which is a perturbation of pure AdS, dual  to a small perturbation  $|\psi(\zeta)\rangle_{\zeta\longrightarrow 0}$ of the ${\rm CFT}_d$ vacuum  $|\psi(0)\rangle$. This is  an asymptotically AdS spacetime.

For higher excited states   $|\psi(\zeta)\rangle$ we can not assume classical supergravity solution since $l_s$ is no longer much less than $L$ and as a consequence stringy corrections of the order $l_s^2$ and higher become important. The geometry (and even the topology) of $AdS_{d+1}$ becomes therefore very different.

An example of a non-trivial dual spacetime is the Schwarzschild-AdS black hole in $d+1$ dimensions given by the metric 

\begin{eqnarray}
 ds^2= -f_M(r)dt^2+\frac{dr^2}{f_M(r)}+r^2d\Omega_{d-2}.
\end{eqnarray}
The function $f_M$ is given by the difference of two pieces (the first being the usual  Schwarzschild term)
\begin{eqnarray}
  f_M(r)=1-\frac{2\mu}{r^{d-3}}+\frac{r^2}{L^2}~,~\mu=\frac{8\pi G_N M}{(d-1){\rm Vol}({\bf S}^{d-1})}~,~\mu_{d=4}=G_NM.
\end{eqnarray}
This is an eternal black hole in $d+1$ dimensions which is asymptotically a pure AdS in contrast to the eternal Schwarzschild black hole  which is asymptotically a flat Minkowski spacetime.

\begin{figure}[H]
\begin{center}
  \includegraphics[angle=-0,scale=0.5]{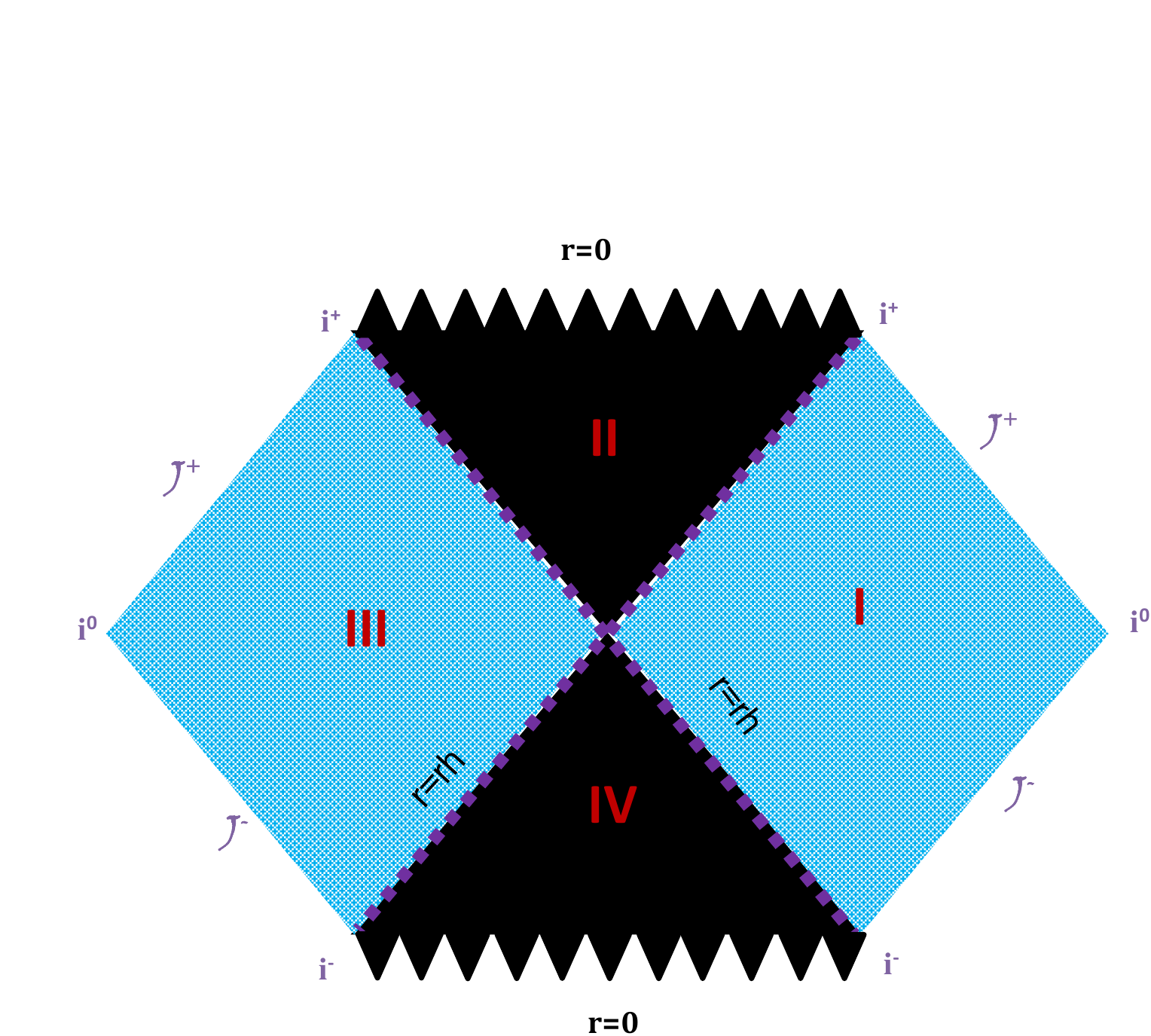}
\end{center}
\caption{The Schwarzschild spacetime.}\label{sch}
\end{figure}

Indeed, if we set $L=\infty$ we obtain  Schwarzschild black hole. This is characterized by the Penrose diagram (\ref{sch}) which summarizes the causal structure of the maximally extended  Schwarzschild spacetime in the Kruskal-Szekeres coordinates $-\infty<R<+\infty$ and $-\infty<T<+\infty$.  The two light-like infinities ${\cal J}^{\pm}$ (where light rays begin and end) and the space-like infinity $i^0$ ($r=\infty$) are the same as those of Minkowski spacetime and hence the eternal Schwarzschild black hole is asymptotically a flat Minkowski spacetime. The time-like trajectories begin and end on the two time-like infinities $i^{-}$ and $i{+}$ respectively which are distinct surfaces from the singularity at $r=0$. The horizon is located at $r_h=(2\mu)^{1/d-3}$. The region II is the interior of the black hole whereas the region I is the exterior. The region III lies also outside the black hole but it is spatially separated and therefore causally disconnected from region I. Region IV is the interior of a white hole, i.e. we can never go there but things can emerge from that region towards region I.

\begin{figure}[H]
\begin{center}
  \includegraphics[angle=-0,scale=0.4]{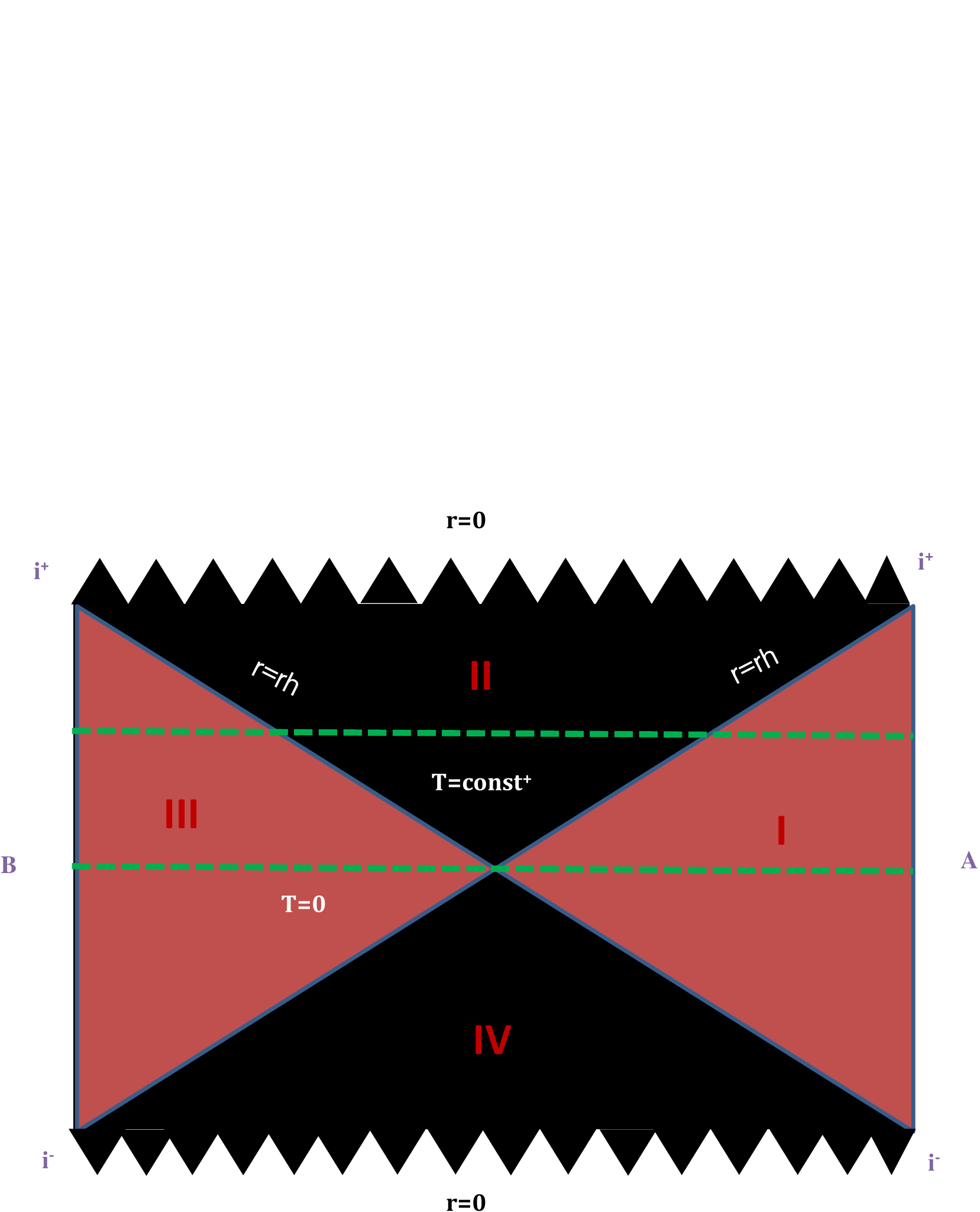}
\end{center}
\caption{The  Schwarzschild-AdS spacetime.}\label{sch1p}
\end{figure}

The Penrose diagram of the   Schwarzschild-AdS spacetime is shown on figure (\ref{sch1p}). The two light-like infinities ${\cal J}^{\pm}$ and the space-like infinity $i^0$ are replaced with the universal covering of global AdS spacetime in both regions I and III. These asymptotic regions are denoted $A$ and $B$ and they are the conformal boundary of AdS spacetime, i.e. ${\bf R}\times {\bf S}^{d-1}$. We have two possible situations:
\begin{itemize}
\item For $r_h<<L$ we can neglect the term $r^2/L^2$ and we end up with an ordinary   Schwarzschild spacetime. The  Schwarzschild black hole at the center will then evaporate due to Hawking emission.
  \item For $r_h>>L$ the Schwarzschild-AdS spacetime is in equilibrium with its Hawking radiation, i.e. emission and absorption rates are equal and the black hole will not evaporate. This is due to the fact that radiation can reach the asymptotic AdS boundary and return in a finite time. 
\end{itemize}

\begin{figure}[H]
\begin{center}
  \includegraphics[angle=-0,scale=0.5]{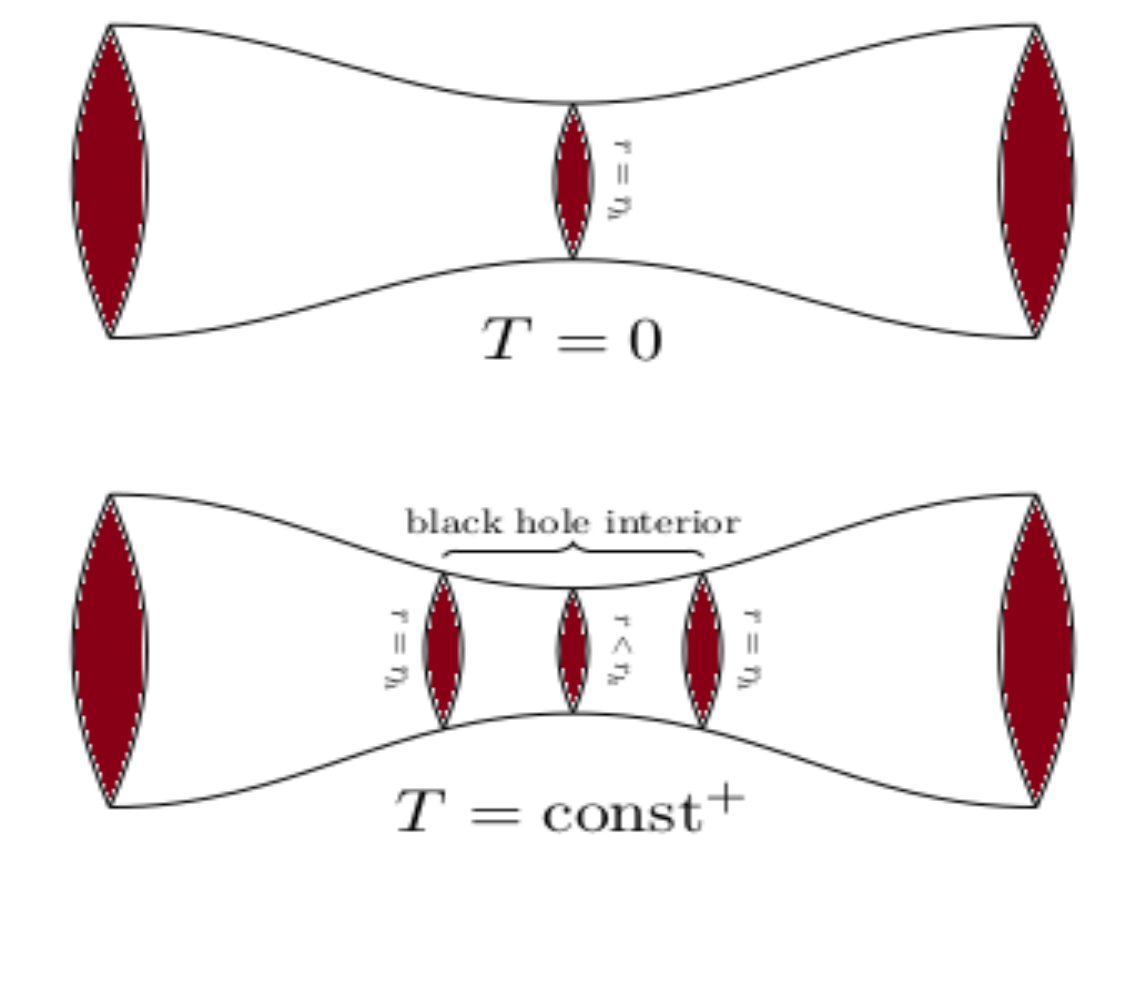}
\end{center}
\caption{The two-sided Schwarzschild-AdS black hole as a wormhole.}\label{sch2}
\end{figure}

The regions I and III are causally disconnected but signals from region I can intersect with signals from region III behind the horizon. This is then a two-sided black hole, i.e. with two different exterior regions, which can also be viewed as a wormhole as depicted in figure (\ref{sch2}) for $T=0$ and $T>0$.   

The Schwarzschild-AdS black hole in $d+1$ dimensions is conjectured in \cite{Maldacena:2001kr} to be dual to the thermofield double state $|\Psi\rangle$ of two identical non-interacting copies of the conformal field theory ${\rm CFT}_d$ living on the cylinder ${\bf R}\times{\bf S}^{d-1}$ (the asymptotic boundaries $A$ and $B$).

The ${\rm CFT}$ dual to the  Schwarzschild-AdS black is defined therefore on $A\cup B$ and as a consequence the quantum system of interest is constituted of two subsystems $Q_A$ and $Q_B$ which contain the degrees of freedom of the local ${\rm CFT}$ living on $A$ and $B$ respectively. The two subsystems $Q_A$ and $Q_B$ can only interact via entanglement.

Let ${\cal H}_A$ and ${\cal H}_B$ be the Hilbert spaces associated with $Q_A$ and $Q_B$ respectively and let $\{|E_i^A\rangle\}$ and $\{|E_i^B\rangle\}$ be the corresponding bases. The thermofield double state $|\Psi\rangle$ dual to the  Schwarzschild-AdS black hole is then given by 
\begin{eqnarray}
  |\Psi\rangle=\frac{1}{\sqrt{Z(\beta)}}\sum_i\exp(-\beta E_i/2)|E_i^A\rangle\otimes|E_i^B\rangle.
\end{eqnarray}
The partition function at inverse temperature $\beta$ is given by
\begin{eqnarray}
 Z(\beta)=\sum_i\exp(-\beta E_i).
\end{eqnarray}
The reduced density matrix for the subsystem $Q_A$ is immediately given by integrating out the degrees of freedom associated with the subsystem $Q_B$, viz
\begin{eqnarray}
  \rho_A=tr_B |\Psi\rangle\langle\Psi|=\frac{1}{\sqrt{Z(\beta)}}\sum_i\exp(-\beta E_i)|E_i^A\rangle\langle E_i^A|.
\end{eqnarray}
The corresponding entanglement entropy is given by
\begin{eqnarray}
  S_A&=&-tr_A \rho_A\log \rho_A\nonumber\\
  &=&\frac{1}{\sum_ie^{-\beta E_i}}\sum_ie^{-\beta E_i}\beta E_i+\log(\sum_ie^{-\beta E_i}).
\end{eqnarray}
This entanglement entropy is always non-zero except in the limit $\beta\longrightarrow \infty$. Indeed, in the limit of zero temperature the reduced density matrix $\rho_A$ approaches $\rho_A=|E_0^A\rangle\langle E_0^A|$ where $E_0$ denotes the energy of the ground state and $E_0=0$. Hence, in this limit only the ground state is occupied and as a consequence the entanglement entropy vanishes. We have then in the limit of zero temperature the behavior
\begin{eqnarray}
  |\Psi\rangle=\frac{1}{\sqrt{Z(\beta)}}\sum_i\exp(-\beta E_i/2)|E_i^A\rangle\otimes|E_i^B\rangle\longrightarrow |\Phi\rangle=|E_0^A\rangle\otimes|E_0^B\rangle.
\end{eqnarray}
This is a product state with zero entanglement entropy describing two completely uncorrelated subsystems $Q_A$ and $Q_B$.

The thermofield double state $|\Psi\rangle$ where $Q_A$ and $Q_B$ are entangled is dual to the Schwarzschild-AdS black hole which is a connected spacetime in which light signals traveling from $A$ and $B$ can intersect. Similarly, the product state $|\Phi\rangle$ where $Q_A$ and $Q_B$ are now disentangled or uncorrelated is dual to a spacetime consisting of the product of two pieces corresponding to two disconnected regions $A$ and $B$, i.e. light signals traveling from $A$ and $B$ can not intersect. Thus, in the limit $\beta\longrightarrow \infty$ entanglement between $Q_A$ and $Q_B$ is removed and correspondingly connectivity between regions $A$ and $B$ is reduced until they become disconnected in the limit of zero temperature. This shows clearly that entanglement between $Q_A$ and $Q_B$ is a necessary condition for classical connectivity between $A$ and $B$.

\begin{figure}[H]
\begin{center}
    \includegraphics[angle=-0,scale=0.6]{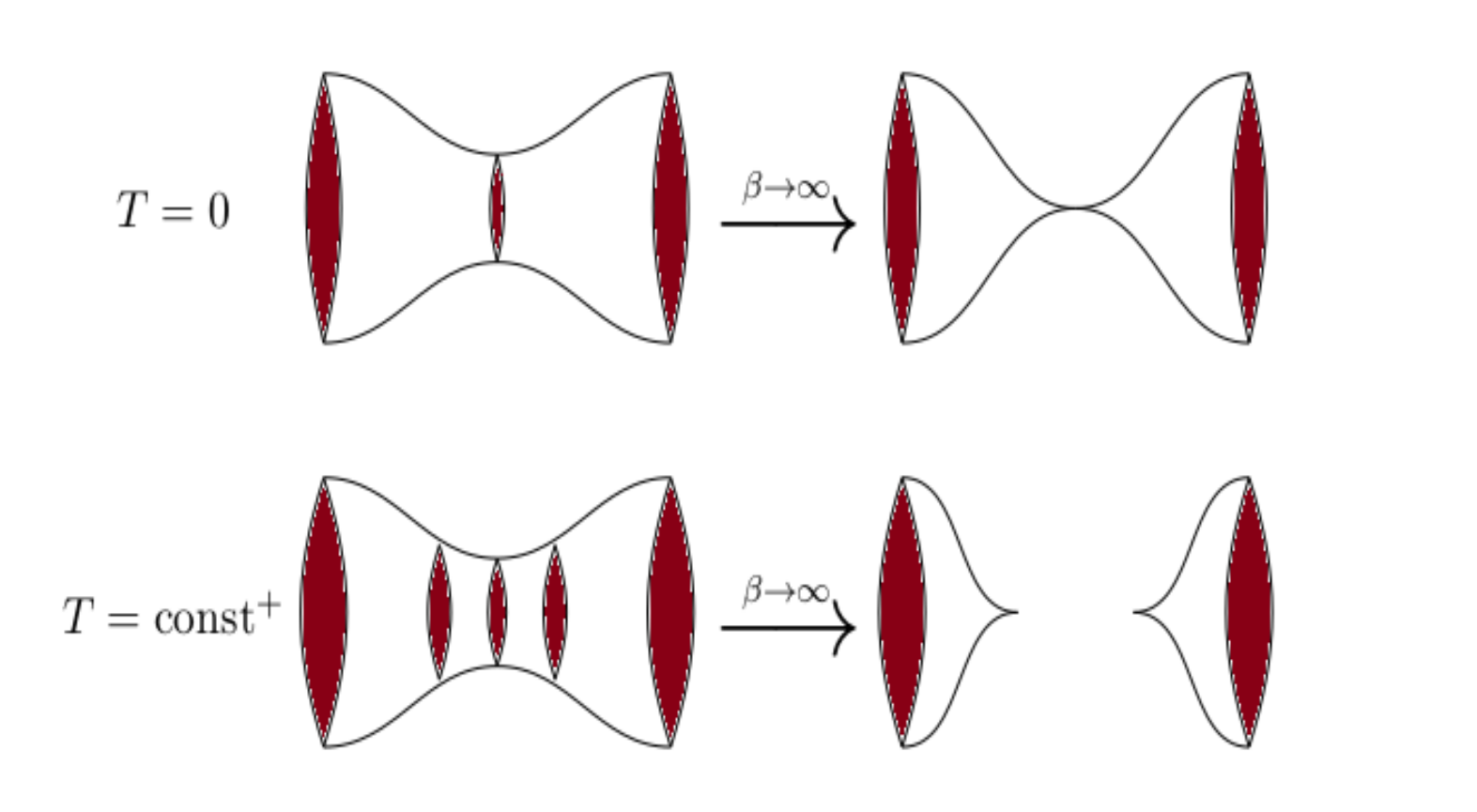}
\end{center}
\caption{The limit $\beta\longrightarrow \infty$ in the Schwarzschild-AdS black hole.}\label{sch3p}
\end{figure}

This picture can also be confirmed from the Ryu-Takayanagi formula $S_{\rm BH}=A_{\rm BH}/4G$. In the limit $\beta\longrightarrow\infty$ in the thermofield double state the entanglement between $Q_A$ and $Q_B$ is removed and thus $S_{\rm BH}\longrightarrow 0$ or equivalently $A_{\rm BH}\longrightarrow 0$. In other words, the limit $\beta\longrightarrow\infty$ in the thermofield double state decreases the connectivity in the dual black hole spacetime as seen on figure (\ref{sch3p}). Indeed, the proper distance between any two points in $A$ and $B$  (encoded in the mutual information between $S_A$ and $S_B$) goes to $\infty$ as the entanglement between $Q_A$ and $Q_B$ goes to zero in the limit $\beta\longrightarrow\infty$.

This relation between quantum entanglement and spacetime connectivity generalizes to any quantum state with a classical spacetime dual.

\subsection{The first law of thermodynamics and CFT}
َAs before we will consider a one-parameter family of CFT states $|\psi(\zeta)\rangle$ living on the boundary of AdS spacetime $AdS_{d+1}$. The dual spacetime of the vacuum state $|\psi(0)\rangle$ of the CFT is assumed to be given by the Poincar\'e patch of the anti-de Sitter spacetime defined by the metric

\begin{eqnarray}
  ds^2= \frac{L^2}{z^2}(dz^2+dx_{\mu}dx^{\mu}).
\end{eqnarray}
The dual spacetime of the perturbed state  $|\psi(\zeta)\rangle$,i.e.  with $\zeta\longrightarrow 0$,  is denoted by ${\cal M}_{\psi}$ and  is a small perturbation of anti-de Sitter spacetime with boundary $\partial {\cal M}_{\psi}$ which is itself a small perturbation of Minkowski spacetime.

Let $B$ some spatial region on the boundary  $\partial {\cal M}_{\psi}$ and let $\bar{B}$ its complement. The accessible region $B$ is a ball shaped region of radius $R$ on the Cauchy surface $\Sigma_{\partial{\cal M}_{\psi}}\in \partial{\cal M}_{\psi}$. The entanglement entropy of this spatial region in the CFT is equal to the von Neumann entropy of the reduced density matrix
\begin{eqnarray}
\rho_B=tr_{\bar{B}}|\psi(\zeta)\rangle\langle\psi(\zeta)|.
\end{eqnarray}
The entanglement entropy is thus given by
\begin{eqnarray}
  S_B=-tr_B\rho_B\log \rho_B.
\end{eqnarray}
We can immediately compute 
\begin{eqnarray}
  \frac{d}{d\zeta}S_B=-tr_B\frac{d}{d\zeta}\rho_B.\log \rho_B.
\end{eqnarray}
We define the so-called modular Hamiltonian by the relation
\begin{eqnarray}
  H_B=-\log \rho_B(\zeta=0).
\end{eqnarray}
Thus
\begin{eqnarray}
  \frac{d}{d\zeta}S_B&=&tr_BH_B\frac{d}{d\zeta}\rho_B+O(\zeta)\nonumber\\
  &=&\frac{d}{d\zeta}tr_BH_B\rho_B+O(\zeta)\nonumber\\
   &=&\frac{d}{d\zeta}\langle H_B\rangle+O(\zeta).
\end{eqnarray}
The expectation value of the modular Hamiltonian is what we call the hyperbolic energy, viz $\langle H_B\rangle=E_B^{\rm Hyp}$. We get then

\begin{eqnarray}
  \frac{d}{d\zeta}S_B   &=&\frac{d}{d\zeta}E_B^{\rm Hyp}.
\end{eqnarray}
This is effectively the first law of thermodynamics $dE=dS$ which holds in the CFT for arbitrary perturbations $\zeta$ of the vacuum state $|\psi(0)\rangle$ and not only for thermal and equilibrium configurations as it is usually the case for the first law of thermodynamics.

\subsection{Holographic or AdS entanglement entropy}
The variation in the entanglement entropy in the CFT is given by  $dS_B/d\zeta$. In the AdS the variation in the entanglement entropy is given by the Ryu-Takayanagi formula, viz
\begin{eqnarray}
  \delta S_B=\frac{\delta A_B}{4G_N}\equiv  \frac{d}{d\zeta}S_B  .
\end{eqnarray}
$A_B$ is the area of the extremal surface $\tilde{B}$ in the bulk such that $\partial \tilde{B}=\partial B$ where the accessible region $B$ is a ball shaped region of radius $R$. This is the holographic interpretation of the entanglement entropy.

Thus, the variation in the entanglement entropy  $dS_B/d\zeta$ of the CFT state $|\psi\rangle$ with a dual spacetime given by pure AdS corresponding to a ball shaped region $B$ is proportional to the variation  $\delta A_B$  in the area of the extremal surface $\tilde{B}$ due to the corresponding perturbation of the AdS space.

Let  $g_{ab}^0$ be the metric of pure AdS corresponding to the vacuum state $|\psi(0)\rangle$ and $g_{ab}$ be the metric of the perturbed AdS corresponding to the excited state $|\psi(\zeta)\rangle$. We have
\begin{eqnarray}
g_{ab}=g_{ab}^0+\delta g_{ab}.
\end{eqnarray}
The perturbation is given by  Fefferman-Graham coordinates
\begin{eqnarray}
\delta g_{ab}= L^2z^{d-2}\bar{h}_{ab}(z,x)=z^{d-2}{h}_{ab}(z,x).
\end{eqnarray}
The surface $\tilde{B}$ is an extension of the spatial region $B$ into the bulk and thus it is a co-dimension two surface characterized by some embedding functions $X^a(\sigma)$ with area $A_B$ given by 

\begin{eqnarray}
A_B(g,X)=\int d^{d-1}\sigma\sqrt{{\rm det}\gamma_{ij}}.
\end{eqnarray}
The induced metric $\gamma_{ij}$ on this surface $\tilde{B}$  is given by 
\begin{eqnarray}
\gamma_{ij}=g_{ab}\frac{\partial X^a}{\partial \sigma^{i}}\frac{\partial X^b}{\partial \sigma^{j}}.
\end{eqnarray}
In pure AdS this area is extremized by some embedding functions $X_{\rm ext}^0$ whereas in perturbed AdS it is extremized by some other  embedding functions $X_{\rm ext}$. We have then the variation in the area of the extremal surface given by
\begin{eqnarray}
  \delta A_B(g,X_{\rm ext})&=&\frac{\delta  A_B(g,X_{\rm ext}^0)}{\delta g}\delta g+\frac{\delta  A_B(g^0,X_{\rm ext})}{\delta X^a}\delta X^a\nonumber\\
  &=&\frac{\delta  A_B(g,X_{\rm ext}^0)}{\delta g}\delta g+O(\delta g^2).
\end{eqnarray}
In this equation we have used the fact that $\delta X$ is of order $\delta g$ and as a consequence the second term is subleading of order $\delta g^2$.  The variation in the area of the extremal surface is thus obtained by holding the embedding functions fixed and varying the spacetime metric.

Hence, the variation in the induced metric which in fact controls directly the variation in $ A_B(g,X_{\rm ext})$ should be obtained by   holding the embedding functions fixed and varying the spacetime metric, viz 
\begin{eqnarray}
  \delta \gamma_{ij}=\delta g_{ab}\frac{\partial X^a_{\rm ext}}{\partial \sigma^{i}}\frac{\partial X^b_{\rm ext}}{\partial \sigma^{j}}.
\end{eqnarray}
In this equation $X_{\rm ext}=X_{\rm ext}^0$ and obviously 
\begin{eqnarray}
  \delta \gamma_{ij}=\frac{d}{d\zeta}\gamma_{ij}|_{\zeta=0}.
\end{eqnarray}
The variation  in the area of the extremal surface is given explicitly in terms of the variation of the induced metric by
\begin{eqnarray}
\delta A_B(g,X_{\rm ext})=\int d^{d-1}\sigma\sqrt{{\rm det}\gamma_{ij}^0}(\frac{1}{2}\gamma_0^{kl}\delta \gamma_{kl}).
\end{eqnarray}
 The embedding functions $X^a_{\rm ext}$ are mappings from the spatial surface $\tilde{B}$ to the bulk whose boundary at $z=0$ is the spatial surface $B$ which is a ball shaped region of radius $R$. Both $B$ and $\tilde{B}$ are co-dimension two surfaces.

For pure AdS the bulk surface $\tilde{B}$ which turns out to be extremal is given by the sphere in $d$ dimensions of radius $R$, i.e.
\begin{eqnarray}
  \vec{x}^2+z^2=R^2.
\end{eqnarray}
This has been shown explicitly for $AdS_3$ and the argument for higher dimensional AdS spacetimes is similar. 

We can now parametrize the extremal surface  $\tilde{B}$ for pure AdS by $\sigma^{i}=x^i$ and we choose the embedding functions  $X^a_{\rm ext}~:~{\bf R}^{d-1}\longrightarrow {\bf R}^{d+1}$ as follows
\begin{eqnarray}
X^0_{\rm ext}=t_0~,~X^1_{\rm ext}=x^1~,~...~,~X^{d-1}_{\rm ext}=x^{d-1}~,~X^d_{\rm ext}=z=\sqrt{R^2-\vec{x}^2}.
\end{eqnarray}
We compute in the radial gauge (defined by $h_{zz}=h_{z\mu}=0$) the variation in the induced metric given by 
\begin{eqnarray}
  \delta \gamma_{ij}=z^{d-2}h_{ij}.
\end{eqnarray}
By using the above embedding functions we can easily compute the unperturbed induced metric
\begin{eqnarray}
  \gamma_{ij}^0&=&g_{ab}^0\frac{\partial X^a}{\partial x^{i}}\frac{\partial X^b}{\partial x^{j}}\nonumber\\
  &=&\frac{L^2}{z^2}(\delta_{ij}+\frac{x_ix_j}{z^2}).
\end{eqnarray}
By using these two last results the variation  in the area of the extremal surface can now be computed explicitly to obtain
\begin{eqnarray}
\delta A_B=\frac{L^{d-3}R}{2}\int_{\tilde{B}} d^{d-1}x\bigg(\delta^{ij}-\frac{1}{R^2}(x-x_0)^i(x-x_0)^j\bigg)h_{ij}.\label{Result0}
\end{eqnarray}
The point $\vec{x}_0$ is the center of the ball shaped region $B$. The variation in the entanglement entropy in the AdS is thus given by
\begin{eqnarray}
\delta S_B=\frac{L^{d-3}R}{8G_N}\int_{\tilde{B}} d^{d-1}x\bigg(\delta^{ij}-\frac{1}{R^2}(x-x_0)^i(x-x_0)^j\bigg)h_{ij}.\label{Result1}
\end{eqnarray}
\subsection{Modular Hamiltonian and hyperbolic energy}
In this section we will follow closely \cite{Casini:2011kv}. First we will define the so-called modular Hamiltonian in the CFT without holography then we will construct the AdS expression.

\subsubsection{Modular Hamiltonian}

We start by defining the causal development (also called domain of dependence) ${\cal D}$ of the region $B$ by the set of all points $p$ for which every inextensible causal (timelike or lightlike) curve through $p$ necessarily intersect $B$. Classically, the field values on ${\cal D}$ are completely determined in terms of the initial field values on $B$. Quantum mechanically, any operator in ${\cal D}$ is determined in terms of the fields in $B$ alone and hence it can be calculated using the reduced density matrix $\rho_B$. The domain of dependence ${\cal D}$ is also the causal development of any other spacelike surface whose boundary is the same as the boundary of $B$. 

The reduced density matrix $\rho_B$ is hermitian and positive semi-definite and thus it can be rewritten in terms of a hermitian operator $H_B$ as
\begin{eqnarray}
\rho_B=\exp(-H_B).
\end{eqnarray}
This operator $H_B$ is called the modular Hamiltonian or entanglement Hamiltonian. Generically, this Hamiltonian is not local since it can not be represented by a local expression in terms of fields defined on $B$. This Hamiltonian  generates the modular group given by the transformations
\begin{eqnarray}
tr\rho{\cal O}=tr\rho U(s){\cal O}U(-s)~,~U(s)=\rho^{is}=\exp(-is H).
\end{eqnarray}
The operator $U(s)$, although it looks like a time evolution transformation along $s$, does not generate a local flow on ${\cal D}$ since $H_B$ is not local. Indeed, the operator defined by ${\cal O}(s)=U(s){\cal O}U(-s)$ is not local, i.e. it is not defined at a point $x$, even if we start with a local operator ${\cal O}$ defined at a point $x$, e.g. ${\cal O}=\phi(x)$.

\subsubsection{Rindler wedge}
\begin{figure}[H]
\begin{center}
\includegraphics[angle=-0,scale=0.6]{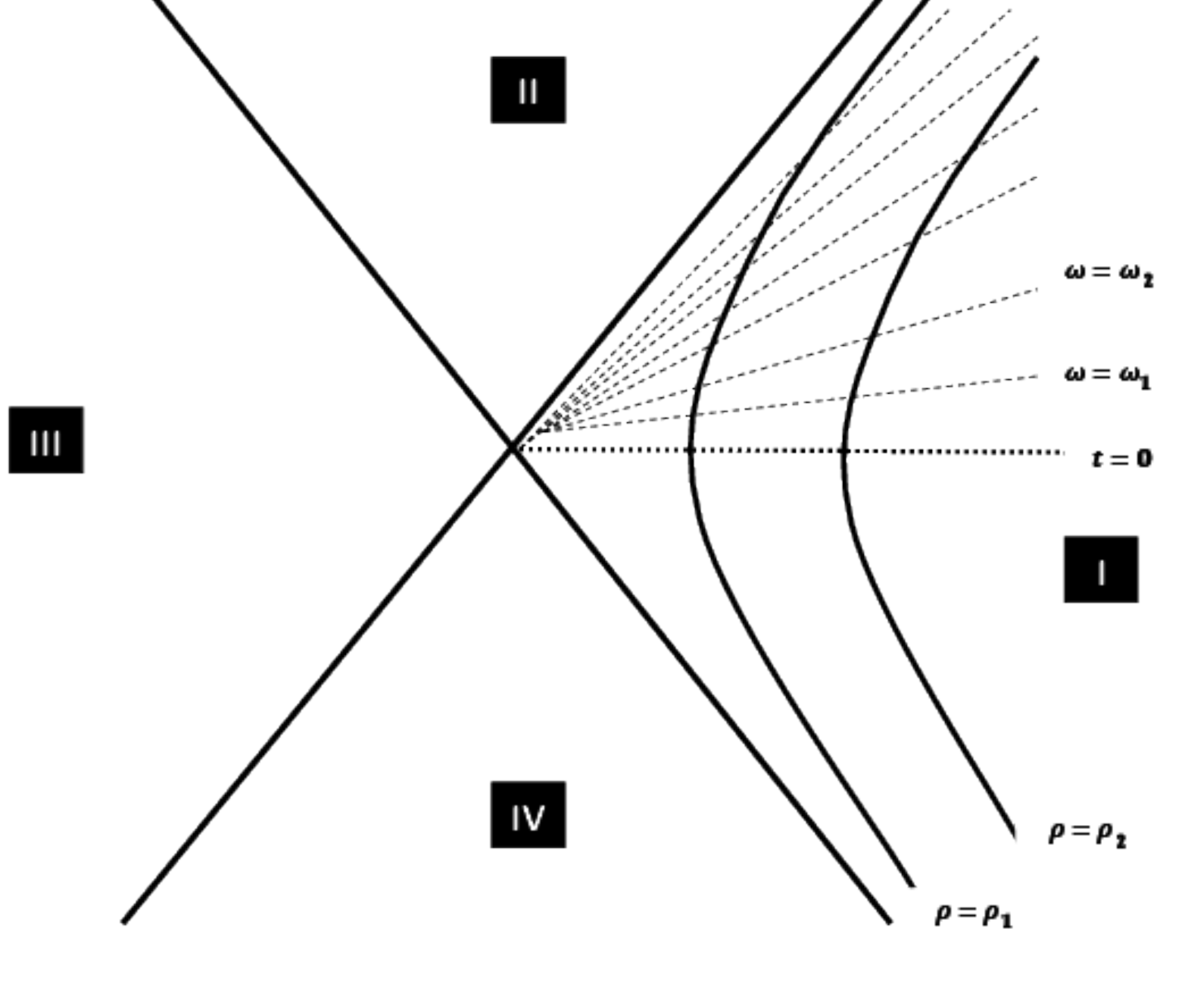}
\end{center}
\caption{Rindler decomposition.}\label{rindler}
\end{figure}

However, there are special circumstances where the modular Hamiltonian is local. The first example is Rindler space ${\cal R}$  \cite{Casini:2011kv}. The metric is given by ($T=\rho\sinh\omega$ and $Z=\rho\cosh\omega$)
\begin{eqnarray}
  ds^2&=&\rho^2d\omega^2-d\rho^2-dX^2-dY^2\nonumber\\
  &=&dT^2-dZ^2-dX^2-dY^2.
\end{eqnarray}
The Rindler wedge is region I which is the part of Minkowski spacetime accessible to a uniformly accelerating observer. Quadrants II and III have
no causal relations with quadrant I. Quadrant IV provides initial data for the Rindler wedge. This is the Rindler decomposition of Minkowski spacetime. See figure (\ref{rindler}).

A translation in the Rindler time $\omega\longrightarrow \omega +c$ corresponds to a Lorentz boost along the $Z$ direction in Minkowski spacetime. The corresponding generator is the Rindler Hamiltonian $H_{\omega}$, which  is precisely the Hamiltonian in quadrant I, given by 
 \begin{eqnarray}
H_{\omega}=\int_{\rho=0}^{\rho=\infty}\rho d\rho dX dY T^{00}(\rho,X,Y).
\end{eqnarray}
 $T^{00}$ is the Hamiltonian density with respect to the Minkowski observer. Note also that in the $T-Z$ plane the lines of constant Rindler time $\omega$ are straight lines through the origin. The proper time separation between these lines is $\delta\tau=\rho\delta\omega$. This is the origin of the $\rho$ factor multiplying $T^{00}$.

The surface $T=0$ is divided into two halves. The first half in region I and the second half in region III. The fields in region I ($Z>0$) act in the Hilbert space ${\cal H}_R$ and those in region III ($Z<0$) act in the Hilbert space ${\cal H}_L$. We have then 
\begin{eqnarray}
\phi(X,Y,Z)=\phi_R(X,Y,Z)~,~Z>0.
\end{eqnarray}
\begin{eqnarray}
\phi(X,Y,Z)=\phi_L(X,Y,Z)~,~Z<0.
\end{eqnarray}
The general wave functional of interest is the pure state
\begin{eqnarray}
\Psi=\Psi(\phi_L,\phi_R).
\end{eqnarray}
This is the ground state of the Minkowski Hamiltonian which can be computed using Euclidean path integrals. We get after some calculation the transition matrix element \cite{Ydri:2017oja}
\begin{eqnarray}
\Psi(\phi_L,\phi_R)&=&\langle \phi_R|\langle\phi_L|\Omega\rangle\nonumber\\
&\propto &\sum_i e^{-\pi E_i}\langle \phi_R|i_R\rangle \langle\phi_L|i^*_L\rangle.
\end{eqnarray}
In other words, we get the ground state 
\begin{eqnarray}
|\Omega\rangle=\frac{1}{\sqrt{Z}}\sum_i e^{-\pi E_i}|i_R\rangle|i^*_L\rangle.
\end{eqnarray}
The entanglement between the left and right wedges is now fully manifest.

However, we want to compute the reduced density matrix $\rho_{A}$ used by observers in the Rindler quadrant (quadrant I) to describe the Minkowski vacuum. This is the density matrix associated with the spatial half-space $Z>0$, $X=Y=0$ corresponding to the time slice $T=0$ which is denoted by $A$. In other words, the Rindler wedge ${\cal R}$ is the causal development of the half-space $A$. We can define immediately the reduced matrix $\rho_{A}$ by the relation 
\begin{eqnarray}
\rho_{A}(\phi_R,\phi_R^{'})&=&\int \Psi^*(\phi_L,\phi_R)\Psi(\phi_L,\phi_R^{'})d\phi_L\nonumber\\
&=&\frac{1}{Z}\sum_ie^{-2\pi E_i}\langle i_R|\phi_R\rangle\langle \phi_R^{'}|i_R\rangle\nonumber\\
&=&\frac{1}{Z}\langle \phi_R^{'}|e^{-2\pi H_{\omega}}|\phi_R\rangle.
\end{eqnarray}
We get then the reduced density matrix 
\begin{eqnarray}
\rho_{A}
&=&\frac{1}{Z}e^{-2\pi H_{\omega}}.
\end{eqnarray}
Thus the fiducial observers in the Rindler wedge ${\cal R}$ see the vacuum as a thermal ensemble with a Maxwell-Boltzmann distribution at a temperature 
\begin{eqnarray}
T_{A}=\frac{1}{2\pi}.
\end{eqnarray}
This is the Unruh effect.
 
The modular flow on the Rindler wedge ${\cal R}$ corresponds therefore to the Rindler time translations $\omega\longrightarrow \omega +c$ and the modular Hamiltonian is given immediately by

\begin{eqnarray}
H_{A}=2\pi H_{\omega}+\log Z.
\end{eqnarray}

\subsubsection{The ball shaped region $B$}  Another very important example in which the modular flow and the modular Hamiltonian can be shown to be  local is a conformal field theory defined on the causal development ${\cal D}$ of a ball shaped region $B$ of radius $R$. It was shown in  \cite{Casini:2011kv} that this causal development ${\cal D}$ is conformally related to the Rindler wedge by a special conformal transformation.

Let us denote now the coordinates on the Rindler wedge by $X^{\mu}$ with metric $ds^2=\eta_{\mu\nu}dX^{\mu}dX^{\mu}$. The Rindler Wedge is the causal development or domain of dependence of the half-space $A$ corresponding to  $X^0=0$,  $X^1>0$ and $X^2=X^3=0$. This causal development can be given by $X^1\geq |X^0|$ or equivalently in terms of the null coordinates $X^{\pm}=X^1\pm X^0$ by $\{X^+\geq 0\}\cap \{X^-\geq 0\}$.

The causal development ${\cal D}$ with coordinates $x^{\mu}$ and metric $ds^2=\Omega^{-2} \eta_{\mu\nu}dx^{\mu}dx^{\mu}$ is obtained from the Rindler edge via the special conformal transformation \cite{Casini:2011kv}
\begin{eqnarray}
x^{\mu}=\frac{X^{\mu}-(X.X)C^{\mu}}{\Omega}+2R^2C^{\mu}~,~C^{\mu}=(0,\frac{1}{2R},0,...,0)~,~\Omega=1-2(X.C)+(X.X)(C.C).\label{sct}\nonumber\\
\end{eqnarray}
This maps the domain of dependence $\{X^+\geq 0\}\cap \{X^-\geq 0\}$ to the domain of dependence $\{x^+\leq R\}\cap \{x^-\leq R\}$ where the null coordinates $x^{\pm}$ are defined by  $x^{\pm}=r\pm t$ where $r=\sqrt{(x^1)^2+...+(x^{d-1})^2}$.

Let us introduce the hyperbolic coordinates on the Rindler wedge by $X^1=\rho\cosh \omega$ and $X^0=\rho\sinh\omega$. We have then $X^{\pm}=\rho\exp(\pm\omega)$. The modular flow on the Rindler wedge is given by the time translations $\omega\longrightarrow \omega^{\prime}=\omega+c$. We will write $c=2\pi s$ since the period is given precisley by the inverse temperature $1/T=2\pi$. Under this modular flow the coordinates transform as $X^{\pm}(0)=X^{\pm}\longrightarrow X^{\pm}(s)=X^{\pm}\exp(\pm 2\pi s)$. Under the special conformal transformation (\ref{sct}) this modular flow on the Rindler wedge transforms to the modular flow on ${\cal D}$ given explicitly by
\begin{eqnarray}
  x^{\pm}(s)=R\frac{R+x^{\pm}-e^{\mp 2\pi s}(R-x^{\pm})}{R+x^{\pm}+e^{\mp 2\pi s}(R-x^{\pm})}.\label{mf}
\end{eqnarray}
We compute for the Rindler wedge ${\cal R}$ the variation of the modular flow
\begin{eqnarray}
  \delta X^{\pm}(s)=\pm 2\pi \delta s  X^{\pm}(s)\Leftrightarrow  \delta X^{1}(s)= 2\pi \delta s  X^{0}(s)~,~\delta X^{0}(s)= 2\pi \delta s  X^{1}(s).
\end{eqnarray}
We focus on the time slice $s=\omega=0$. we get immediately the equations
\begin{eqnarray}
\delta X^1=\delta\rho=0~,~\delta X^0=\rho\delta\omega=2\pi\rho \delta s.
\end{eqnarray}
The corresponding generator $H_{A}$ is then given by 
 \begin{eqnarray}
H_{A}=2\pi \int d\rho dX dY \rho T^{00}(\rho,X,Y)+{\rm constant}.
\end{eqnarray}
 We go through the same steps for the caual development ${\cal D}$.  We compute immediately for ${\cal D}$ the variation of the mdoular flow
 \begin{eqnarray}
  \delta x^{\pm}(s)=\frac{\pm 4\pi R \delta s e^{\mp 2\pi s}(R^2-(x^{\pm})^{2})}{\big(R+x^{\pm}+e^{\mp 2\pi s}(R-x^{\pm})\big)^2}.
\end{eqnarray}
We focus on the ball $B$ of radius $R$ corresponding to the time slice $s=t=0$ and hence $x^{\pm}=r$. The variation of the mdoular flow becomes
 \begin{eqnarray}
  \delta x^{\pm}=\pm\frac{ \pi  \delta s (R^2-r^{2})}{R}.
 \end{eqnarray}
 This gives immediately the shift
 \begin{eqnarray}
  \delta r=0~,~\delta t=\frac{ 2\pi  \delta s (R^2-r^{2})}{2R}.\label{infD}
 \end{eqnarray}
The corresponding generator $H_B$ is then given by 
 \begin{eqnarray}
H_B=2\pi \int_B d^{d-1}x \frac{R^2-r^{2}}{2R} T^{00}(x)+{\rm constant}.\label{HB}
 \end{eqnarray}
A careful derivation of $H_B$ is given in \cite{Casini:2011kv}. The special conformal transformation (\ref{sct}), which is a symmetry of Minkowski spacetime, is associated with a unitary operator $U_0$ in the CFT which leaves the vacuum invariant, viz $U_0|0\rangle=|0\rangle$. This unitary operator acts on primary operators as

\begin{eqnarray}
\phi(x)=\Omega^{\Delta}U_0\phi(X)U_0^{-1}.
 \end{eqnarray}
This unitary operator must then act on the density matrices as follows
\begin{eqnarray}
  \rho_B&=&U_0\rho_A U_0^{-1}\nonumber\\
  &=&\frac{1}{Z}e^{-2\pi U_0H_{\omega}U_0^{-1}}\nonumber\\
  &=&\frac{1}{Z}e^{-H_{B}}. 
\end{eqnarray}
Next, if $U_{\cal R}$ is the quantum operator which generates the modular flow on the Rindler wedge ${\cal R}$ then the quantum operator which generates the modular flow on ${\cal D}$ must be given by
\begin{eqnarray}
U_{\cal D}(s)=U_0U_{\cal R}(s)U_0^{-1}.
\end{eqnarray}
We can then act on primary operators as follows
\begin{eqnarray}
  U_{\cal D}(s)\phi(x[s_0])U_{\cal D}(-s) &=&U_0U_{\cal R}(s)U_0^{-1}\phi(x[s_0])U_0 U_{\cal R}(-s)U_0^{-1}\nonumber\\
  &=&\Omega^{\Delta}(x[s_0])U_0U_{\cal R}(s)\phi(X[s_0])U_{\cal R}(-s)U_0^{-1}\nonumber\\
  &=&\Omega^{\Delta}(x[s_0])U_0\phi(X[s_0+s])U_0^{-1}\nonumber\\
    &=&\Omega^{\Delta}(x[s_0])\Omega^{-\Delta}(x[s_0+s])\phi(x[s_0+s]).
\end{eqnarray}
The pre-factor $\Omega^{\Delta}(x[s_0])\Omega^{-\Delta}(x[s_0+s])$ evaluated on the surface $t=0$ is equal to $1$ at the order $\delta s$ and hence it does not contribute to the infinitesimal shift  (\ref{infD}). The modular Hamiltonian is therefore given indeed by (\ref{HB}).
\subsubsection{Hyperbolic energy}  The hyperbolic energy is the expectation value of the modular Hamiltonian. And we are interested in the first order variation in the $\zeta$ of the hyperbolic energy.

If we assume that the ball shaped region $B$ of radius $R$ is centered at $\vec{x}_0$ then the hyperbolic energy must be given by 

 \begin{eqnarray}
E_{B}^{\rm Hyp}=2\pi \int_B d^{d-1}x \frac{R^2-(\vec{x}-\vec{x}_0)^{2}}{2R} \langle T^{00}(x)\rangle.
 \end{eqnarray}
 For an infinitesimal ball $B$ we can safely replace $ \langle T^{00}(x)\rangle$ with  $ \langle T^{00}(x_0)\rangle$. We can then easily compute
 \begin{eqnarray}
   \delta E_B&=&\frac{d}{d\zeta}{\Big|}_{\zeta=0}E_{B_{\rm inf}}^{\rm Hyp}\nonumber\\
   &=&2\pi \int_B d^{d-1}x \frac{R^2-(\vec{x}-\vec{x}_0)^{2}}{2R} \frac{d}{d\zeta}{\Big|}_{\zeta=0}\langle T^{00}(x)\rangle\nonumber\\
   &=&2\pi  \frac{d}{d\zeta}{\Big|}_{\zeta=0}\langle T^{00}(x_0)\rangle\int_B d^{d-1}x \frac{R^2-(\vec{x}-\vec{x}_0)^{2}}{2R}\nonumber\\
   &=&\frac{2\pi R^d S_{d-2}}{d^2-1} \frac{d}{d\zeta}{\Big|}_{\zeta=0}\langle T^{00}(x_0)\rangle,
 \end{eqnarray}
where $S_{d-2}=\int d\Omega_{d-2}$ is the area of the unit sphere ${\bf S}^{d-2}$.

However, by using the first law of thermodynamics, the holographic interpretation of the entanglement entropy and the Ryu-Takayanagi formula we obtain
 \begin{eqnarray}
   \frac{d}{d\zeta}{\Big|}_{\zeta=0}E_{B_{\rm inf}}^{\rm Hyp}&=& \frac{d}{d\zeta}{\Big|}_{\zeta=0}S_{B_{\rm inf}}\nonumber\\
   &=&{\rm lim}_{R\longrightarrow 0}\delta S_B\nonumber\\
   &=&{\rm lim}_{R\longrightarrow 0}\frac{\delta A_B}{4G_N}.
 \end{eqnarray}
 We employ now the  holographic result (\ref{Result0}) to obtain 

 \begin{eqnarray} 
   \frac{d}{d\zeta}{\Big|}_{\zeta=0}\langle T^{00}(x_0)\rangle=\frac{L^{d-3}d}{16\pi G_N}\delta^{ij}h_{ij}(x_0,z=0).
 \end{eqnarray}
 We have thus another crucial holographic result given by
  \begin{eqnarray} 
   \frac{d}{d\zeta}{\Big|}_{\zeta=0}\langle T^{00}(x)\rangle=\frac{L^{d-3}d}{16\pi G_N}\delta^{ij}h_{ij}(x,z=0).\label{Result3}
  \end{eqnarray}
  By substituting back into the variation of the hyperbolic energy for a ball shaped region $B$ of radius $R$ we obtain the holographic interpretation of the second side of the first law of thermodynamics, viz
  \begin{eqnarray}
   \delta E_B
  &=&\frac{L^{d-3}d}{16 G_N}\int_B d^{d-1}x \frac{R^2-(\vec{x}-\vec{x}_0)^{2}}{R} \delta^{ij}h_{ij}(x,z=0).\label{Result2}
  \end{eqnarray}

  \subsection{Einstein's equations}

In the remainder we will consider a  ball shaped region $B$ of radius $R$ centered around the origin, i.e. $\vec{x}_0=0$. This is a spatial surface at the time slice $t=0$.
  
The first law of thermodynamics in the CFT is given by 
  \begin{eqnarray}
  \frac{d}{d\zeta}S_B   &=&\frac{d}{d\zeta}E_B^{\rm Hyp}.
  \end{eqnarray}
  The holographic interpretation of the two sides of this equation is given by the two equations (\ref{Result1}) and (\ref{Result2}), viz 
\begin{eqnarray}
\frac{d}{d\zeta}S_B\equiv \delta S_B=\frac{L^{d-3}R}{8G_N}\int_{\tilde{B}} d^{d-1}x\bigg(\delta^{ij}-\frac{1}{R^2}x^ix^j\bigg)h_{ij}(x,z),
\end{eqnarray}
where $\tilde{B}$ is the extremal surface in the bulk such that $\partial \tilde{B}=\partial B$ and $z$ should be understood as $z=\sqrt{R^2-\vec{x}^2}$. And
\begin{eqnarray}
  \frac{d}{d\zeta}E_B^{\rm Hyp}\equiv  \delta E_B
  &=&\frac{L^{d-3}d}{16 G_N}\int_B d^{d-1}x \frac{R^2-r^{2}}{R} \delta^{ij}h_{ij}(x,z=0).
\end{eqnarray}
The first law of thermodynamics in the AdS is then given by the non-local constraints
\begin{eqnarray}
  \int_{\tilde{B}} d^{d-1}x\bigg(R^2\delta^{ij}-x^ix^j\bigg)h_{ij}(x,z)=\frac{d}{2}\int_{B} d^{d-1}x (R^2-r^{2}) \delta^{ij}h_{ij}(x,z=0).\label{nlc}\nonumber\\
\end{eqnarray}
Remark that this equation relates the metric perturbation on the boundary $z=0$ to the metric perturbation in the bulk $z\neq 0$. The central claim  is that these non-local equations are precisely the Einstein's equations linearized around AdS   \cite{Lashkari:2013koa,Faulkner:2013ica}.

Also we have seen that the modular Hamiltonian of a  ball shaped region $B$ of radius $R$ evaluated at time $t=0$ is given by the following expression 
 \begin{eqnarray}
H_B=2\pi \int_B d^{d-1}x \frac{R^2-r^{2}}{2R} T^{00}(x)+{\rm constant}.
 \end{eqnarray}
The constant is chosen such that the trace of the density matrix is equal $1$.  The corresponding modular flow (\ref{mf}) is a symmetry which is generated by the conformal timelike Killing vector of Minkowski spacetime given by
 \begin{eqnarray}
\xi=\frac{\pi}{R}\bigg((R^2-t^2-r^2)\partial_t-2tx^i\partial_i\bigg).
 \end{eqnarray}
 We associate to this a bulk Killing vector $\xi_B$ which obviously approaches $\xi$ as we approach the AdS boundary $z=0$. This is given explicitly by
  \begin{eqnarray}
\xi_B=\frac{\pi}{R}\bigg((R^2-z^2-t^2-r^2)\partial_t-2t(z\partial_z+x^i\partial_i)\bigg).
  \end{eqnarray}
  This is indeed a symmetry which preserves the metric, i.e. an isometry, since the Lie derivative of the metric with respect to $\xi_B$ is zero, viz ${\cal L}_{\xi_B}g_{AdS}=0$. The corresponding conserved quantity is the so-called canonical energy associated with the Killing vector $\xi_B$.  This Killing vector $\xi_B$ also vanishes on the Rindler horizon, i.e. on the extremal surface  $\tilde{B}$ in the unperturbed AdS given by $\vec{x}^2+z^2=R^2$.

\begin{figure}[H]
\begin{center}
  \includegraphics[angle=-0,scale=0.6]{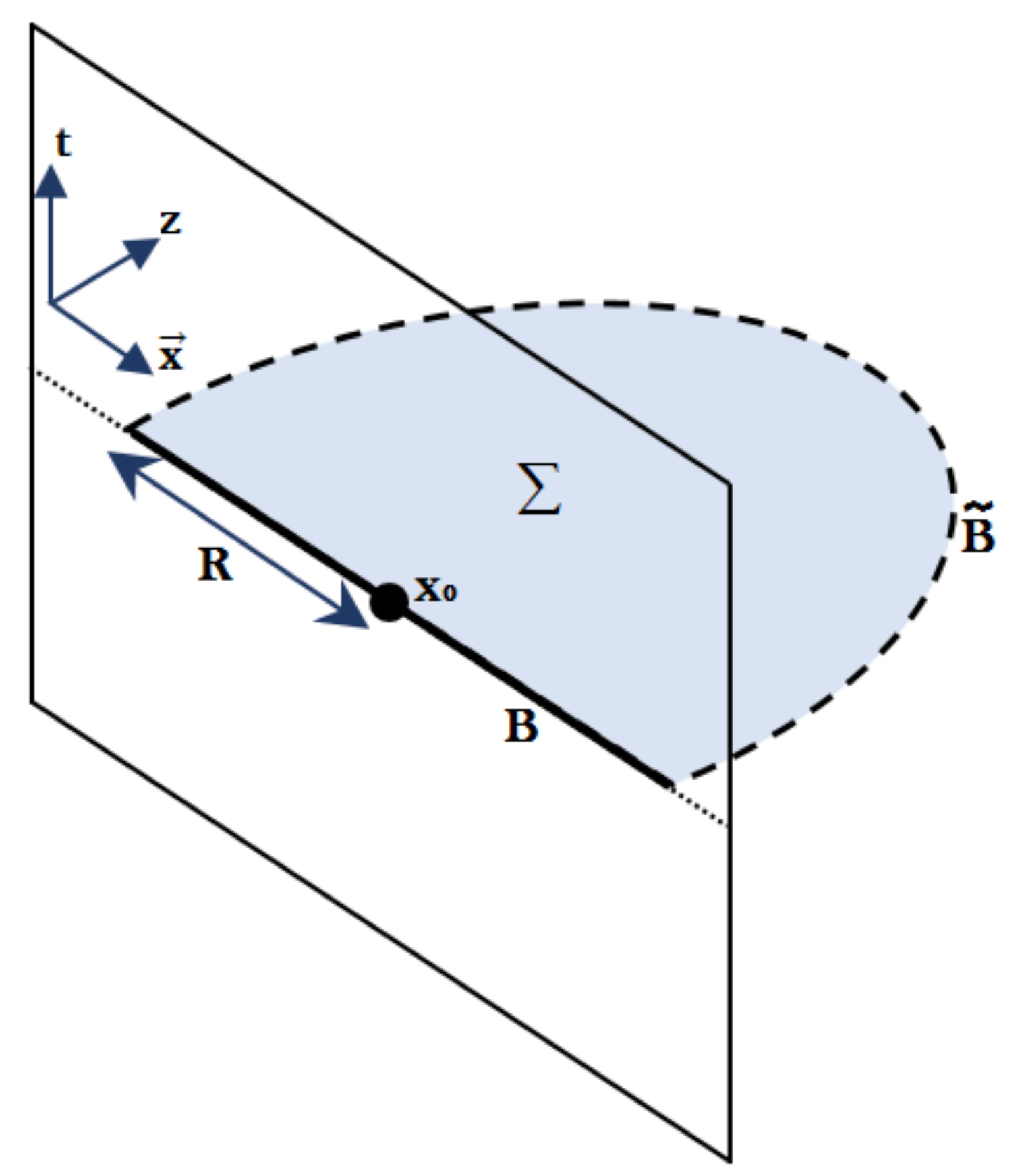}
\end{center}
\caption{The volume  $\Sigma$  bounded by $B\cup\tilde{B}=\partial \Sigma$.}\label{BBtilde}
\end{figure}

Clearly the function $\xi_B$ is defined in the spatial region $\Sigma$ between $B$ and $\tilde{B}$, i.e. $\Sigma$ is the volume bounded by $B\cup\tilde{B}=\partial \Sigma$. In other words, $\Sigma$ is a co-dimension $d-$dimensional surface in the bulk. See figure (\ref{BBtilde}). The volume form on this region $\Sigma$ is given by 
\begin{eqnarray}
 \epsilon_{a_1}=\frac{1}{d!}\epsilon_{a_1a_2...a_{d+1}}dx^{a_2}\wedge...\wedge dx^{a_{d+1}},
\end{eqnarray}
where $\epsilon_{a_1...a_{d+1}}$ is the anti-symmetric tensor in $d+1$ dimensions normalized such that  $\epsilon_{z t x^1...x^{d-1}}=\sqrt{-g}$.  Since the region $\Sigma$ is a spatial surface all components but $\epsilon^{t}$ vanish and as a consequence the volume form is simply given by
\begin{eqnarray}
  {\rm vol}_{\Sigma}=\epsilon^{t}.
\end{eqnarray}

The approach to convert the above non-local constraints (\ref{nlc}) to local equations is similar to the way we convert the integral form of Maxwell's equations to differential forms where we mainly use Stokes' theorem. ُIndeed, the non-local constraints (\ref{nlc}) can be converted into the local Einstein's equations if for every choice of $B$ there exists a differential form $\chi_B$ such that
\begin{eqnarray}
 \delta S_B=\int_{\tilde{B}}\chi_B.
\end{eqnarray}
\begin{eqnarray}
 \delta E_B=\int_{{B}}\chi_B.
\end{eqnarray}
The non-local constraints $\delta E_B=\delta S_B$ become therefore
\begin{eqnarray}
  \int_{{B}}\chi_B=\int_{\tilde{B}}\chi_B& \Rightarrow &\int_{{B}}\chi_B-\int_{\tilde{B}}\chi_B=0\nonumber\\
  & \Rightarrow &\int_{\partial \Sigma}\chi_B=0\nonumber\\
  & \Rightarrow &\int_{ \Sigma}d\chi_B=0.
\end{eqnarray}
The form $\chi_B$ is obtained by means of the Iyer-Wald formalism \cite{Iyer:1994ys}. Explicitly the form $\chi_B$ can be shown to satisfy
\begin{eqnarray}
 d\chi_B=2\xi_B^0\delta E_{00} {\rm vol}_{\Sigma}.
\end{eqnarray}
In this equation $\delta E_{00}$ is time-time component of the Einstein equations linearized about AdS spacetime given explicitly by
\begin{eqnarray}
 \delta E_{00} =\frac{z^d}{2L^2}\bigg(\partial_z^2h_i^i+\frac{d+1}{z}\partial_zh_i^i+\partial_j\partial^j h_i^i-\partial^i\partial^jh_{ij}\bigg).
\end{eqnarray}
Hence, the first law of thermodynamics in the CFT, in which the energy is understood as hyperbolic energy and the entropy is understood as entanglement entropy,  yields under holographic extension, for ball shaped regions in slices of constant time, a set of non-local constraints in the bulk which are precisely the time-time component of the Einstein equations linearized about AdS spacetime, i.e.
\begin{eqnarray}
\delta E_{00}=0.
\end{eqnarray}
To obtain the other components of the Einstein equations  we repeat the same steps for ball shaped regions in general frames of reference. We obtain then
\begin{eqnarray}
 \delta E_{\mu\nu}=0.
\end{eqnarray}
The remaining equations $\delta E_{zz}=\delta E_{z\mu}=0$ are in fact constraints in the dual geometry which are equivalent to the conservation and tracelessness of the energy-momentum tensor $T_{\mu\nu}$ in the CFT at the boundary. This can be verified explicitly by considering infinitesimal balls.

 In summary, we have thus shown that the first law of thermodynamics for ball shaped regions in the CFT  is exactly equivalent to linearized Einstein equations in the gravity dual theory. This is a very concrete mechanism for the emergence of spacetime geometry from quantum entanglement.

\appendix

\section{Exercises}
\paragraph{Exercise $1$:} Show that in anti-de Sitter the proper time to go from the center $r=0$ to the boundary $r=1$ and back is finite equal to $\tau=\pi$. What do you conclude? 
\paragraph{Exercise $2$:}
Show that the boundary $z=0$ of $AdS_{d+1}$ in the Poincare patch becomes ${\bf R}^d$ in the Euclidean rotation whereas the horizon $z=\infty$ shrinks to a point. What do you get by adding the point $z=\infty$ to the boundary ${\bf R}^d$ and what the resulting compactified Euclidean $AdS_{d+1}$?.
\paragraph{Exercise $3$:}
\begin{enumerate}
\item Show that the generators of $SO(2,1)$ in the global coordinates $\tau$, $r$ are given by the differential operators 
\begin{eqnarray}
L_2^0=\partial_{\tau}.
\end{eqnarray}
 \begin{eqnarray}
L_1^0=\sin\tau\sin r \partial_{\tau}-\cos\tau \cos r\partial_r.
\end{eqnarray}
\begin{eqnarray}
L_2^1=-\cos\tau\sin r \partial_{\tau}-\sin\tau \cos r\partial_r.
\end{eqnarray}
\item Verify that they satisfy the algebra 
\begin{eqnarray}
[L_2^0,L_1^0]=-L_2^1~,~[L_2^0,L_2^1]=L_1^0~,~[L_1^0,L_2^1]=L_2^0.
\end{eqnarray}
\item Show that the lowering and raising operators (special conformal and momentum generators) $K$ and $P$ take in the global coordinates $\tau$, $r$ the form
 \begin{eqnarray}
K=\exp(-i\tau)\big(i \sin r \partial_{\tau}-\cos r\partial_r\big).
\end{eqnarray}
\begin{eqnarray}
P=-\exp(i\tau)\big(i \sin r \partial_{\tau}+\cos r\partial_r\big).
\end{eqnarray}
\item Show that the Casimir operator for the conformal group $SO(2,1)$ which is precisely the Klein-Gordon Laplacian is given by
  \begin{eqnarray}
\nabla&=&D^2+\frac{1}{2}(PK+KP)\nonumber\\
&=&\cos^2r(-\partial_{\tau}^2+\partial_r^2).
\end{eqnarray}
\end{enumerate}

\paragraph{Exercise $4$:}
Show that the  diffeomorphisms
\begin{eqnarray}
(\tilde{\sigma}^1, \tilde{\sigma}^2)\longrightarrow (a\tilde{\sigma}^1+b \tilde{\sigma}^2,c\tilde{\sigma}^1+d \tilde{\sigma}^2).
\end{eqnarray}
are compatible with the equivalence relation 
\begin{eqnarray}
(\tilde{\sigma}^1, \tilde{\sigma}^2)\sim (\tilde{\sigma}^1+n, \tilde{\sigma}^2+m)~,~n,m\in {\bf Z}.  
\end{eqnarray}
Determine the domain of the coefficients $a$, $b$, $c$ and $d$. By requiring invertibility and one-to-one determine the group structure corresponding to the above transformations.
\paragraph{Exercise $5$:}
\begin{itemize}
\item Show that 
\begin{eqnarray}
\int 1=\int 2i dz\wedge d\bar{z}=\int 4\tau_2d\sigma^1\wedge d\sigma^2=4\tau_2.
\end{eqnarray}

\item Show that the Fourier transform of the function 
\begin{eqnarray}
f(n^{\prime}r)=\exp\bigg(-2\pi\frac{1}{\tau_2}(n^{\prime}r-\tau_1 nr)^2\bigg)
\end{eqnarray}  
is 
\begin{eqnarray}
\tilde{f}(p)=\sqrt{\frac{\tau_2}{2}}\exp\bigg(2\pi i\tau_1 n rp -\frac{1}{2}\pi\tau_2 p^2\bigg).
\end{eqnarray}  

\item Show that the partition function of the conformal field theory of free bosons on the torus is given by 
\begin{eqnarray}
\int {\cal D}\Phi\exp(-S[\Phi])&=&(q\bar{q})^{-c/24}tr q^{L_0}\bar{q}^{\bar{L}_0}.
\end{eqnarray}   
Verify its modular invariance under the modular transformation $\tau\longrightarrow -1/\tau$.
\end{itemize}

\paragraph{Exercise $6$:}

Show that the dilatation operator, the translation and Lorentz generators, and the special conformal generators are represented on scalar fields with scaling dimension $\Delta$ by the differential operators
\begin{eqnarray}
D\Phi=-i(x^{\mu}\partial_{\mu}+\Delta)\Phi.
\end{eqnarray}
\begin{eqnarray}
P_{\mu}\Phi=-i\partial_{\mu}\Phi.
\end{eqnarray}
\begin{eqnarray}
M_{\mu\nu}\Phi=i(x_{\mu}\partial_{\nu}-x_{\nu}\partial_{\mu})\Phi+S_{\mu\nu}\Phi.
\end{eqnarray}
\begin{eqnarray}
K_{\mu}\Phi=(-2i\Delta x_{\mu}-x^{\nu}S_{\mu\nu}-2ix_{\mu}x^{\nu}\partial_{\nu}+ix^2\partial_{\mu})\Phi.
\end{eqnarray}
Work for simplicity in three dimensions. Show that the transformation law of the scalar field $\Phi$ under conformal transformations is given by 
\begin{eqnarray}
\Phi(x)\longrightarrow \Phi^{\prime}(x^{\prime})=|\frac{\partial x^{\prime}}{\partial x}|^{-\Delta/d}\Phi(x).
\end{eqnarray} 

\paragraph{Exercise $7$:}
Show that under special conformal transformation we have the behavior 
\begin{eqnarray}
x^{\mu}\longrightarrow x^{\prime \mu}=\frac{x^{\mu}+b^{\mu}x^2}{1+2bx+b^2x^2}\Rightarrow 
 r_{12}^{\prime 2}=\frac{r_{12}^2}{(1+2bx_1+b^2x^2_1)(1+2bx_2+b^2x^2_2)}.
\end{eqnarray} 
\begin{eqnarray}
|\frac{\partial x^{\prime}}{\partial x}|=\frac{1}{(1+2bx+b^2x^2)^d}.
\end{eqnarray} 
\paragraph{Exercise $8$:}
Show that the requirement of invariance under translations, rotations, scalings and special conformal transformations constrains the three-point function of a conformal field theory such that
\begin{eqnarray}
\langle \Phi_1(x_1)\Phi_2(x_2)\Phi_3(x_3)\rangle=\frac{C_{123}}{r_{12}^{\Delta_1+\Delta_2-\Delta_3}r_{13}^{\Delta_1+\Delta_3-\Delta_2}r_{23}^{\Delta_2+\Delta_3-\Delta_1}}.
\end{eqnarray} 

\paragraph{Exercise $9$:} Show that for a scalar field in Euclidean $AdS_{d+1}$ the canonical momentum with respect to $z$ is given by
\begin{eqnarray}
      \Pi=\eta \sqrt{g}\phi g^{zz}\partial_z\phi.
\end{eqnarray}
\paragraph{Exercise $10$:} Show that the required counter term to renormalize the action (\ref{rt1}) is given by

\begin{eqnarray}
  S^{}_{\rm ct}
  &=&\frac{\eta}{2}\eta_1 \int_{} \sqrt{\gamma}d^dx \phi^2\nonumber\\
  &=&\frac{\eta}{2}\eta_1 L^{d}\int_{} \frac{d^{d}k}{(2\pi)^d}\bigg(A(k)A(-k)\epsilon^{-2\nu}+2 A(k)B(-k)\bigg).
\end{eqnarray}
Determine the value of $\eta_1$.

\paragraph{Exercise $11$:} Show that the two-point function (\ref{rt5}) takes in position space the form
\begin{eqnarray}
\langle{\cal O}(x){\cal O}(0)\rangle=\frac{2\nu L^{d-1}\eta}{\pi^{d/2}}\frac{\Gamma(\frac{d}{2}+\nu)}{\Gamma(-\nu)}\frac{1}{|x|^{2\Delta}}.
\end{eqnarray}
Use the identity
\begin{eqnarray}
\int \frac{d^dk}{(2\pi)^d}\exp(ikx)k^n=\frac{2^n}{\pi^{d/2}}\frac{\Gamma(\frac{d+n}{2})}{\Gamma(-\frac{n}{2})}\frac{1}{|x|^{d+n}}.
\end{eqnarray}

\paragraph{Exercise $12$:} 
Show that  $\ln tr q^{L_0+\bar{L}_0}$ admits an expansion in positive powers of $q$ due to the requirement of positive dimensions of all fields (locality). Show then that the contributions from  $\ln tr q^{L_0+\bar{L}_0}$ to the entanglement entropy are exponentially suppressed. Recall that $q=\exp(-\kappa L)$.

\paragraph{Exercise $13$:}

The thermofield double state $|\psi\rangle$ dual to the  Schwarzschild-AdS black hole is given by 
\begin{eqnarray}
  |\Psi\rangle=\frac{1}{\sqrt{Z(\beta)}}\sum_i\exp(-\beta E_i/2)|E_i^A\rangle\otimes|E_i^B\rangle.
\end{eqnarray}
Compute the reduced density matrix for the subsystem $Q_A$ and its entanglement entropy. Show that in the limit $\beta\longrightarrow\infty$ we have the behavior 
\begin{eqnarray}
  |\Psi\rangle=\rangle\longrightarrow |\Phi\rangle=|E_0^A\rangle\otimes|E_0^B\rangle.
\end{eqnarray}

\paragraph{Exercise $14$:}
 The entanglement entropy of a spatial region $B$ in the CFT is equal to the von Neumann entropy of the reduced density matrix
\begin{eqnarray}
\rho_B=tr_{\bar{B}}|\psi(\zeta)\rangle\langle\psi(\zeta)|.
\end{eqnarray}
Show that
\begin{eqnarray}
  \frac{d}{d\zeta}S_B=-tr_B\frac{d}{d\zeta}\rho_B.\log \rho_B.
\end{eqnarray}

\paragraph{Exercise $15$:}
\begin{enumerate}
\item Show that
\begin{eqnarray}
  \frac{\delta  A_B(g^0,X_{\rm ext})}{\delta X^a}\delta X^a
\end{eqnarray}
is of order $\delta g^2$.
\item Show that the variation  in the area of the extremal surface is given in terms of the variation of the induced metric by
\begin{eqnarray}
\delta A_B(g,X_{\rm ext})=\int d^{d-1}\sigma\sqrt{{\rm det}\gamma_{\mu\nu}^0}(\frac{1}{2}\gamma_0^{\rho\lambda}\delta \gamma_{\rho\lambda}).
\end{eqnarray}
\item Show that in the radial gauge the variation in the induced metric given by 
\begin{eqnarray}
  \delta \gamma_{ij}=z^{d-2}h_{ij}.
\end{eqnarray}
\item Verify that the unperturbed induced metric is given by
\begin{eqnarray}
  \gamma_{ij}^0
  &=&\frac{L^2}{z^2}(\delta_{ij}+\frac{x_ix_j}{z^2}).
\end{eqnarray}
\item Show that the variation  in the area of the extremal surface is given by
\begin{eqnarray}
\delta A_B=\frac{L^{d-3}R}{2}\int_{\tilde{B}} d^{d-1}x\bigg(\delta^{ij}-\frac{1}{R^2}(x-x_0)^i(x-x_0)^j\bigg)h_{ij}.
\end{eqnarray}

\end{enumerate}

\paragraph{Exercise $16$:}
\begin{enumerate}
\item Verify that under the special conformal transformation (\ref{sct}) the metric transforms as
\begin{eqnarray}
  ds^2=\eta_{\mu\nu}dX^{\mu}dX^{\mu}=\Omega^{-2} \eta_{\mu\nu}dx^{\mu}dx^{\mu}.
\end{eqnarray}

\item ٍShow that the domain of dependence $\{X^+\geq 0\}\cap \{X^-\geq 0\}$ of the Rindler wedge maps under the special conformal transformation (\ref{sct}) to the domain of dependence $\{x^+\leq R\}\cap \{x^-\leq R\}$ of ${\cal D}$ where the null coordinates $x^{\pm}$ are defined by  $x^{\pm}=r\pm t$ where $r=\sqrt{(x^1)^2+...+(x^{d-1})^2}$.

\item Show that the modular flow on the Rindler wedge given by $X^{\pm}(s)=X^{\pm}\exp(\pm 2\pi s)$ transforms under the special conformal transformation (\ref{sct}) to the modular flow on ${\cal D}$ given explicitly by
\begin{eqnarray}
  x^{\pm}(s)=R\frac{R+x^{\pm}-e^{\mp 2\pi s}(R-x^{\pm})}{R+x^{\pm}+e^{\mp 2\pi s}(R-x^{\pm})}.
\end{eqnarray}
\end{enumerate}

\paragraph{Exercise $17$:}
By employ the  holographic result (\ref{Result0}) verify that

  \begin{eqnarray} 
   \frac{d}{d\zeta}{\Big|}_{\zeta=0}\langle T^{00}(x)\rangle=\frac{L^{d-3}d}{16\pi G_N}\delta^{ij}h_{ij}(x,z=0).
  \end{eqnarray}
  
 \paragraph{Exercise $18$:}
\begin{enumerate}
  \item Show that the time-time component of the Einstein equations linearized about AdS spacetime is given  by
\begin{eqnarray}
 \delta E_{00} =\frac{z^d}{2L^2}\bigg(\partial_z^2h_i^i+\frac{d+1}{z}\partial_zh_i^i+\partial_j\partial^j h_i^i-\partial^i\partial^jh_{ij}\bigg).
\end{eqnarray}
Use the Fefferman-Graham metric.
\item By employing the Iyer-Wald formalism show that the form $\chi_B$ satisfies
\begin{eqnarray}
 d\chi_B=2\xi_B^0\delta E_{00} {\rm vol}_{\Sigma}.
\end{eqnarray}
\item Show that the form $\chi_B$ is given explicitly by
  \begin{eqnarray}
    \chi_B{\Big|}_{\Sigma}=\frac{z^d}{16\pi G_N}\bigg\{\epsilon^t_z\bigg(\frac{2\pi z}{R}+\frac{d}{z}\xi^t+\xi^t\partial_z\bigg)h^i_i
    +\epsilon^t_i\bigg[\bigg(\frac{2\pi x^i}{R}+\xi^t\partial^i\bigg)h^j_j-\bigg(\frac{2\pi x^j}{R}+\xi^t\partial^j\bigg)h^i_j\bigg]\bigg\}.\nonumber\\
  \end{eqnarray}
\item Verify explicitly that
\begin{eqnarray}
 \delta S_B=\int_{\tilde{B}}\chi_B.
\end{eqnarray}
\begin{eqnarray}
 \delta E_B=\int_{{B}}\chi_B.
\end{eqnarray}  
\end{enumerate}

\end{document}